\documentclass[11pt]{article}
\pdfoutput=1

\usepackage{cancel,slashed}
\usepackage{bbm}
\usepackage{bm}
\usepackage{amsmath,amsfonts,amssymb,mathtools,nicefrac,nccmath,cases}
\usepackage{graphicx}
\usepackage{cite}
\usepackage{skak}

\usepackage{dsfont}
\usepackage{latexsym}
\usepackage{mathrsfs}
\usepackage{color}
\usepackage{mdframed}
\usepackage{adjustbox}
\usepackage[hyperfootnotes=false,linktocpage]{hyperref}
\usepackage{transparent}
\usepackage[hang,flushmargin]{footmisc}
\usepackage{blkarray}
\usepackage{multirow}
\usepackage{empheq}
\usepackage{comment}
\usepackage{wasysym}
\usepackage{tikzsymbols}

\usepackage{caption}
\usepackage{subcaption}
\usepackage[normalem]{ulem}

\newcommand{\bmat}{\left(\begin{array}}
\newcommand{\emat}{\end{array}\right)}

\def\a {\alpha}
\def\b {\beta}
\def\q {\mathsf{q}}

\def\1{{\bf 1}}
\def\2{{\bf 2}}
\def\3{{\bf 3}}
\def\4{{\bf 4}}
\def\6{{\bf 6}}

\def\targ#1#2{\genfrac{[}{]}{0pt}{}{#1}{#2}}
\def\targ2#1#2{\genfrac{}{}{0pt}{}{#1}{#2}}

\definecolor{mygr}{rgb}{0,0.6,0}
\definecolor{mygrey}{rgb}{0,0.1,0.2}
\definecolor{myblue}{rgb}{0,0.5,0.9}
\definecolor{myblue2}{rgb}{0,0.5,0.5}
\definecolor{myblue3}{rgb}{0,0.7,0.9}
\definecolor{myblue4}{rgb}{0,0.6,0.6}
\definecolor{myorange}{rgb}{1,0.5,0}
\definecolor{mypurple}{rgb}{0.6,0,1}
\definecolor{mygolden}{rgb}{1,0.8,0.2}
\definecolor{mycyan}{rgb}{0,1,1}
\definecolor{mymagenta}{rgb}{1,0,1}
\definecolor{mykiwi}{rgb}{0.8,1,0.5}
\definecolor{mybrown}{cmyk}{0.14, 0.42, 0.56, 0.2}
\definecolor{myturq}{cmyk}{0.99, 0, 0.2, 0.4}
\definecolor{myaubergine2}{cmyk}{0.4, 0.5, 0, 0.1}
\definecolor{myaubergine}{cmyk}{0.6,0.85,0,0}
\definecolor{CycleGreen}{cmyk}{0.52,0,1,0}
\definecolor{CycleBrown}{cmyk}{0, 0.4, 0.9, 0.2}

\DeclareFontFamily{U}{rcjhbltx}{}
\DeclareFontShape{U}{rcjhbltx}{m}{n}{<->rcjhbltx}{}
\DeclareSymbolFont{hebrewletters}{U}{rcjhbltx}{m}{n}

\DeclareMathSymbol{\lamed}{\mathord}{hebrewletters}{108}
\DeclareMathSymbol{\mem}{\mathord}{hebrewletters}{109}
\DeclareMathSymbol{\ayin}{\mathord}{hebrewletters}{96}
\DeclareMathSymbol{\tsadi}{\mathord}{hebrewletters}{118}
\DeclareMathSymbol{\qof}{\mathord}{hebrewletters}{113}
\DeclareMathSymbol{\resh}{\mathord}{hebrewletters}{114}
\DeclareMathSymbol{\pe}{\mathord}{hebrewletters}{112}
\DeclareMathSymbol{\pesofit}{\mathord}{hebrewletters}{80}
\DeclareMathSymbol{\samekh}{\mathord}{hebrewletters}{115}
\DeclareMathSymbol{\tav}{\mathord}{hebrewletters}{116}
\DeclareMathSymbol{\vav}{\mathord}{hebrewletters}{119}
\DeclareMathSymbol{\het}{\mathord}{hebrewletters}{120}
\DeclareMathSymbol{\yod}{\mathord}{hebrewletters}{121}
\DeclareMathSymbol{\zayin}{\mathord}{hebrewletters}{122}
\DeclareMathSymbol{\alephdot}{\mathord}{hebrewletters}{128}
\DeclareMathSymbol{\tsadisofit}{\mathord}{hebrewletters}{90}
\DeclareMathSymbol{\shin}{\mathord}{hebrewletters}{152}

\newcommand{\Mpl}{M_{\textrm{Pl}}}
\def\CL {{\cal L}}
\def\CN {{\cal N}}
\def\CM {{\cal M}}
\def\CK {{\cal K}}

\def\CF {{\cal F}}
\def\CR {{\cal R}}
\def\CQ {{\cal Q}}
\def\CI {{\cal I}}

\def\CP {{\cal P}}

\def\be{\begin{equation}}
\def\ee{\end{equation}}
\def\bea{\begin{eqnarray}}
\def\eea{\end{eqnarray}}
\def\bes{\begin{subequations}}
\def\ees{\end{subequations}}

\def\eps{{\epsilon}}
\def\oh{\frac{1}{2}}

\def\re{\mbox{Re}\, }
\def\im{\mbox{Im}\, }

\def\om{\omega}

\usepackage{multicol}
\usepackage{float}
\usepackage{caption}
\def\p {{\partial}}

\def\g {{\gamma}}
\def\G {{\Gamma}}

\def\CO {{\cal O}}

\newenvironment{eqn}{\begin{equation}\begin{aligned}}{\end{aligned}\end{equation}\noindent}
\newenvironment{eqn*}{\begin{equation*}\begin{aligned}}{\end{aligned}\end{equation*}\noindent}

\makeatletter
\newsavebox\myboxA
\newsavebox\myboxB
\newlength\mylenA

\newcommand*\xoverline[2][0.75]{%
\sbox{\myboxA}{$\m@th#2$}%
\setbox\myboxB\null
\ht\myboxB=\ht\myboxA%
\dp\myboxB=\dp\myboxA%
\wd\myboxB=#1\wd\myboxA
\sbox\myboxB{$\m@th\overline{\copy\myboxB}$}
\setlength\mylenA{\the\wd\myboxA}
\addtolength\mylenA{-\the\wd\myboxB}%
\ifdim\wd\myboxB<\wd\myboxA%
   \rlap{\hskip 0.5\mylenA\usebox\myboxB}{\usebox\myboxA}%
\else
    \hskip -0.5\mylenA\rlap{\usebox\myboxA}{\hskip 0.5\mylenA\usebox\myboxB}%
\fi}
\makeatother

\begin{document}
\pagestyle{plain}

\makeatletter
\@addtoreset{equation}{section}
\makeatother
\renewcommand{\theequation}{\thesection.\arabic{equation}}

\pagestyle{empty}
\rightline{IFT-UAM/CSIC-26-008}
\rightline{EFI-26-1}
\vspace{0.5cm}
\begin{center}
\Huge{{Curvature divergences and gravity decoupling \\ in Calabi--Yau rigid limits}
\\[10mm]}
\Large{Alberto Castellano,$^{1,2}$ Fernando Marchesano$^{3}$ and Lorenzo Paoloni$^{3,4}$}\\[12mm]
\small{
${}^{1}$ Enrico Fermi Institute $\&$ Leinweber Institute for Theoretical Physics, \\ University of Chicago, Chicago, IL 60637, USA \\[2mm]
${}^{2}$ Kavli Institute for Cosmological Physics, University of Chicago, Chicago, IL 60637, USA \\[2mm]
${}^{3}$ Instituto de F\'{\i}sica Te\'orica UAM-CSIC, c/Nicol\'as Cabrera 13-15, 28049 Madrid, Spain \\[2mm] 
${}^{4}$ Departamento de F\'{\i}sica Te\'orica, Universidad Aut\'onoma de Madrid, 28049 Madrid, Spain 
\\[10mm]} 
\small{\bf Abstract} \\[5mm]
\end{center}
\begin{center}
\begin{minipage}[h]{15.0cm} 

Four-dimensional $\CN=2$ supergravity theories become rigid in gravity-decoupling limits. We study this effect for  type II string compactifications on general Calabi--Yau manifolds, focusing on vector-multiplet trajectories whose endpoints exhibit axionic shift symmetries. This comprises field excursions of both finite- and infinite distance, but the latter display specific features due to the appearance of light towers of extremal BPS states, in agreement with Swampland principles. We show that vector multiplets split into two sets: those with gravitational and with rigid mutual interactions, and that only a subset of the latter---dubbed core RFT---can fully decouple from gravity. We characterise the core RFT in terms of the axionic shift symmetry, and derive decoupling criteria based on kinetic and Pauli interaction mixing. Our framework is illustrated in large complex structure, conifold-like, and Seiberg--Witten limits. In the last case, Pauli mixing obstructs decoupling whenever the dyonic and extremal BPS towers appear at the same scale. Across all these examples, the decoupling from gravity is signalled by a divergent moduli-space scalar curvature.

\end{minipage}
\end{center}
\newpage
\setcounter{page}{1}
\pagestyle{plain}
\renewcommand{\thefootnote}{\arabic{footnote}}
\setcounter{footnote}{0}


\tableofcontents

\newpage

\section{Introduction and summary}
\label{s:intro}

One of the most celebrated insights of string compactifications is that they provide a clean,  geometric picture describing the decoupling of field theories from gravity. In this scheme, one first localises the internal wavefuncions of a field theory sector in a particular region of the compactification manifold, using the properties of D-branes, warping and/or the geometry of shrinkable cycles. Then one takes a limit such that all other wavefunctions of the EFT become spread out across the extra dimensions, including that of the graviton. Assuming that EFT couplings are computed by wavefunction overlap, this implies that those that mix localised and non-localised fields tend to zero along the limit, prompting the decoupling. 

While this compelling picture has led to several ground-breaking ideas, it is important to realise that it is of microscopic nature, and moreover only valid in those regimes where 10d or 11d supergravity gives a good approximation. A general,  overarching macroscopic picture of gravity decoupling is still lacking. The standard wisdom indicates that, from the EFT perspective, one should be sending $\Mpl/\Lambda_{\rm QFT} \to \infty$, where $\Lambda_{\rm QFT}$ is the cut-off scale of the decoupling field theory sector. However, it does not tell us anything about the nature of $\Lambda_{\rm QFT}$, nor how to engineer such a limit. In fact, it is not even clear if the above  mechanism based on localisation maps into one or several decoupling schemes when translated into macroscopic terms.

In this sense, 4d $\CN=2$ effective field theories (EFTs) provide an optimal framework for developing such a general, macroscopic understanding of gravity decoupling. On the one hand, this is the setting in which the program of engineering quantum field theories from string compactifications was initially developed   \cite{Kachru:1995fv,Klemm:1996bj,Katz:1996fh}. On the other hand, the  4d EFT Lagrangian structure offers a sharp distinction between $\CN=2$ supergravity theories and $\CN=2$ rigid field theories (RFTs) in which gravity is absent. This distinction is particularly clear in the vector multiplet sector, where the dynamics can be formulated entirely in terms of special Kähler geometry \cite{Seiberg:1994rs,Andrianopoli:1996cm,Katz:1996fh,Lerche:1996ni,Katz:1997eq,Freed:1997dp,Billo:1998yr,Gunara:2013rca,Lauria:2020rhc}, and where several approaches to describing the decoupling of gravity have already been proposed.

In this work, we aim to lay the groundwork for a general effective field theory  description of gravity decoupling. As a laboratory, we consider 4d $\CN=2$ EFTs obtained from compactifying type II string theory on Calabi--Yau (CY) three-folds, focusing on gravity-decoupling limits within their vector multiplet moduli space. Our strategy will be to combine both the machinery of special Kähler geometry and the insights of the Swampland Programme \cite{Vafa:2005ui} (see \cite{Brennan:2017rbf,Palti:2019pca,vanBeest:2021lhn,Grana:2021zvf,Agmon:2022thq,Harlow:2022ich,VanRiet:2023pnx} for detailed reviews) in order to elucidate the general features of this decoupling. More precisely, since many of these limits correspond to trajectories of infinite distance in field space, the physics behind the Swampland Distance Conjecture (SDC) \cite{Ooguri:2006in} plays a central r\^ole in our discussion. Particularly relevant are the techniques on mixed Hodge structures that have been used to prove the conjecture in this setting \cite{Grimm:2018ohb,Grimm:2018cpv,Corvilain:2018lgw} (see also \cite{Hassfeld:2025uoy,Monnee:2025ynn}). The scope of these results in fact goes beyond infinite-distance limits, and encompasses any asymptotic limit which develops an axionic shift symmetry at its endpoint. To this approximate 0-form global symmetry one can associate a monodromy group that acts non-trivially on the EFT gauge sector, and that allows one to organise all couplings of the vector multiplet fields in a perturbative expansion. 

From the viewpoint of special geometry, the standard approach to describe vector multiplet couplings in 4d $\CN=2$ supergravity is in terms of a prepotential $\CF$, which is a homogeneous degree-two function of the projective coordinates $X^I$, $I= (0, i)$ where $i=1,\dots, n_V,$ labels the  vector multiplets \cite{Candelas:1990pi}. In a given  moduli space region one chooses a reference coordinate $X^0$ with a scale $m_*$ associated to it, and describes the EFT Coulomb branch in terms of $z^i = X^i/X^0$. Then $m_*$ and $|z^i| m_*$ are the masses of would-be BPS particles with unit electric charge under the U(1)$_I$ on this branch. In the standard picture of gravity-decoupling \cite{Seiberg:1994rs}, one starts by identifying the electric charge of $X^0$ with that of the graviphoton, and $m_* \simeq \Mpl$. Then one sends this scale to infinity with respect to the rest, or equivalently takes the limit $z^i \to 0$. Consequently, around this point, one should obtain a rigid field theory with prepotential given by $\mathscr{F}(z^i) \equiv \CF/(X^0)^2$. 

While this prescription is suitable to describe RFT limits at finite distance in moduli space, it cannot be directly applied to infinite-distance singularities. The reason is that, as predicted by the Distance Conjecture, there is an infinite tower of states becoming light as one approaches the limit. Such states are BPS and extremal \cite{Gendler:2020dfp,Bastian:2020egp}, which means that the U(1)'s under which they are charged cannot be part of a rigid theory. In other words, Swampland principles imply that certain vector multiplets remain inherently gravitational along infinite-distance degenerations. Therefore, to decouple from gravity, one should decouple from these additional U(1)'s as well. 

Even if this makes certain gravity-decoupling limits more involved, it also hints on how to formulate them  in terms of special geometry. Indeed, one may now identify $m_*$ with the scale of the lightest SDC tower, which is both the mass of an actual BPS state as well as a maximal 4d EFT cut-off scale. This was the choice made in \cite{Marchesano:2023thx} to study infinite-distance limits in the large-volume regime of type IIA CY compactifications, where $m_* = m_{\rm D0}$ was taken to be the D0-brane mass. One direct advantage of this choice is that the standard large-volume prepotential $\CF_{\rm LV}$ is tailored for this reference scale, in the sense that to describe a decoupling RFT with fields $\{z^\mu\} \subset \{z^i\}$ and cut-off $m_*$, one should simply restrict $\mathscr{F}(z^i) \equiv \CF_{\rm LV}/(X^0)^2$ to $\{z^\mu\}$, and treat the rest of fields as non-dynamical parameters. The subset $\{z^\mu\}$ was found to be specified by topological data of the limit, as a set of shrinkable divisors that do not couple to the leading field theory direction. Finally, unless the decoupling RFT displays 4d $\CN=4$ supersymmetry \cite{Marchesano:2024tod}, it was found that the moduli space scalar curvature diverges along the limit. The Curvature Criterion \cite{Marchesano:2023thx} posits that this occurs for all moduli spaces compatible with Quantum Gravity. 

As pointed out in \cite{Castellano:2024gwi}, most of these features extend to other infinite-distance limits, as long as one identifies $m_*$ with the SDC tower scale. For instance, the rigid prepotential $\mathscr{F}_{\rm RFT}$ is again obtained by restricting $\CF/(X^0)^2$ to a subset of fields $\{z^\mu\}$, and the most involved part of the problem is to identify such $\{z^\mu\}$ or, equivalently, those vector multiplets that belong to the decoupling RFT. To discard some, one may look at the charge-to-mass ratio $\gamma_\mathsf{q}$ of BPS particles  across the lattice $\Gamma \ni \mathsf{q}$ of quantised charges. In general, those states with $\gamma_\mathsf{q} = \CO(1)$ along the limit are expected to form a cone of extremal BPS particles \cite{Alim:2021vhs}, which as already said identifies a subsector of U(1)'s as gravitational. Recriprocally, when the charge-to-mass ratios of certain BPS states diverge along the limit it is because there exists a vector multiplet with rigid mutual interactions \cite{Castellano:2024gwi}. However, this does not mean that such a {\em rigid U(1)} decouples from gravity. To do so, it must have negligible kinetic mixing as well as mixing interactions with all the U(1)'s belonging to the gravitational sector. As we discuss in detail, these conditions turn out to be quite restrictive, and prevent many of the rigid U(1)'s from actually decoupling from gravity. 

Nevertheless, just as in the type IIA large-volume case, one would expect a simple recipe of topological nature which selects a subset of rigid U(1) that can decouple form the gravitational sector. As it turns out, such a prescription is provided by the monodromy group associated to the limit. Indeed, the duality group of the supergravity theory acts non-trivially on the graviphoton, rotating the latter among some of the $n_V +1$ quantised U(1)'s in the Coulomb branch surrounding the limit endpoint. A gravity-decoupled U(1) must have no component along the graviphoton direction, which means that it must not belong to the monodromy orbit described by the former. This implies a block-diagonal structure for the monodromy group, when acting both in the space of quantised electro-magnetic U(1)'s and in the lattice  $\Gamma$ of charged particles. The block which rotates the graviphoton contains the gravitational U(1)'s as well as those rigid fields that cannot decouple from them. The remaining block comprises the gauge subsector that can in fact decouple from gravity and which, following \cite{Blanco:2025qom}, we dub as core RFT:
\vspace{0mm}
\begin{mdframed}[leftmargin=1cm, rightmargin=1cm]
\begin{center}
{\em The core RFT is specified by a subset of rigid U(1)'s closed both under \\ electro-magnetic duality and the action of the monodromy group.}
\end{center}
\end{mdframed}
\vspace{2mm}

To see what this means in practice, let us consider the vector multiplet sector of type IIB  compactified on a Calabi--Yau threefold $Y_3$. Along the Coulomb branch, the set of U(1)'s can be described in terms of a symplectic basis of three-cycles $(A^J, B_J) \in H_3(Y_3,\mathbb{Z})$ such that, e.g., $A^J$ can be identified with the electric fields and $B_J$ with their magnetic duals. Closure under electro-magnetic duality means that we consider a subset of pairs $\{A^\lambda, B_\lambda\}_{\lambda \in \Lambda}$. Rigidity of a given pair $(A^\lambda, B_\lambda)$ means that the mutual couplings of their U(1)'s are described by a rigid Lagrangian or, equivalently, that the charge-to-mass ratios of 4d particles made from D3-branes wrapping both of these cycles blow up along the limit. Finally, closure under monodromy means that the lattice $\Gamma_{\rm cRFT} \equiv \langle \{A^\lambda, B_\lambda\}_{\lambda \in \Lambda}\rangle \subset H_3(Y_3,\mathbb{Z})$ is invariant under the monodromy group action on three-cycles. From these properties, one deduces the block-diagonal structure of the monodromy group, with respect to both $\Gamma_{\rm cRFT}$ and its symplectic complement. 

We show that this definition of the core RFT correctly reproduces the gravity-decoupling limits identified in \cite{Marchesano:2023thx,Marchesano:2024tod}, by analysing the mirror-dual large complex structure (LCS) points in type IIB string theory. Beyond this regime, we apply our framework to investigate both finite- and infinite-distance limits away from the LCS singularity. As stressed in \cite{Bastian:2021eom,Bastian:2021hpc}, such limits typically involve metric essential instantons---exponential corrections that must be incorporated in order to prevent singular behaviour in the EFT couplings. We argue that metric essential instantons are intimately connected to RFT limits. In particular, their presence implies the existence of BPS states whose masses are parametrically below the reference scale $m_*$ along the limit, together with a charge-to-mass ratio that diverges. Moreover, the data encoded by these instantons allows one to construct a lattice $\Gamma_{\rm ess}$ satisfying all the defining properties of a core RFT, thereby suggesting the existence of a gravity-decoupling sector.

We illustrate these concepts through explicit examples of conifold, coni–LCS, and Seiberg--Witten (SW) limits, all of which exhibit metric essential instantons and a core RFT built from them. In each case, the absence of kinetic mixing between the core RFT and the rest of the EFT arises automatically, a feature that becomes manifest upon choosing a symplectic frame adapted to the core RFT definition. In contrast, decoupling at the level of interactions is more subtle, as revealed by a careful analysis of SW limits. We find that, along certain limits, the SW sector fails to decouple from the rest of the EFT at the level of Pauli interactions. Notably, these trajectories realise those regimes wherein the moduli space curvature does not diverge, in precise agreement with the Curvature Criterion. The underlying reason is that, in these asymptotic regimes, the scalar curvature can be interpreted as a sum of physical Pauli couplings, revealing a direct correlation between curvature divergences and the Pauli hierarchies  required for decoupling. 

Remarkably, the decoupling of the Seiberg--Witten sector can also be understood in terms of towers of states. More precisely, one should compare the tower of SW dyonic states, associated to the scale $\Lambda_g$, with the species cut-off $\Lambda_{\rm sp}$, which in this case corresponds to an emergent heterotic string scale \cite{Lee:2019wij}. When the quotient $\Lambda_{\rm sp}/\Lambda_g$ diverges it is because the dyonic tower, which represents the rigid UV completion of the SW sector, lies parametrically below the species scale, and then the decoupling takes place. In spite of this suggestive picture, one should stress that a divergence in $\Lambda_{\rm sp}/\Lambda_g$ can only be thought as a sufficient condition for decoupling, but not a necessary one. Indeed, LCS emergent string limits display both $\Lambda_{\rm sp}/\Lambda_g \sim \CO(1)$ as well as gravity decoupling, whenever the core RFT is the dimensional reduction of a Little String Theory. In this regard, notice that according to the Curvature Criterion a divergence in the scalar curvature is also a sufficient condition for decoupling. Adding our results, it should, however, also become necessary if we exclude core RFTs with extended $\CN=4$ supersymmetry. 

The rest of the paper is organised as follows. In section \ref{s:CYlimits}, we review asymptotic limits in the vector multiplet moduli space of type IIB CY compactifications, as well as how to describe rigid limits along the lines of \cite{Castellano:2024gwi}. We discuss the necessary  conditions for a rigid theory to decouple from the rest of the EFT, as well as how they relate to the moduli space curvature. We argue that metric essential instantons naturally engineer rigid limits where gravity decoupling typically occurs, which motivates the definition of core RFT in terms of monodromy data. In section \ref{s:LCS}, we apply such a definition to large complex structure limits, and reinterpret previous results in the literature from our current perspective. In section \ref{s:conifoldlikesingus}, we analyse conifold and coni-LCS singularities, which contain metric essential instantons. We show that the core RFT always decouples from gravity and the scalar curvature diverges. In section \ref{s:swpoints}, we study in detail two explicit examples of Seiberg--Witten limits. Here the core RFT coincides with the SW sector, as expected, but its decoupling from the rest of the EFT turns out to be non-trivial. To detect when the decoupling happens one may either consider divergences in the scalar curvature, in agreement with the Curvature Criterion, or compare the species scale to the SW tower of dyons, which must be parametrically below the former. We finally draw our conclusions in section \ref{s:conclu}.

Several technical details have been relegated to the appendices. Appendix \ref{ap:special} contains the background on special geometry necessary for the discussion in the main text. Appendix \ref{ap:SWp}, on the other hand, includes some complementary material concerning Seiberg--Witten limits. 


\section{Calabi--Yau rigid limits and essential instantons}
\label{s:CYlimits}

As stressed in \cite{Grimm:2018ohb,Grimm:2018cpv,Corvilain:2018lgw}, 
finite- and infinite-distance limits in type II Calabi--Yau vector multiplet moduli spaces can be classified through a unified framework, which highlights the r\^ole of monodromies and axionic shift symmetries around such moduli space singularities. In this section, we consider limits of this sort which are rigid, meaning that they contain a rigid field theory subsector: a subset of vector multiplets whose  mutual couplings are asymptotically governed by a 4d $\CN=2$ rigid Lagrangian. Such limits have been described in \cite{Marchesano:2023thx,Marchesano:2024tod,Castellano:2024gwi} within the large complex structure/volume regime. To extend this analysis to other regions one must consider limits with  essential instantons 
\cite{Bastian:2021eom,Bastian:2021hpc}, while at the same time use the machinery of special geometry and the  significance of special coordinates in the physics of BPS particles. With all these ingredients at hand, we will propose a general description in terms of the monodromy action for what we dub the \emph{core RFT}, which is the subsector of the RFT that can actually decouple from gravity. As will be illustrated later on with explicit examples, that the core RFT decouples from gravity is not automatic, and it is typically captured by a divergence in the moduli space curvature.

\subsection{The vector multiplet moduli space and its curvature}

Let us consider type II string theory compactified on a Calabi--Yau threefold. The Lagrangian describing the vector multiplet sector of the resulting 4d ${\cal N}=2$ EFT reads \cite{Ferrara:1988ff,Andrianopoli:1996cm,Lauria:2020rhc}
\begin{eqn}
\label{SVM}
S_{\rm 4d}^{\rm VM}  =  \frac{1}{2\kappa_{4}^2} \int_{\mathbb{R}^{1,3}} R \star \mathbbm{1} - 2 g_{i\bar{j}} dz^i \wedge \star d\bar{z}^{\bar{j}}  +  \frac{1}{2} \int_{\mathbb{R}^{1,3}} \CI_{IJ} F^I \wedge \star F^J + \CR_{IJ} F^I \wedge F^J\,  ,    
\end{eqn}
where similar conventions to \cite{Freedman_VanProeyen_2012} are being followed. The complex fields $z^i$ parametrise the EFT vector multiplet moduli space, with $g_{i\bar{j}}$ a special K\"ahler metric. For type IIB string theory compactified on a Calabi--Yau $Y_3$, the K\"ahler potential for this metric is defined in terms of its holomorphic three-form $\Omega$ as \cite{Strominger:1990pd}
\be
\kappa_4^2 K = - \log \left(i\int_{Y_3} \Omega (z) \wedge \bar{\Omega} (\bar{z})  \right)\, .
\label{Kahlerpot}
\ee
Similarly, one can describe the rest of the couplings in terms of the periods of $\Omega$ and their derivatives. In practice, it is more convenient to do so using special coordinates, which at any given local patch can always be found for a choice of symplectic basis of three-cycle classes $(A^I, B_J) \in H_3(Y_3,\mathbb{Z})$ \cite{Craps:1997gp}. Such a basis satisfies $A^I \cdot A^J = B_I \cdot B_J = 0$, and $A^I \cdot B_J = \delta^I_J$,  $I = (0,i) = 0, 1, \dots,  h^{2,1}(Y_3) = n_V$, with its Poincar\'e dual basis of harmonic three-forms $(\beta^I, \alpha_J)$ verifying $\int_{Y_3} \alpha_J \wedge \beta^I  = \delta^I_J$.\footnote{\label{fnote:PoincareDual}Our conventions for the Poincaré dual $\text{PD}\, [a]$ of a three-cycle $a \in H_3(Y_3, \mathbb{R})$ are such that 
\be
\int_{a} \omega = \int_{Y_3} \omega \wedge \text{PD}\, [a] \, ,\notag
\label{alternativePoincaredual}
\ee
for any closed three-form $\omega$.} Upon expanding $\Omega$ on the latter, one finds
\be
\Omega(z) = X^I (z) \alpha_I - \CF_J (z) \beta^J\, , 
\label{Omega}
\ee
with $\CF_I = \p_{X^I} \CF$. Here $\CF$ is a holomorphic, homogeneous degree-two function of the periods $X^I = \int_{A^I} \Omega$, dubbed the prepotential, which encodes all the couplings of the Lagrangian \eqref{SVM}. 

In this context, special coordinates are defined as $z^i = X^i/X^0$, and parametrise the vector multiplet moduli space. The corresponding metric takes the form
\be
g_{i\bar{j}} =  k_i k_{\bar{j}} + 2|X^0|^2 e^{K} \im \CF_{ij}\, ,
\label{IIBmetric}
\ee
where $\CF_{IJ} = \p_{X^I} \p_{X^J} \CF$, and we have introduced
\be
k_i = -i e^K \int_{Y_3} \p_{z^i} \Omega \wedge \bar{\Omega}\, .
\label{monok}
\ee
Notice that the factors of $X^0$ cancel out, so that \eqref{IIBmetric} can be seen as a function of $(z^i, \bar{z}^{\bar{j}})$. The rest of the couplings in \eqref{SVM} read
\begin{align}
\mathcal{N}_{IJ} &\equiv \CR_{IJ} + i \CI_{IJ} = \bar{\CF}_{IJ} + 2i\, \frac{\im \CF_{IK}\im \CF_{JL} X^K X^L}{\im \CF_{MN} X^M X^N }  \, ,
\label{calNIJ}
\end{align}
where $\CR_{IJ}, \CI_{IJ} \in \mathbb{R}$ should be again understood as functions of $(z^i, \bar{z}^{\bar{j}})$. Of particular importance for our discussion will be the supergravity gauge kinetic function $\CI_{IJ}$, which encodes the behaviour of the different gauge couplings along each limit. Its rigid field theory version is \cite{Cecotti:2015wqa}
\begin{equation}
    N_{IJ} \equiv 2\im \CF_{IJ} = 2\im \p_{X^I} \p_{X^J} \CF\, ,
    \label{NIJ}
\end{equation}
which will also play a non-trivial r\^ole in the following. Finally, the vector multiplet moduli space curvature in Planck units reads \cite{Strominger:1990pd,Andrianopoli:1996cm}
\be
R = - 2 n_V(n_V + 1) + 2 |X^0|^6 e^{2K} g^{i\bar{j}}g^{k\bar{l}}g^{m\bar{n}}  \CF_{ikm} \bar{\CF}_{\bar{j}\bar{l}\bar{n}}\, ,
\label{IIBscalar}
\ee
where $\CF_{IJK} \equiv \p_{X^I} \p_{X^J} \p_{X^K} \CF$. As discussed in Appendix \ref{ap:special}, one can alternatively express the curvature in terms of \eqref{NIJ} as
\be
R = - 2 n_V(n_V + 1) + 2 e^{-K} N^{IJ}N^{KL}N^{RS}\mathcal{F}_{IKR}\bar{\mathcal{F}}_{\bar J\bar L\bar S}\, .
\label{IIBscalarN}
\ee
Clearly $R (z,\bar{z})$ is bounded from below, and the only way in which one can generate a divergence is when some term(s) within the sum in the right hand side of \eqref{IIBscalar} blow up. One may for instance consider points where the cubic derivative of the prepotential $\CF_{IJK}$ develops poles, and/or the field space metric degenerates. Alternatively, \eqref{IIBscalarN} suggests to consider strong coupling singularities, or instead limits where some gauge couplings remain constant and  $e^{K}|X^0|^2 \to 0$. 

This last scenario was the one explored in \cite{Marchesano:2023thx}, in the mirror context to  large complex structure (LCS) infinite-distance limits. It was found that curvature divergences are always sourced by a term of the form $R_{\rm rigid}e^{-K}|X^0|^{-2}$, where $R_{\rm rigid}$ is the  curvature of a  vector multiplet subsector that asymptotes to a rigid theory along the limit, and decouples from gravity. This motivated the Curvature Criterion \cite{Marchesano:2023thx}, which posits that infinite-distance curvature divergences signal the existence of a gravity-decoupling EFT sector. In this paper we will argue that a very different kind of vector multiplets limits, namely those which feature metric essential instantons \cite{Bastian:2021eom,Bastian:2021hpc}, also obey the Curvature Criterion. For this we first need to link metric essential instantons to the 4d $\CN=2$ rigid field theory limits defined in \cite{Castellano:2024gwi}, which we now turn to discuss. 

\subsection{Rigid limits}
\label{ss:rigid}

One of the advantages of the special coordinates is that they give us a direct connection to the notion of rigid field theory subsectors of the EFT. Indeed, the vector-multiplet Lagrangian of a rigid 4d $\CN=2$ field theory is also specified by some holomorphic function $\mathscr{F}(z^\mu)$ \cite{Grimm:1977xp,Breitenlohner:1981sm}. In this case the gauge kinetic function is simply given by $\mathscr{F}_{\mu\nu} \equiv \p_{z^\mu} \p_{z^\nu} \mathscr{F}$, whereas the K\"ahler potential reads $\mathscr{K} = 2\im \left[ \bar{z}^{\bar{\mu}} \p_{z^\mu} \mathscr{F} \right]$, from which the rigid relation $G_{\mu\bar{\nu}} = 2 \im \mathscr{F}_{\mu\nu}$ follows. 

Following \cite{Castellano:2024gwi}, one can identify an RFT subsector of a 4d $\CN=2$ supergravity EFT by considering the subset of special coordinates  $\{ z^\mu \} \subset \{z^i \}$ for which the (hermitian) condition\footnote{\label{fnote:hermitiancondition}This and other relations of the same kind that appear throughout this work should be understood to hold when contracted with any pair of (complex conjugate) vectors along the indicated field space directions.}
\be
 e^{-K} |X^0|^{-2} k_\mu k_{\bar{\nu}} \ll N_{\mu\nu}\, ,
\label{cond}
\ee 
holds. If one then expresses the kinetic terms for $\{ z^\mu \}$ in units of the following reference scale 
\be
m_* = e^{K/2} |X^0| \Mpl\, ,
\label{mstar}
\ee
with $\Mpl = \sqrt{8\pi}/\kappa_4$, one recovers a rigid metric. Moreover, for the kind of limits that we are interested in this work, and which develop a continuous axionic shift symmetry at their endpoint, one can show that \eqref{calNIJ} may be approximated by
\begin{eqn}
    \CN_{ij} \simeq \bar{\CF}_{ij}-i \varepsilon|X^0|^{-2}e^{-K} k_{i} k_{\bar{j}}\, ,
    \label{calNlimit}
\end{eqn}
when projected along the U(1) directions $I,J=i,j\neq 0$. Notice that in the previous expression we introduced the quantity $\varepsilon$, which can be argued to be a moduli-dependent object whose complex modulus is, in addition, upper bounded by 1; see Appendix \ref{ap:special} for details on this point. In particular, in all the examples considered in this work, we find that it always asymptotes to (at most) a constant along the corresponding limit.

Therefore, due to \eqref{calNlimit}, we have that $\CN_{\mu\nu} \simeq  \bar{ \CF}_{\mu\nu}$ is attained whenever \eqref{cond} holds. That is, we find that the slice of moduli space that corresponds to the field directions $\{ z^\mu \}$ is described by a 4d $\CN=2$ Lagrangian with rigid kinetic terms and a cut-off scale $m_*$. We dub such limits as rigid or RFT limits \cite{Castellano:2024gwi}.

Note that a rigid Lagrangian for the field subsector $\{z^\mu\}$ describes rigid mutual interactions, but it does not guarantee that such a component decouples from gravity nor from other vector multiplets. The limit \eqref{cond} is just a kinematic statement about a slice of moduli space. It does not prevent kinetic mixing with other sectors of the EFT, nor suppresses interactions with them. In order to discuss those additional conditions that are necessary for an actual decoupling, it is useful to first reconsider the meaning of RFT limits in terms of charged BPS particles.

\subsection{RFT limits and BPS particles}

Since the condition \eqref{cond} is covariant, one could in principle check it in any kind of coordinate system. However, it becomes particularly significant when evaluated for special coordinates, since then each index $\mu$ is related to a quantised three-form period, and as such it is directly connected to the physics of BPS particles and the U(1) fields under which they are charged. Indeed, in our setup a BPS particle is made of a D3-brane wrapping a special Lagrangian three-cycle $\Sigma_\mathsf{q}$, which homologically is an integer linear combination of the basis elements $(A^I, B_J)$:
\begin{eqn}
    [\Sigma_\mathsf{q}] = q_I [A^I] - p^I [B_I] \in H_3(Y_3, \mathbb{Z})\, .
    \label{Piq}
\end{eqn}
Here $\mathsf{q} = (p^I, q_I)^{\mathsf{T}} \in \mathbb{Z}^{2 +2n_V}$ is the period vector of (minus) the Poincaré dual of $[\Sigma_\mathsf{q}]$ and it contains the quantised gauge charges of the particle under the field strengths $(F^I, G_I)$, with $G_I \equiv \delta \CL / \delta F^I$ the magnetic dual of $F^I$ \cite{Ceresole:1995ca}. The intersection of two three-cycles is
\begin{eqn}
[\Sigma_{\mathsf{q}}] \cdot [\Sigma_{\mathsf{q}'}] = \mathsf{q}^{\mathsf{T}} \eta\, \mathsf{q}'\, ,\qquad  \eta = \begin{pmatrix}
0  & \mathbb{I}  \\ -\mathbb{I} & 0
\end{pmatrix}\, , 
\label{eta}
\end{eqn}
in terms of which one can express the central charge $Z_\mathsf{q}$ and mass of a BPS particle as
\be
m_\mathsf{q} = |Z_\mathsf{q}|\, \Mpl\, , \quad Z_\mathsf{q} = \int_{[\Sigma_{\mathsf{q}}]} \Omega(z) = e^{\kappa_4^2 K/2} \left( q_I X^I- p^I \CF_I \right) = e^{\kappa_4^2K/2} {\bf \Pi}^{\mathsf{T}} \eta\, \mathsf{q}\, . 
\label{massD3}
\ee
Notice that in this framework the reference scale \eqref{mstar} is nothing but the mass of a specific BPS particle $m_* = m_{\rm D3}(A^0)
$. As pointed out in \cite{Marchesano:2023thx}, for LCS infinite-distance limits it is natural to identify $m_*$ with the leading SDC tower scale, which defines a UV cut-off for a 4d RFT that remains dynamical along the limit.\footnote{Note that this differs from other familiar approaches wherein one identifies $m_{\rm D3}(A^0) =  2 m_{\rm D3}(B_0) \simeq \Mpl$ \cite{Seiberg:1994rs}.} In the following sections we show that the same strategy can be applied for infinite-distance rigid limits away from the LCS regime, while obtaining the RFT prepotential directly from the supergravity one, as claimed in \cite{Castellano:2024gwi}. 

The squared physical charge of a particle with charge $\mathsf{q}$ is microscopically given by
\begin{eqn}
    \CQ_\mathsf{q}^2 = \oh \int_{Y_3} (p^I\a_I - q_I\b^I) \wedge \star (p^I\a_I - q_I\b^I)\, .
    \label{Hodgest}
\end{eqn}
whereas from the 4d EFT viewpoint we have \cite{Ceresole:1995ca}
\be
\CQ^2_\mathsf{q}  \equiv - \oh \,{\mathsf{q}}^{\mathsf{T}} \begin{pmatrix} \mathbb{I}  & - \CR\\  & \mathbb{I} \end{pmatrix} \begin{pmatrix} \CI  \\  & \CI^{-1} \end{pmatrix} 
\begin{pmatrix} \mathbb{I}  \\ - \CR & \mathbb{I} \end{pmatrix}{\mathsf{q}} \, .
\label{chargeD3}
\ee
The rigid version of the charge, dubbed $\CQ_{\rm rigid}^2$, is obtained from  \eqref{chargeD3} by replacing $\mathcal{N}_{IJ} \to \bar{\CF}_{IJ}$. 

\noindent The physical charge of a state $\q$ reflects the structure of gauge couplings in the 4d Lagrangian. Indeed, employing the democratic formulation\footnote{The pseudo-action \eqref{eq:democratic4d} must be supplemented with the duality constraint $\star G_I= - \mathcal{I}_{IJ} F^I + \CR_{IJ} \star F^J$.} one can write the gauge piece of \eqref{SVM} as \cite{Bergshoeff:2001pv,Herraez:2018vae,Marchesano:2022axe}
\begin{eqn}\label{eq:democratic4d}
 S_{\rm 4d}^{\rm VM}  \supset \int_{\mathbb{R}^{1,3}}  \mathsf{F}^{\mathsf{T}} \begin{pmatrix} \mathbb{I}  & - \CR\\  & \mathbb{I} \end{pmatrix} \begin{pmatrix} \CI  \\  & \CI^{-1} \end{pmatrix} 
\begin{pmatrix} \mathbb{I}  \\ - \CR & \mathbb{I} \end{pmatrix} \wedge \star {\mathsf{F}}\,  ,    \qquad \text{with} \qquad \mathsf{F}= \begin{pmatrix}F^I \\ G_I \end{pmatrix}\, .
\end{eqn}
Let us now describe a quantised vector boson direction as
\begin{eqn}
    U(1)_\q \equiv q_I F^I -  p^I G_I =\mathsf{F}^\mathsf{T} \eta\, \q \, ,
    \label{U1q}
\end{eqn}
which amounts to select the U(1) whose internal wavefuntion is the Poincar\'e dual of \eqref{Piq}. The gauge coupling of such $U(1)_\q$ matches the physical charge \eqref{chargeD3} of a particle with charge vector given precisely by $\q$. Moreover, the projection of $U(1)_\q$ into the graviphoton is captured by the central element $i \bar{Z}_{\q}$ introduced in eq. \eqref{massD3} above. Finally, using the identity  \cite{Ceresole:1995ca}
\begin{eqn}
    \CQ^2_\q = |Z_\q|^2 + \CQ_{\mathsf{q},{\rm rigid}}^2\, ,
    \label{QZQrig}
\end{eqn}
one concludes that whenever the charge-to-mass ratio of a BPS particle 
\be\label{eq:charge2massdef}
\gamma_\mathsf{q}  \equiv \frac{\CQ_\mathsf{q} \Mpl}{m_\mathsf{q} }\, ,
\ee
becomes very large, it is because $\CQ_\q \simeq \CQ_{\rm rigid}$. The physical interpretation is that, when the charge-to-mass ratio of some BPS particle is very large, gravity becomes irrelevant in the (tree-level) Feynman diagrams that implement the no-force condition between identical BPS particles. The cancellation of forces then involves just the gauge charge and the exchange of moduli, as in a rigid 4d $\CN=2$ supersymmetric theory. Equivalently, the interaction of the state $\q$ with the graviphoton becomes negligible compared to its self-coupling via the $U(1)_{\q}$ sector, and so this piece of the Lagrangian reduces to the one of a rigid field theory. 

On the other hand, one of the main results of the theory of mixed Hodge structures, exploited extensively in \cite{Grimm:2018ohb,Grimm:2018cpv,Corvilain:2018lgw}, is that one can easily classify the asymptotic behaviour of $\CQ^2_\mathsf{q}$ along finite- and infinite-distance vector multiplet Calabi--Yau limits. For this, one needs the monodromy group $M = \{ e^{k^iP_i} , k^i \in \mathbb{Z}\}$ associated to the latter, where $e^{P_i}$ are commuting unipotent operators with integer entries. The log-monodromy matrices $P_i$ then determine a weight filtration
\begin{eqn}
    0 \subset W_{-3} \subset W_{-2} \subset W_{-1} \subset W_0 \subset W_1 \subset W_2 \subset W_3 = \Gamma\, ,
    \label{weight}
\end{eqn}
for each value of $i$, on which they act as  $P_i W_\ell \subset W_{\ell-2}$. Here, $\Gamma = \{\mathsf{q} = (p^I, q_I)^{\mathsf{T}}\}$ is the lattice of quantised charges, and the $W_\ell$ are sublattices built from the $P_i$, see e.g. \cite{Grimm:2018cpv} for their precise definition. Each vector $\mathsf{q}$ belongs to one or several of these sublattices, and from this information one can reconstruct the asymptotic behaviour of its charge along the limit. In particular, let us consider a single-modulus limit parametrised as $T = b + it \to i\infty$, with $b$ the periodic coordinate associated to the monodromy. If $\ell$ is the smallest value for which $\mathsf{q} \in W_\ell$ holds, we have that 
\begin{eqn}
    \CQ^2_\mathsf{q} \sim t^{\ell}\, .
    \label{chargeasym}
\end{eqn}
Also, the intersection of two three-cycles satisfies the following constraint:
\begin{eqn}
    [\Sigma_{\mathsf{q}_{\ell_1}}] \cdot [\Sigma_{\mathsf{q}_{\ell_2}}] =  \mathsf{q}^{\mathsf{T}}_{\ell_1} \eta\, \mathsf{q}_{\ell_2} \neq 0 \qquad \text{only if} \qquad \ell_1 + \ell_2 \geq 0\, ,
    \label{interw}
\end{eqn}
where $\ell_i$ is the smallest integer for which $\mathsf{q}_{\ell_i} \in W_{\ell_i}$, $i=1,2$. Finally, one can obtain information about the asymptotic behaviour of \eqref{massD3} using how the holomorphic three-form $\Omega$ decomposes in terms of the filtration \eqref{weight} along the limit. In general, there is a smallest $w$ such that $\Omega$ is contained within $W_w$. The range of this number is  $w=0,1,2,3$, which characterises the type of singularity as well as the asymptotic behaviour of the K\"ahler potential 
\begin{eqn}
    e^{-\kappa_4^2 K} \sim |X^0|^2 t^w\, .
    \label{Kahlerw}
\end{eqn}
The minimal value $\ell$ such that $\mathsf{q} \in W_\ell$ will set an upper bound for the asymptotic behaviour of $m_\mathsf{q}$, given that in general $m_\mathsf{q}/\Mpl \leq \CQ_\mathsf{q}$. Regarding the reference scale $m_* = m_{\rm D3}(A^0)$, when $\mathsf{q}_{A^0}$ represents a state of the leading SDC tower one has that $\mathsf{q}_{A^0} \in W_{-w} = W_{\ell_{\rm min}}$. Similarly, for finite distance limits we also choose $\mathsf{q}_{A^0} \in W_0$, although now $ W_{\ell_{\rm min}} = W_{-1}$.

To link this to the physics of rigid limits, let us consider a particle with charge $\q$ such that
\be
p^I = 0 \quad \text{for} \quad I \notin \{\mu\} \, , \qquad q_I = \re \CF_{I \mu} p^\mu \, .
\label{chcharges}
\ee
Following \cite{Castellano:2024gwi}, one can see that the condition \eqref{cond} translates into
\begin{eqn}
3 \frac{m_\mathsf{q}^2}{\Mpl^2}  \ll 2\frac{m_\mathsf{q}^2}{\Mpl^2} + \CQ_{\mathsf{q},{\rm rigid}}^2  =    \CQ^2_\mathsf{q} \implies \gamma_\mathsf{q}^2 \gg 3\, .
\label{magblows}
\end{eqn}
 If it exists, such a particle corresponds to a BPS monopole stripped out from the electric charges induced by the (rigid) Witten effect.\footnote{This cancellation of electric charges amounts to the relation $q_I = \re \CF_{I i} p^i$, which can be satisfied exactly for certain choices of moduli fields, or to arbitrary accuracy for generic ones. In practice, at weak coupling we will not need to impose the cancellation in \eqref{chcharges} to detect RFT directions.} If we consider a limit where the physical charge and mass of \eqref{chcharges} have a different asymptotic behaviour, then $\gamma_\mathsf{q} \to \infty$ and \eqref{cond} will be satisfied. Similarly, other BPS particles in the theory whose charge-to-mass ratio blows up along the limit will impose constraints on the 4d EFT Lagrangian. For instance, for an electrically charged particle $\mathsf{q} = ( 0, \ldots, q_{i})^{\mathsf{T}}$ one has (c.f. footnote \ref{fnote:hermitiancondition})
\begin{eqn}
    \gamma_{\mathsf{q}}^2 \gg 3 \ \iff \ e^K X^i \bar{X}^{\bar{j}}  \ll N^{ij}\, .
\label{condual}
\end{eqn}
Then, using  the following identity \cite{Lauria:2020rhc}
  \begin{eqn}
        g^{i\bar j}f_i^I \bar{f}_{\bar j}^J&=N^{IJ}+e^{K}X^I\bar{X}^J =-\frac12\mathcal{I}^{IJ}-e^{K}\bar{X}^IX^J\, ,
            \label{invf}
\end{eqn}
one deduces that $\CI^{ij} \simeq - 2N^{ij}$, obtaining gauge kinetic terms whose inverse resemble those of a rigid field theory. This can be considered as a complementary result to $\CI_{\mu\nu} \simeq -\oh N_{\mu\nu}$, obtained from \eqref{calNlimit} and \eqref{cond}, but it is important to notice that they are not equivalent to each other. They become so if one assumes that $N_{IJ}$ has a block-diagonal form, with all the mixed gauge kinetic terms $N_{\mu I}$ with $I \notin \{\mu\}$ being negligible. To see this more directly, one may exploit the electro-magnetic symmetries of the charge-to-mass ratio introduced in \eqref{eq:charge2massdef}. In particular, one can rewrite it as 
\begin{eqn}
\g_{\mathsf{q}}^2 =2 + e^{-K} \frac{\bar{r}_I N^{IJ} r_J}{\bar{r}_I \bar{X}^I X^J r_J}\, , \qquad \text{with} \ r_I  = q_I - {\cal F}_{IJ} p^J\, .
\label{gammar}  
\end{eqn}
Multiplying $r_I$ by an overall complex number $\xi$ leaves the charge-to-mass ratio invariant. It is easy to see that the electric and magnetic states that appear in \eqref{condual}  and \eqref{magblows} have a vector $r_I$ that is real and purely imaginary, respectively. Thus one can relate both kinds of charge-to-mass ratio by taking $\xi = \pm i$. In other words, the conditions \eqref{cond} and \eqref{condual} happen simultaneously. However, the electric and magnetic charges of those charge-to-mass ratios that satisfy $\gamma_\q^2 \gg 3$ are related as $q_{I} = \oh N_{I\mu}p^{\mu}$. Hence, only when $N_{IJ}$ is block diagonal the indices in eqs. \eqref{cond} and \eqref{condual} actually coincide. As we discuss in the next subsection, this is one of the conditions required to decouple the rigid vector multiplets U(1)$_\mu$ from the rest of the 4d EFT. 

Since given a vector charge $\mathsf{q}$ one can always perform a symplectic transformation to bring it to either of these two forms, one concludes that
\begin{eqn}
    e^{-K}|X^0|^{-2} g_{\mu\bar{\nu}} \to N_{\mu\nu} \ \text{for some} \  \{z^\mu\} \iff \gamma_{\mathsf{q}} \to \infty \ \text{for some} \ \mathsf{q}\, ,
    \label{RFTlimit}
\end{eqn}
which corresponds to the definition of RFT limit put forward in \cite{Castellano:2024gwi}.  Let us stress that this statement can be made when the $\{z^\mu\}$ are a subset of special coordinates, but not in general. This is because special coordinates are directly connected to a basis of quantised three-cycles, and hence are tailored to describe the physics of BPS particles and the U(1)'s under which they are charged. As we will see below, other set of coordinates that are familiar from infinite-distance limits are more suitable to describe the physics of BPS strings, and so one must carefully relate both sets to correctly analyse the physics associated to RFT sectors. 

The picture that emerges from the above discussion is that along asymptotic Coulomb branch limits there are roughly two kinds of behaviours for the mutual interactions of a U(1), which are detected by the appropriate charge-to-mass ratio. Using the definition \eqref{U1q},  whenever $\gamma_\q \to \infty$ we say that $U(1)_{\eta\q}$ is a rigid U(1), in the sense that its mutual interactions are governed by a piece of the Lagrangian that flows to a rigid theory. If instead $\gamma_\q$ remains finite, the corresponding piece of the Lagrangian remains inherently gravitational along the  trajectory, and we say that $U(1)_{\eta\q}$ is of gravitational type. In this case the  BPS particle with charge $\mathsf{q}$ is oftentimes extremal \cite{Alim:2021vhs}, and so $\gamma_\mathsf{q}$ can be interpreted as an extremality factor which, due to the Weak Gravity Conjecture, is expected to remain of $\CO(1)$. In this regard, the results of \cite{Gendler:2020dfp,Bastian:2020egp} (see also \cite{Castellano:2024gwi}) indicate that, for infinite-distance limits, the gravitational U(1)'s always include the gauge bosons associated to the SDC tower and to the direction of the limit, a statement that below we will reinterpret in terms of the monodromy group action. Of course, this distinction between rigid and gravitational U(1)'s may not be physically relevant if one cannot decouple them from each other. In practice, this is achieved by selecting a subsector of the rigid U(1)'s that satisfy certain additional requirements, which we now turn to discuss.

\subsection{Decoupling conditions}
\label{ss:decoupling}

As pointed out above, the RFT condition \eqref{cond} can be understood as the diverging charge-to-mass ratio of a `pure' monopole. Since the mass of such a state is its graviphoton charge, one can rephrase \eqref{cond} as the fact that for certain would-be BPS monopoles\footnote{Neither to define $\gamma_\mathsf{q}$ nor the RFT condition \eqref{RFTlimit} it is necessary that the charge $\mathsf{q}$ hosts a stable BPS particle. However, to our knowledge in all limits where the RFT decouples from gravity such BPS particles do exist.} the graviphoton charge is negligible. However, as we just discussed, along a given limit there are further vector multiplets that should be regarded as being of gravitational nature. Hence, for an RFT to fully decouple from gravity one should  also require that the charge under all those gravitational U(1)'s vanishes. As explained in Appendix \ref{ap:special}, the charge of a magnetic state of the form \eqref{chcharges} under the vector multiplet U(1)$_i$ is proportional to $e^{-K/2}|X^0|^{-1}g_{i\bar{\mu}}p^\mu$, and so to decouple an RFT from the gravitational sector one should avoid significant kinetic mixing with any of its gauge fields. In addition, interactions that couple the RFT with the gravitational sector must be appropriately suppressed. 

While this notion of decoupling boils down to standard field theory wisdom, it is important to stress that not all rigid U(1)'s satisfy it. In general, one finds that only a subset of rigid U(1)'s can decouple from the gravitational sector in a significant region of moduli space. We will denote such subsector as cRFT, anticipating the definition of core RFT that will be given in section \ref{ss:core}. In short, we have that the following conditions must be satisfied:

\begin{itemize}
    \item[-] There is no significant kinetic mixing between the cRFT and other scalar field directions. Similarly, there is no gauge kinetic mixing between the U(1)'s of the cRFT and the rest.

    Since $g_{i \bar{j}}$ and $-\CI_{IJ}$ are symmetric and positive definite matrices, one can translate these requirements as the condition that, if $\{z^\lambda, z^\rho, \dots\} \subset \{z^\mu\}$ describe the decoupling RFT directions, both  $|X^0|^2 e^K g^{\lambda\bar{\rho}}$ and $-\oh \CI^{\lambda\rho}$ reduce to the inverse of $N_{\lambda\rho}$ along the rigid limit \cite{Castellano:2024gwi}. It follows from the discussion below \eqref{condual} that the condition for $\CI_{IJ}$ is satisfied if $N_{IJ}$ is block diagonal with respect to the cRFT sector, that is $N_{\lambda I} \simeq 0$ with $I \notin \{\lambda, \rho,\dots\}$. Regarding the metric, one can use the Sherman-Morrison formula \cite{Sherman-Morrison1,Sherman-Morrison2} to write
    \begin{eqn}
    |X^0|^2 e^K g^{\bar{k} j}=  \left(N^{-1}\right)^{k j}-  \frac{\left(N^{-1}\right)^{k m}k_m k_{\bar{n}} \left(N^{-1}\right)^{nj}}{e^K|X^0|^2 +  k_{\bar{n}} \left(N^{-1}\right)^{n m} k_m}\, .
    \label{morry}
    \end{eqn}
    with $\left(N^{-1}\right)^{ij} N_{jk} = \delta^i_k$. Hence, when $N_{IJ}$ is approximately block diagonal, $\left(N^{-1}\right)^{\lambda \rho}$ reduces to the inverse of $N_{\lambda \rho} \simeq e^{-K} |X^0|^{-2} g_{\lambda \bar{\rho}}$, as per \eqref{cond}. We thus find that both requirements boil down to absence of mixing in the (rigid) gauge kinetic function. 

    \item[-] There are no significant interactions mixing the cRFT and other sectors. In particular, the gaugino-gauge boson couplings encoded in the 4d $\CN=2$ Pauli terms should not mix the core RFT with other U(1)'s significantly. More precisely, we have that \cite{Andrianopoli:1996cm}
    \begin{eqn}
     \mathcal{L}_{\rm Pauli} \supset -\frac14\CP_{Ijk} (F^{I}_-)_{\mu \nu}\bar{\lambda}^{j A} \gamma^{\mu \nu} \lambda^{k B} \epsilon_{AB} + h.c.\, ,
      \label{Pauli}
    \end{eqn}
    where $\CP_{Ijk}\equiv 2 \CI_{IJ}D_jf_k^J$ and we defined the (anti-)self-dual forms $F_\pm^I = \frac12 (F^I \mp i \star_4 F^I)$, in terms of which the gauge kinetic terms read $\mathcal{L}_{\rm kin} \supset \oh \overline{\CN}_{IJ} F^I_- \wedge F^J_{-} + h.c.$

    One can rewrite these Pauli couplings as (see Appendix \ref{ap:special})
    \begin{eqn}
    \CP_{Ijk} =  2i e^{-K/2}  f_j^J f_k^K \CF_{JKL} N^{LM} \CI_{MI}=2ie^{K/2}(X^0)^2\CF_{jkL}N^{LM}\CI_{MI}\, ,
    \label{Pauli2}
    \end{eqn}
    where the second equality uses the properties of the special coordinates.\footnote{\label{fnote:Pauli}Pauli terms also read $i\CP_{Ijk} = e^{-K/2}\CF_{IJK} f^J_if^K_j \left(\delta^I_L-\frac{(N\bar X)_L}{(\bar X N\bar X)}\bar X^I\right) = e^{-K/2}\CF_{IJK} f^J_if^K_j \left(\delta^I_L-ie^{K/2}T_L\bar X^I\right)$, with the piece in parenthesis being a projector that removes the graviphoton direction $T_I= 2i e^{K/2} \CI_{IJ}X^J$. Therefore, the couplings shown in \eqref{Pauli} should be understood as interactions involving exclusively the vector multiplets. Pauli terms that mix the gravitino with gauginos are proportional to the gauge kinetic mixing \cite{Ceresole:1995ca}.} The normalised Pauli terms for a subsector $\Lambda$ of the EFT with no mixing with the rest reads
    \begin{eqn}
    \lVert \CP \rVert^2_{\Lambda} \equiv  - \CI^{IJ} g^{j\bar{m}} g^{k\bar{n}} \CP_{Ijk} \bar{\CP}_{J\bar{m}\bar{n}} \, .
    \label{NormPauli}
    \end{eqn}
    where the sum runs over the indices in $\Lambda$. For instance, we may use the previous decoupling condition to take $\Lambda = $ cRFT. Absence of Pauli mixing between the cRFT and the rest of the EFT can then be formulated as follows
    \begin{eqn}
    \lVert \CP \rVert^2_{\in\text{cRFT}} \gg  \lVert \CP \rVert^2_{\rm mixed}\, ,
    \label{nomixing2}
    \end{eqn}
    where the rhs contains a sum of normalised Pauli couplings that include indices both within and without the core RFT. Notice that absence of (gauge) kinetic mixing between the core RFT and the rest of the EFT does not suppress certain contributions to $\lVert \CP \rVert^2_{\rm mixed}$, and thus \eqref{nomixing2} is in fact independent from the previous condition. 

\end{itemize}

Remarkably, this second requirement is directly connected to the asymptotic behaviour of the moduli space curvature. In fact, one may write the Riemann tensor as
\begin{equation}
    R_{i\bar j k\bar l}=g_{i\bar j}g_{k\bar l}+g_{i\bar l}g_{k\bar j}-e^{2K}|X^0|^6 \CF_{ikm}\bar \CF_{\bar j\bar l\bar n}g^{m\bar n }=g_{i\bar j}g_{k\bar l}+g_{i\bar l}g_{k\bar j} + \oh   \CP_{M ik} \bar{\CP}_{N\bar{j}{\bar{l}}} \CI^{M N}\, ,
    \label{curvPauli}
\end{equation}
where to reach the second equality we used $\CF_{IJK} X^K = 0$ and eq. \eqref{invf}, see Appendix \ref{ap:special} for more details. Thus, it follows that the scalar curvature can be written as
\begin{eqnarray}
   \label{IIBscalarP} 
    R &= &- 2 n_V(n_V + 1) + \lVert \CP \rVert^2_{\text{EFT}} \\
    & =&  - 2 n_V(n_V + 1) + \lVert \CP \rVert^2_{\in\rm cRFT} + \lVert \CP \rVert^2_{\rm mixed}+\lVert \CP \rVert^2_{\notin\rm cRFT} 
    \, ,
\nonumber
\end{eqnarray}
that is, as a constant term plus the normalised sum of all Pauli terms entering the 4d $\mathcal{N}= 2$ EFT.  This provides a very clear physical interpretation for the moduli-dependent piece of the field space curvature, namely as a sum of physical interactions among the different vector multiplets. In the second line, $\lVert \CP \rVert^2_{\rm mixed}$ is defined as in \eqref{nomixing2}, and $\lVert \CP \rVert^2_{\notin\rm cRFT}$ as in \eqref{NormPauli} but with the indices running exclusively over fields outside of the core RFT.

Reciprocally, these expressions give a geometric meaning to the decoupling conditions above. To see this, consider a limit with a one-dimensional RFT whose field direction is given by $z^\mu$. The Riemann tensor along the latter reads 
\begin{equation}
    R_{\mu\bar \mu \mu\bar \mu} \simeq  \frac{m_*^4}{\Mpl^4} \left( 2N_{\mu \mu}^2 - |X^0|^2\CF_{\mu\mu m}\bar \CF_{\bar \mu \bar \mu\bar n}\,g^{m\bar n }\right)\, , \qquad \text{with} \quad g^{\mu\bar{\mu}} \simeq \frac{\Mpl^2}{m_*^2} \left(N^{-1}\right)^{\mu\mu}\, .
    \label{curvRFT}
\end{equation}
If the first decoupling condition is met, we can identify $\left(N^{-1}\right)^{\mu\mu}$ with $N_{\mu\mu}^{-1}$, and one of the terms of the sum contributes as 
\begin{eqn}
    \frac{m_*^2}{\Mpl^2} R_{\mu\bar \mu \mu\bar \mu}^{\rm rigid} \, ,\qquad \text{with} \quad R_{\mu\bar \mu \mu\bar \mu}^{\rm rigid} = - |X^0|^2|\CF_{\mu\mu \mu}|^2 \left(N^{-1}\right)^{\mu\mu} = - \oh N_{\mu\mu}^2 R_{\rm rigid} \, ,
\end{eqn}
where $R_{\rm rigid} \equiv 2 |X^0|^2 G^{\mu\bar{\nu}} G^{\rho\bar{\sigma}} G^{\tau\bar{\eta}}  \CF_{\mu\rho\tau} \bar{\CF}_{\bar{\nu}\bar{\sigma}\bar{\eta}}$ is the standard RFT curvature, measured in units of the reference scale $m_*$. Hence, whenever $(\Mpl^2/m_*^2)R_{\rm rigid} \gg 4$ we can safely neglect the first contribution to $R_{\mu\bar \mu \mu\bar \mu}$ in \eqref{curvRFT}, which thus shall be written as a sum of Pauli terms:
\begin{equation}
    R_{\mu\bar \mu \mu\bar \mu} \simeq \oh \CP_{M \mu\mu} \bar{\CP}_{N \bar{\mu}{\bar{\mu}}} \CI^{M N} \, , \qquad \text{with} \quad  \oh   |\CP_{\mu\mu 
    \mu}|^2 \CI^{\mu \mu} \simeq \frac{m_*^2}{\Mpl^2} R_{\mu\bar \mu \mu\bar \mu}^{\rm rigid}\, .
    \label{curvRFTPauli}
\end{equation}
In this context, the contribution to $\lVert \CP \rVert^2_{\rm mixed}$ in \eqref{nomixing2} comes from the terms in the sum \eqref{curvRFTPauli} such that $M\neq \mu \neq N$. Therefore, imposing the second decoupling condition implies that
\begin{eqn}
     R_{\mu\bar \mu \mu\bar \mu} \simeq  \frac{m_*^2}{\Mpl^2} R_{\mu\bar \mu \mu\bar \mu}^{\rm rigid}\, ,
\end{eqn}
recovering the RFT prediction for the curvature, up to the appropriate change of units. Again, note that the first decoupling condition may suppress those terms in \eqref{curvRFTPauli} with $M =\mu$ and $N\neq \mu$, but those with $M \neq \mu \neq N$ could still contribute non-negligibly to $R_{\mu\bar \mu \mu\bar \mu}$. If that is the case, the curvature along the slice of moduli space that we associate with the RFT will differ significantly from the RFT prediction, unless  additional fields are also treated as dynamical. That is, there would not be an actual decoupling between rigid and non-rigid sectors. 

In general, to avoid this possibility one should require that the sectional curvature along any cRFT direction matches the RFT prediction, namely
\begin{eqn}
    \oh \CP_{M \mu\nu} \bar{\CP}_{N\bar{\rho}\bar{\sigma}} \CI^{MN} \simeq  \oh \CP_{\lambda \mu\nu} \bar{\CP}_{\kappa \bar{\rho}\bar{\sigma}} \CI^{\lambda \kappa}\, , 
\end{eqn}
where all Greek indices belong to the cRFT. This, in particular, implies that
\begin{eqn}
    \rVert \CP \lVert^2_{\in \text{cRFT}} \gg \lVert \CP \rVert^2_{{\rm mixed}'}\, ,
    \label{condmixprime}
\end{eqn}
where $\lVert \CP \rVert^2_{{\rm mixed}'}$ contains a subset of the Pauli mixing terms that appear in the rhs of \eqref{nomixing2}. More precisely, we have a sum of the form \eqref{NormPauli} with gaugino indices $\mu,\nu,\rho,\sigma$ within the cRFT and at least one gauge index $I,J$ outside of it. While this requirement is weaker than \eqref{nomixing2}, {in the examples analysed below it turns out to be sufficient to guarantee decoupling.} 

In section \ref{ss:core} we will propose a definition of a rigid subsector, dubbed core RFT, which is precisely the one that can satisfy both decoupling conditions, and which will play the r\^ole of cRFT above. Its definition will be formulated in terms of the action that the monodromy group of the limit has on the U(1)'s with quantised particle charges $\mathsf{q}$. To motivate such a definition of core RFT from a physics viewpoint, it proves useful to first discuss the meaning of essential instantons, and how they fit within the general scheme of RFT limits.

\subsection{Essential and metric essential instantons}
\label{ss:essential}

The concept of essential instantons was coined in \cite{Bastian:2021eom}, based on the properties of $\Omega$ around singular moduli space boundaries. In particular, one finds that near both finite- and infinite-distance singularities the period vector associated to \eqref{Omega} takes the form 
\begin{eqn}
    {\bf \Pi}  = 
    \begin{pmatrix}
        X^I \\ \CF_J 
    \end{pmatrix} 
    =
e^{T^i P_i} \left( {\bf a}_0 + \sum_{r_i} {\bf a}_{r_1\dots r_n}\, e^{2\pi i r_i T^i} \right)\, ,
\label{periodv}
\end{eqn}
where the coordinates $T^i = b^i + i t^i$ describe the singular locus as $t^i = \infty$. Circling around the singularity once corresponds to the discrete shift symmetry $b^i \to b^i +1$, that implements the monodromy ${\bf \Pi} \to e^{P_i} {\bf \Pi}$ on the period vector, where $e^{P_i}$ are unipotent matrices with integer entries. The coordinates $T^i$ only describe the slice of moduli space transverse to the singularity, while there could be further variables that are treated as spectators, on which ${\bf a}_0, {\bf a}_{r_1\dots r_n}$ may depend. Finally, ${\bf a}_0$ is such that  $P^{w}{\bf a}_0 \neq 0$ but $P^{w+1}{\bf a}_0 =0$, where $P=\sum_i P_i$ and the index $w=0,1,2,3,$ describes the type of degeneration, as in the discussion around \eqref{Kahlerw}.

Following \cite{Bastian:2021eom}, one has a limit with essential instantons whenever ${\bf \Pi}_{\rm pol} \equiv e^{T^iP_i} {\bf a}_0$ together with its derivatives do not generate the full (complex) three-form cohomology $H^3(Y_3, \mathbb{C})$, even if the singularity is of codimension $n_V$, as we assume in the following. Algebraically this means that, if we take the  vector space ${\bf V}_{\rm pol}^{\mathbb{C}}$ built as the complex span of  ${\bf a}_0$, $P_i{\bf a}_0, P_iP_j{\bf a}_0,P_iP_jP_k{\bf a}_0$, the presence of essential instantons implies that $\dim_{\mathbb{C}} {\bf V}_{\rm pol}^{\mathbb{C}} < 2(n_V+2)$. Notice that, by construction, ${\bf V}_{\rm pol}^{\mathbb{C}}$ is independent of the moduli space point around the singularity and that, in particular, it is invariant under the monodromy group $e^{k^i P_i}$, $k^i \in \mathbb{N}$. Since by  consistency $P_i^{\mathsf{T}} \eta + \eta P_i =0$, the same is true for its so-called annihilator $\hat{\bf Q}_{\rm ess}^{\mathbb{C}}$, made of complex vectors $\hat{\mathsf{q}}_{\mathbb{C}} \in \hat{\bf Q}_{\rm ess}^{\mathbb{C}}$ such that $\hat{\mathsf{q}}_{\mathbb{C}}^{\mathsf{T}} \eta {\bf v} = 0$, $\forall \,{\bf v} \in {\bf V}_{\rm pol}^{\mathbb{C}}$. In other words, essential instantons select a fixed monodromy-invariant subspace of $H^3(Y_3, \mathbb{C})$ such that any of its elements $\zeta$ satisfies $\int_{Y^3}  \Omega_{\rm pol}  \wedge \zeta = 0$ around the degeneration point, where $\Omega_{\rm pol}$ corresponds to the approximate period vector  ${\bf \Pi}_{\rm pol}  \equiv  e^{T^i{P}_i}  {\bf a}_0$. This property is very suggestive from the viewpoint of rigid theories because, if we map $\zeta$ to its Poincar\'e dual three-cycle $\q_\zeta \in \hat{\bf Q}_{\rm ess}^{\mathbb{C}} \in H_3(Y_3, \mathbb{C})$, we have a central charge $Z_{\q_\zeta}$ that is exponentially suppressed with respect to others, and hence a mechanism to generate hierarchical masses and charge-to-mass ratios. However, since the physical set of central charges \eqref{massD3} is based on elements of $H_3(Y_3,\mathbb{Z})$ and not of $H_3(Y_3, \mathbb{C})$,  it is in general not obvious if or how the presence of essential instantons reflects on the BPS mass spectrum of the compactification.

Things become more transparent in the case of metric essential instantons. Those were defined in \cite{Bastian:2021hpc} as the subset of essential instantons without which the metric becomes degenerate near the singularity. To detect their presence one may thus check whether the complex span ${\bf V}_{\rm met}^{\mathbb{C}} \equiv \langle{\bf a}_0, P_i{\bf a}_0\rangle$ is such that $\dim_{\mathbb{C}}{\bf V}_{\rm met}^{\mathbb{C}} < n_V +1$. Whenever this happens, there is a Dolbeaut class of $h^{2,1}$-forms that is absent from ${\bf V}_{\rm met}^{\mathbb{C}}$, which is the ultimate source of the metric degeneracy. From this definition, one can argue that the following result also holds
\begin{eqn}
    c^i P_i {\bf a}_0 = 0\, ,
    \label{metricond}
\end{eqn}
with an independent choice of $c^i \in \mathbb{C}$ for each metric essential instanton. Indeed, projected down to the space of (3,0)- and (2,1)-forms, the metric essential instanton condition reads 
\begin{align}
\label{firstmet}
\left[c^i  \p_i \Omega_{\rm pol} + c^0 \Omega_{\rm pol}\right]_{(3,0)} & = \left(c^0-c^i k_{{\rm pol}, i} \right) \Omega_{\rm pol} = 0\, , \\
\left[c^i  \p_i \Omega_{\rm pol} + c^0 \Omega_{\rm pol}\right]_{(2,1)} & =  c^i\left( \p_i\Omega_{\rm pol} + k_{{\rm pol}, i} \Omega_{\rm pol} \right) = 0\, ,
\label{secondmet}
\end{align}
with $c^0 \in \mathbb{C}$ and $k_{{\rm pol},i}$ defined as in \eqref{monok} upon replacing $\Omega \to \Omega_{\rm pol}$. The K\"ahler potential \eqref{Kahlerpot} derived from $\Omega_{\rm pol}$ is of the form $K_{\rm pol} = - \log \left({\cal P} (\im T^i)\right)$, with $\CP$ some polynomial function of degree $w$. With all this at hand, one concludes that the first equation \eqref{firstmet} only has solutions if $c^0 =0$, namely if $c^ik_{{\rm pol},i} = 0$. Applying this to \eqref{secondmet}, we have that \eqref{metricond} follows immediately.\footnote{Using the framework of \cite{Grimm:2018cpv}, one can also arrive at \eqref{metricond} by noticing that ${\bf a}_0$ and $P_i{\bf a}_0$ live in different complex subspaces $I^{p,q}$ of the Deligne splitting.}

Consequently, the field direction $T^i \propto c^i$ does not appear in ${\bf \Pi}_{\rm pol}  \equiv  e^{T^i{P}_i}  {\bf a}_0$, and only does so in \eqref{periodv} through exponentials. This is indeed what one expects from an instanton effect, since from these periods one can reconstruct the whole EFT Lagrangian as well as the spectrum of BPS masses. The above interpretation, however, constrains the allowed coefficients $c^i$ to be of the form $c n^i$, with $c \in \mathbb{C}$, $n^i \in \mathbb{Z}$. Otherwise, the field direction $T^i \propto c^i$ will not contain a periodic coordinate  and, from a physical viewpoint, it would not make sense as an exponential correction.  In other words, defining a $P_\mu \equiv n_\mu^i P_i$ for each metric essential instanton we find that
\begin{eqn}
    {\bf a}_0 \in \ker P_\mu\, .
    \label{metriker}
\end{eqn}
This result allows one to link the presence of metric essential instantons to specific features of the BPS mass spectrum, both for finite- and infinite distance limits. Since the discussion in each of these cases becomes slightly different, we will analyse them separately in what follows.

\subsubsection*{Finite distance limits}\label{sss:EssentialFiniteDistance}

For finite distance limits we have that $w=0$ and so \eqref{metriker} holds by assumption. In this sense, the above discussion can be interpreted as indicating that metric essential instantons are always present for this kind of limits. Moreover, one can draw a clear picture on how these instantons manifest in the spectrum of BPS masses. For this, let us consider the weight filtration that corresponds to this type of degenerations. One has (see, e.g. \cite{Grimm:2018cpv})
\begin{equation} \label{WI}
\begin{array}{ccl}
  W_1  & = & \Gamma\,,\\
  \cup &   &\\
  W_0  & = & \ker\,P\,,\\
  \cup &   &\\
  W_{-1}  & = & \mathrm{Im}\,P\,,\\
\end{array}
\end{equation}
where for simplicity we have assumed a single nilpotent generator $P$, which as expected satisfies  $P W_\ell \subset W_{\ell-2}$. Each of the elements of the filtration should be thought as a sublattice of $\Gamma \equiv \{\mathsf{q} = (p^I, q_I)^{\mathsf{T}}\} \cong H_3(Y_3,\mathbb{Z})$, namely the charge lattice for particle states. The physical charge of these objects is directly given by \eqref{chargeasym}, with a polynomial behaviour that is insensitive to the presence of metric essential instantons. These are detected by looking at the BPS mass \eqref{massD3} for each of the above sublattices. First, it follows from the condition $P {\bf a}_0 = 0$ that $T$ does not appear in \eqref{periodv} at the polynomial level, and so any element of $\Gamma$ has a mass of the form
\begin{eqn}
   m_{\mathsf{q}} = \left|\mathsf{q}^\mathsf{T} \eta \left( {\bf a}_0 + (\mathbb{I}+TP) \sum_{k} {\bf a}_{k} e^{2\pi i k T} \right) \right| m_*\, .
   \label{massessI}
\end{eqn}
Namely a BPS mass that, in units of $m_*$, depends on $T$ through exponentials, and that it asymptotes to $a_\mathsf{q} m_*$ along the limit, with $a_\mathsf{q} \equiv |\mathsf{q}^\mathsf{T} \eta {\bf a}_0 |$. This quantity in fact distinguishes two very different behaviours for \eqref{massessI}. An element $\mathsf{q} \in \Gamma$ such that $a_{\mathsf{q}} =0$ corresponds to a BPS particle whose mass becomes parametrically lighter than the reference scale $m_*$ when $T\to i \infty$. Instead, an element for which $a_{\mathsf{q}} \neq 0$ gives rise to a particle whose BPS mass is asymptotically of the order of $m_*$. It is the first of these two cases that one may identify with the physics expected from essential instantons, in the sense that it reproduces the behaviour of the central charge $Z_{\q_\zeta}$ observed at the beginning of this section. 

It is easy to see that precisely those elements within $W_{-1} = \im P$ fall into this class, since one can always write them as $\q_{-1} = P \q_1$, for some $\q_1 \in W_1 = \Gamma$. Hence, one finds
\begin{eqn}
    a_{\q_{-1}} = |\q_{-1}^\mathsf{T} \eta {\bf a}_0| = |\q_1^\mathsf{T} \eta P {\bf a}_0| = 0\, ,
    \label{vaniaq}
\end{eqn}
where we used that $P^{\mathsf{T}} \eta + \eta P =0$ (alternatively, one could have employed \eqref{interw} to arrive at the same conclusion). More precisely, we have a mass of the form 
\begin{eqn}
   m_{\mathsf{q}_{-1}} = \left|\mathsf{q}_{-1}^\mathsf{T} \eta \left( \sum_{k} {\bf a}_{k} e^{2\pi i k T} \right) \right| m_*\, ,
   \label{massessIel}
\end{eqn}
with no polynomial dependence on $T$ whatsoever. This not only means that the BPS mass $m_{\q_{-1}}$ of the states $\q_{-1} \in W_{-1}$ is exponentially suppressed in Planck units,\footnote{By this we mean with respect to the periodic coordinate $T$ used in \eqref{periodv}.} but also that its charge-to-mass ratio $\gamma_{\q_{-1}}^2$ grows like $e^{2t} /t$, even if its squared physical charge vanishes asymptotically. This is a general result associated to the physics of metric essential instantons: A BPS mass that decreases exponentially in units of either $m_*$ of $\Mpl$ always implies a charge-to-mass ratio that grows exponentially fast, thereby signaling the presence of an RFT limit. 

It remains to see which other states exhibit a diverging charge-to-mass ratio. One can readily recognise that those elements belonging to $W_0$ but not to $W_{-1}$, which we denote as $\q_0 \in W_0\setminus W_{-1}$, generically have asymptotically finite charge-to-mass ratios. Indeed, due to \eqref{chargeasym} their physical charge tends to a constant, whilst their mass is also finite since they are usually such that $a_{\q_0} \neq 0$. The charge vectors of $\q_1 \in W_1 \setminus W_0$ will, on the other hand, always possess a diverging ratio $\gamma_{\q_1}$, given that their mass is at most constant whereas their physical charge diverges. The specific behaviour for $\gamma_{\q_1}$ will therefore rely on whether $a_{\q_1}$ vanishes or not, which in turn depends on the specific form of ${\bf a}_0$ (see section \ref{ss:conifold} for more on this). 

We thus have that the elements in $W_{-1}$ and $W_1 \setminus W_0$ form an RFT subsector of the limit. In fact, these two sets can be related by the action of the log-monodromy generator $P$, which maps $P (W_1 \setminus W_0) \subset W_{-1}$. The physical interpretation is that $W_1 \setminus W_0$ contains the magnetic duals of the elements in  $W_{-1}$. More precisely, if we consider $\q_{-1} \in W_{-1}$ to be an electrically charged particle under a rigid U(1), its preimage $P^{-1} (\q_{-1}) \subset W_1 \setminus W_0$ will include the tower of dyonic states associated to it. One can then select a magnetic dual of $\q_{-1}$ by choosing\footnote{That this can always be done follows from the polarisation condition of the limiting mixed Hodge structure.} an element $\q_1 \in W_1\setminus W_0$ with minimal Dirac pairing $\q_1 \cdot \q_{-1} = \q_1^\mathsf{T} \eta \q_{-1}$ and, within this set, restricting to the charge that minimises $a_{\q_1}$. As a result, the RFT subsector of U(1)$_\lambda$'s associated to metric essential instantons is represented by a set of electro-magnetic pairs  $\{(\q_{-1})_\lambda, (\q_1)_\lambda\}_\lambda$ that corresponds to a basis of $W_{-1}$. These vectors generate a lattice that we dub as $\Gamma_{\rm ess}$, which is the physical analogue of the vector space $\hat{\bf Q}_{\rm ess}^\mathbb{C}$ previously defined. Just like that space, $\Gamma_{\rm ess}$ is invariant under the action of the monodromy group, and so is its symplectic complement which we denote by $\Gamma_{\rm ess}^\perp$. Indeed, the latter represents the elements of $W_0 \setminus W_{-1}$, wherein the monodromy group acts trivially. We thus recover a split of the charge lattice of the form $\Gamma = \Gamma_{\rm ess}^\perp \oplus \Gamma_{\rm ess}$, such that the monodromy takes the block diagonal form
\begin{eqn}
    e^{k P} = 
    \begin{pmatrix}
       \mathbb{I} & 0 \\ 0 &   M_{\rm ess}
    \end{pmatrix}\, ,
\end{eqn}
where $M_{\rm ess}$ represents the action of the latter group on the metric essential instanton sector.

\subsubsection*{Infinite distance limits}\label{sss:EssentialInfiniteDistance}

Infinite distance limits are in general more cumbersome, in the sense that their weight filtration has more terms that need to be considered. Still, the fact that \eqref{metriker} holds for some integer combination of $P_i$ allows to build a sublattice $\Gamma_{\rm ess}$ similar to the one constructed in the finite distance case, which reproduces the key properties that we associate to metric essential instantons. 

Let us, for instance, consider the case of $w=3$ limits. Their weight filtration reads \cite{Grimm:2018cpv}:
\begin{equation} \label{WIV}
\begin{array}{ccl}
  W_3  & = & \Gamma,\\
  \cup &   &\\
  W_2  & = & \ker\,P^3,\\
  \cup &   &\\
  W_1  & = & \ker\,P^2  + \mathrm{Im}\,P,\\
  \cup &   &\\
  W_0  & = & \ker\,P  + \mathrm{Im}\,P \cap\ker\,P^2,\\
  \cup &   &\\
  W_{-1}  & = & \mathrm{Im}\,P \cap\ker\,P  + \mathrm{Im}\,P^2,\\
  \cup &   &\\
  W_{-2}  & = & \mathrm{Im}\,P^2 \cap\ker\,P,\\
  \cup &   &\\
  W_{-3}  & = & \mathrm{Im}\,P^3 , 
\end{array}
\end{equation}
for $P = \sum_i P_i$. Here, $W_{-2} =W_{-3}$, $W_2=W_1$, and ${\bf a}_0$ is identified with an element of $W_3 \setminus W_2$, since we have $P^3 {\bf a}_0 \neq 0$. If we assume that the weight filtration associated to $P_\mu$ is compatible with \eqref{WIV} in the sense of \cite{kerr2019}, then the BPS mass of certain states is significantly restricted.

Notice that for any weight filtration of this sort, the sublattice $W_{-1}$ is comprised by two types of charge vectors. The first ones are obtained upon taking an element $\q_3$ of $W_3\setminus W_2$ and acting with $P^2$ on it. It is clear, though, that a further application of $P$ leads to a non-trivial state in $W_{-3}$. On the other hand, for this class of limits $W_{-3}$ is given by a one-dimensional lattice associated with the SDC tower \cite{Monnee:2025ynn}, whilst $W_3\setminus W_2$ contains its magnetic dual charges. Both sets of states are extremal, i.e. they have $\gamma_\q = \CO(1)$, and are thus unrelated to essential instantons or rigid sectors. Using this, one concludes that the elements within $\im P^2 \subset W_{-1}$ must have a physical charge and BPS mass of the form
\begin{eqn}
    \frac{m_{\q_{-1}}}{\Mpl}  = \left|\mathsf{q}_{3}^\mathsf{T} \eta P^2(\mathbb{I}+T^iP_i) {\bf a}_{0} + \ldots\; \right| \frac{m_*}{\Mpl} \sim t^{-\frac12}\, ,\qquad \CQ_\mathsf{q} \sim t^{-\frac12}\, ,
\end{eqn}
thereby yielding the same $\gamma_\q = \CO(1)$ as those belonging to the Distance Conjecture tower.\footnote{More generally, one can show that the asymptotic charge-to-mass ratio satisfies $\gamma^2_{P \mathsf{q}} \gtrsim \gamma^2_{\mathsf{q}}$,  with $P$ the log-monodromy operator, while the full monodromy action $\exp(P)$ does not change $\gamma^2_\mathsf{q}$ asymptotically.}

The second kind of charge vectors in $W_{-1}$ arise from taking an element $\q_1 \in \ker P^2 \setminus W_0$ and acting with $P$ on it. More concretely, let us consider such a state for the weight filtration constructed out of $P_\mu$. We then have that $\q_{-1} = P_\mu\q_1$, and its associated BPS mass reads
\begin{eqn}
    \frac{m_{\q_{-1}}}{m_*} = \left|\q_{-1}^\mathsf{T} \eta {\bf \Pi}\right| = \left|\q_{1}^\mathsf{T} \eta P_\mu {\bf \Pi}\right| = \left|\mathsf{q}_{-1}^\mathsf{T} \eta  \sum_{r_i} {\bf a}_{r_1\dots r_n} e^{2\pi i r_i T^i} \right|\, ,
    \label{massessIVel}
\end{eqn}
where we have chosen the charge vector $\q_{-1}$ such that $\q_{-1} \in \cap_{i} \ker P_i$. Notice that we recover the same structure as in \eqref{massessIel}, so in this sense these elements are the analogues of those in $W_{-1}$ for finite distance limits. In particular, one retrieves again an exponentially suppressed mass in units of $m_*$, and if we now consider the charge-to-mass ratio of this state we find 
\begin{eqn}
    \gamma_{\q_{-1}} = \frac{m_*}{m_{\q_{-1}}} \frac{\Mpl}{m_*} \CQ_{\q_{-1}}\, .
\end{eqn}
The behaviour of $\gamma_{\q_{-1}}$ is always dominated by $m_*/m_{\q_{-1}}$, which grows exponentially. The polynomial dependence of both $\Mpl/m_* \sim t^{w/2}$ and $\CQ_{\q_{-1}}$ cannot thus compete with the exponential growth, and so $\gamma_{\q_{-1}}$ always diverges, signaling the presence of an RFT sector.  

Just as for finite-distance limits, to all elements of this sort we can associate a magnetic dual $\q_1 \in P_\mu^{-1}(\q_{-1})$ and from there build the lattice $\Gamma_{\rm ess}$. A priori the behaviour for the mass and charge of the magnetic state is more involved than before, since it depends on the weights $\ell_i$ that $\q_1$ has with respect to the rest of the generators $P_{i \neq \mu}$. However, as long as we select $\mathsf{q}_1$ such that $P_i \q_1 \propto \q_{-1}$, $\forall i \neq \mu$, we should obtain a mass of the form
\begin{eqn}
   m_{\mathsf{q}_1} = \left|\mathsf{q}_1^{\mathsf{T}} \eta \left( {\bf a}_0 + \left( \mathbb{I}+T^iP_i\right) \sum_{r_i} {\bf a}_{r_1\dots r_n} e^{2\pi i r_i T^i} \right) \right| m_*\, ,
   \label{massessIVmag}
\end{eqn}
recovering the finite-distance structure \eqref{massessI}. However, contrary to what happened in the finite distance case, for singularites of codimension larger than one the polynomial growth in \eqref{massessIVmag} may win over the exponential suppression along certain limits. As a result, it is no longer true that $m_{\q_1}/m_*$ asymptotes to a constant for any trajectory such that $t^i \to \infty$, and in some cases it may even diverge. Although the ratio $\gamma_{\q_1}$ still blows up in this case,\footnote{Indeed, since $\q_1\in W_1 \setminus W_0$ for the full filtration, $\CQ_{\q_1}^2$ is linear on the moduli $t^n$ that appear in the polynomial piece. Also, since  $P_i^2 \q_1 =0$, the mass asymptotes as most to $m_{\q_1}^2 \sim t^{-w+2} e^{-t} \Mpl^2$. Hence, $\gamma_{\q_1}$ always diverges.} this effect can have significant physical consequences, particularly so in emergent string limits where $m_*$ is identified with the species scale, as we will illustrate in the Seiberg--Witten examples of section \ref{s:swpoints}. 

It follows that we can straightforwardly construct a sublattice $\Gamma_{\rm ess}$ of elements with diverging charge-to-mass ratios that is invariant under the monodromy group action. Furthermore, since it is a symplectic sublattice, its complement $\Gamma_{\rm ess}^\perp= \lbrace \q \in \Gamma\, |\, \q^{\mathsf{T}} \eta \q'=0,\, \forall \q' \in \Gamma_{\rm ess}\rbrace$ is also symplectic and moreover satisfies
\begin{eqn}
    \q^{\mathsf{T}} \eta \q'=0 \implies \left(e^{k^iP_i} \q \right)^{\mathsf{T}} \eta \q' =\q^{\mathsf{T}} \eta e^{-k^iP_i} \q'=0\, , \qquad \text{where}\quad \q' \in \Gamma_{\rm ess}\, ,
\end{eqn}
This implies, in turn, that the monodromy group exhibits a block-diagonal structure with respect to $\Gamma_{\rm ess}$ and its symplectic complement $\Gamma_{\rm ess}^\perp$ in the full lattice $\Gamma$. To sum up, we again find that the monodromy group action on $\Gamma = \Gamma_{\rm ess}^\perp \oplus \Gamma_{\rm ess}$ reads
\begin{eqn}
    e^{k^iP_i} = 
    \begin{pmatrix}
       M_{\rm grav} & 0 \\ 0 &   M_{\rm ess}
    \end{pmatrix}\, ,
    \label{blockMIV}
\end{eqn}
where $M_{\rm ess}$ represents the action of the monodromy on $\Gamma_{\rm ess}$ and $M_{\rm grav}$ that on $\Gamma_{\rm ess}^\perp$. This time, however, the latter will be generically non-trivial. Finally, a similar strategy can be applied to construct $\Gamma_{\rm ess}$ in $w=2$ and $w=1$ infinite-distance limits. 

One final comment is in order. In general, the monodromy matrix that one obtains by circling around the singularity is of the form
\begin{eqn}
    {\cal T}_i = {\cal T}_i^{(s)} {\cal T}_i^{(u)}\, ,
    \label{monodec}
\end{eqn}
where ${\cal T}_i^{(s)}$ is of finite order and ${\cal T}_i^{(u)}$ is unipotent. Following \cite{Grimm:2018ohb}, the discussion above centers on the unipotent piece of the monodromy, in the sense that in \eqref{periodv} we take $P_i = \log {\cal T}_i^{(u)}$. Therefore, the statement that a period (or subset of periods) is monodromy-invariant should be understood as so under the unipotent piece ${\cal T}_i^{(u)}$, according to the usual practice in the literature. Notice, however, that finite order monodromies do play an important r\^ole when describing RFT subsectors. Indeed, in the presence of an order $k$ semisimple part one may have that one of the quantised periods is invariant under the action of ${\cal T}_i^{(u)}$ but picks a $\mathbb{Z}_k$ phase upon acting with the full operator \eqref{monodec}. For instance, it might be that
\begin{eqn}
    X^\mu \to e^{\frac{2\pi i}{k}} X^\mu\, .
    \label{phaseX}
\end{eqn}
If that is the case, by the reasoning above the magnetic period will have the expansion \eqref{massessIVmag}, but with $a_{\q_1} =0$, since otherwise  \eqref{phaseX} cannot be true.\footnote{This follows from $[P_i, \mathcal{T}^{(s)}]=0$, which implies that the decomposition of $\Gamma$ in $\mathcal{T}^{(s)}$-eigenspaces is compatible with the weight filtration. Thus, the magnetic periods $\mathcal{F}_\mu$ must also transform as \eqref{phaseX}, hence forcing $a_{\q_1} =0$.}  This means that we will have a magnetic state whose central charge is also controlled by the instanton corrections in \eqref{periodv}, and whose charge-to-mass ratio diverges as fast as its electric dual. In general, we find that the presence of states that are only charged under finite-order monodromies imply an RFT limit. We will illustrate this particular feature in some of the examples of sections \ref{s:conifoldlikesingus} and \ref{s:swpoints}, but leave a more general discussion for the future.

\subsubsection*{Metric essential instantons and special coordinates}\label{sss:specialcoords&instantons}

The discussion above relies on the framework of mixed Hodge structures, different from the one of special geometry exploited previously to describe the physics of RFT limits. In the following, we would like to discuss how these two pictures can be combined, and more precisely how the results above can be understood from the viewpoint of special coordinates.  

The definition of metric essential instanton implies, in particular, that the K\"ahler potential $K_{\rm pol} = - \log \left({\cal P} (\im T^i)\right)$ constructed from the polynomial approximation ${\bf \Pi}_{\rm pol} = e^{T^iP_i} {\bf a}_0$ leads to a degenerate metric along the moduli space $\{T^i\}$ transverse to the singularity. It follows that, in the full metric tensor, some eigenvalues of $g_{T^i \bar{T}^j}$ are suppressed as
\begin{eqn}
    \eps = e^{-2\pi t^i m_i}\, ,
\end{eqn}
for some choice of $m_i \in \mathbb{Z}_{\geq0}$, while the remaining ones depend polynomially on the $t^i$. This hierarchical eigenvalue structure is very suggestive from the viewpoint of the Curvature Criterion, because the exponentially suppressed ones will give rise to curvature divergences through \eqref{IIBscalar}, unless the cubic derivatives $\CF_{ijk}$ that they couple to exhibit a stronger exponential suppression. 

Notice, however, that the variables $\{T^i\}$ do not correspond to special coordinates, with the exception of LCS limits. In general, these periodic fields are well-suited to describe the physics of 4d axionic BPS strings \cite{Lanza:2020qmt,Lanza:2021udy,Lanza:2022zyg}, but not that of BPS particles. Hence, to understand the implications of metric essential instantons for RFT limits it is  more adequate to translate the above statements in terms of special coordinates. From the results of Appendix \ref{ap:special}, one finds 
\begin{equation}\label{detgvsdetN}
    \det g_{T^i\bar{T}^j}=-(|X^0|^2e^K)^{n_V+1}\left|\det\left(\frac{\partial z^i}{\partial 
    T^j}\right)\right|^2\det N_{IJ}\, ,
\end{equation}
where $z^i = X^i/X^0$ are special coordinates and $N_{IJ}$ is the symmetric  matrix defined in \eqref{NIJ}, which has signature $(n_V, 1)$ with the time-like direction given by $N_{IJ} X^I \bar{X}^J = -e^{-K}$. Hence, if $\det g_{T^i\bar{T}^j} = \CO (\eps)$, this will either reflect in the Jacobian determinant or in the corresponding positive eigenvalues of $N_{IJ}$. The latter possibility is, however, unlikely, since if $N_{\mu\mu}$ is such an eigenvalue it means that $k_\mu = \CO(\eps)$, thus \eqref{cond} is satisfied and one should identify this coordinate with an RFT direction. However, one then falls into a contradiction, since $\CO(\eps) \simeq N_{\mu\mu} \simeq - 2\CI_{\mu\mu}$, which is instead polynomial on $\{t^i\}$.

Consequently, let us consider the first case. The fact that $\det g_{T^i\bar{T}^j} = \CO (\eps)$ implies that at least for some electric period $X^\mu \neq X^0$ we must have 
\begin{eqn}
   z^\mu =  X^\mu/X^0 = a^\mu  + \CO(\eps) \, ,
    \label{elemetric}
\end{eqn}
with $a^\mu \in \mathbb{C}$. That is, the BPS mass $m_{\rm D3}(A^\mu)$ of the  electric state $\q_{A^\mu}$ will be of the form
\begin{eqn}
   m_{\mathsf{q}} = \left|\mathsf{q}^{\mathsf{T}} \eta \left( {\bf a}_0 + e^{T^iP_i} \sum_{r_i} {\bf a}_{r_1\dots r_n} e^{2\pi i r_i T^i} \right) \right| m_*\, ,
   \label{massess}
\end{eqn}
with $a_{\q_{A^\mu}} = |a^\mu|$. Whenever $a^\mu \neq 0$ the mass $m_{\q_{A^\mu}}$ flows to the reference scale $m_* =m_{\rm D3}(A^0)$, while if $a^\mu =0$ it will become parametrically lighter. If we now identify this electric state with the $\q_{-1}$ ones considered above, we see that we are in the second case. Namely, our results fix $a_{\q_{A^\mu}} = 0$ and replace $e^{T^iP_i} \to \mathbb{I}$. It follows that we have an RFT limit exhibiting a charge-to-mass ratio $\gamma_{\q_{A^\mu}}$ that diverges exponentially fast for this electrically charged particle.

As for its magnetic counterpart, we have that $\CF_\mu = \CF_{\mu I}X^I$, and so its expansion into polynomial and exponential terms will depend on the second derivatives of the prepotential $\CF_{IJ}$. However, for a decoupling RFT sector only the following structure seems physically acceptable
\begin{eqn}
    \CF_{\mu}/X^0 = b_\mu + \CO(\eps) \, ,
    \label{magmetric}
\end{eqn}
where $b_\mu \in \mathbb{C}$. Otherwise, the magnetic RFT state will behave asymptotically as $\CF_\mu \simeq \CF_{\mu n} X^n$, with $X^n \sim X^0 T^n $ one of the electric periods  in ${\bf \Pi}_{\rm pol}$. This would mean that the BPS mass of this state is specified by fields with Planckian vevs. Generically, this can be avoided by imposing $\CF_{\mu n} = \CO(\eps)$ which leads to \eqref{magmetric} and to $N_{\mu n} = \CO(\eps)$. In other words, gravity decoupling not only demands the approximate block-diagonal structure for $N_{IJ}$ discussed in section \ref{ss:decoupling}, but more precisely that $N_{\mu n} = \CO(\eps)$ for any electric period of the form $z^n = X^n/X^0 \sim  T^n \to i\infty$, which by assumption belongs to the gravitational sector of the limit. As a consequence of \eqref{magmetric}, we again have a mass of the form \eqref{massess} which, upon identifying the magnetic state $\gamma_{\q_{B_\mu}}$ with the charge $\q_1$ above, we can further constrain by replacing $e^{T^iP_i} \to \mathbb{I} + T^i P_i$. As explained, since this is state is never at the bottom of the weight filtration (namely $\q_1 \notin W_{-w}$), its charge-to-mass ratio $\gamma_{\q_{B_\mu}}$ always diverges, independently of whether $b^\mu$ vanishes or not.

Note that this discussion fixes significantly the general form of the K\"ahler potential. Roughly speaking, we have a set of special coordinates $z^n \sim X^n/X^0 \sim T^n$ that correspond to the independent electric periods that appear in the polynomial approximation ${\bf \Pi}_{\rm pol} \equiv e^{T^iP_i} {\bf a}_0$. The rest are exponential coordinates $z^\mu \sim e^{k T^\mu}$ on which the approximate K\"ahler potential $K_{\rm pol}$ cannot depend, or else it would not be possible to have $k_{{\rm pol}, \mu} = 0$ in a region of moduli space. As a result, we recover an expression of the form
\begin{eqn}
    K = -\log \left(\CP_{\rm pol}(\im T^i) - \CQ_{\rm ess}(T^i, \bar{T}^i, z^\mu, \bar{z}^\mu) \right) \, ,
\end{eqn}
where $\CP_{\rm pol}$ is the polynomial that appears in $K_{\rm pol}$, while ${\CQ}_{\rm ess}$ is the piece that arises from the essential instanton corrections. The latter admits an expansion of the type
\begin{eqn}
    \CQ_{\rm ess}/\CP_{\rm pol} =  N_{\mu\nu} z^\mu \bar{z}^\nu  + \im (A_\mu z^\mu) + \ldots\, ,  
\end{eqn}
Let us remark that this yields an expression for the K\"ahler potential that fits within the general one provided in e.g. \cite{Bastian:2021hpc}, although slightly more precise. 

Finally, notice that the behaviour \eqref{magmetric} is a  consequence of a block-diagonal structure \eqref{blockMIV} for the monodromy group $M$ on $\Gamma_{\rm ess}^\perp \oplus \Gamma_{\rm ess}$. Indeed, if the monodromy action is block-diagonal, the element $N_{\mu k}$ with indices $\mu$ and $k$ in two different blocks cannot have any power-like dependence on the saxionic fields $t^i$ that describe the limit. Additionally, it forbids a polynomial or constant coupling of the form  $N_{\mu n}$, with $\q_{A^\mu} \in \Gamma_{\rm ess}$ and $\q_{A^n} \in \Gamma_{\rm ess}^\perp$ such that $X^n/X^0 \sim T^n \to i\infty$, which in turn leads to \eqref{magmetric}. As a result, most of these mixing elements are negligible with respect to $N_{\mu\nu}$, which due to \eqref{calNlimit} and the RFT condition \eqref{cond}, satisfies $N_{\mu\nu} \simeq - 2 \CI_{\mu\nu}$ and so depends polynomially on the $t^i$. That is, the block structure of $N_{IJ}$ is simplified. Reciprocally, the form \eqref{magmetric} ensures that the monodromy action only mixes this magnetic period with electric periods of the kind \eqref{elemetric}, as it happened for the pair $\q_{1}, \q_{-1} \in \Gamma_{\rm ess}$.  We thus see that the physics of gravity-decoupling RFTs in limits with metric essential intantons is tightly related to the construction of the sublattice $\Gamma_{\rm ess}$, as well as to the block-diagonal structure of the monodromy group with respect to it.

\subsection{The core RFT and the Curvature Criterion}
\label{ss:core}

We have seen how, for limits with metric essential instantons, there is a natural RFT sector of the theory  related to a symplectic monodromy-invariant sublattice of charges $\Gamma_{\rm ess}$ with diverging charge-to-mass ratios. This implies that the the monodromy group that describes the singularity has the block-diagonal structure 
\begin{eqn}
    e^{k^iP_i} = 
    \begin{pmatrix}
       M_{\rm grav} & 0 \\ 0 &   M_{\rm ess}
    \end{pmatrix}\, ,
    \label{blockgen}
\end{eqn}
with respect to $\Gamma = \Gamma_{\rm ess}^\perp \oplus \Gamma_{\rm ess}$, where $\Gamma_{\rm ess}^\perp$ is the symplectic complement of  $\Gamma_{\rm ess}$. This structure constrains the rigid gauge kinetic terms $N_{IJ}$ along the lines required by the first decoupling condition discussed in section \ref{ss:decoupling}. In particular, it imposes that the magnetic states of the rigid theory have a mass dependence of the form \eqref{massess}, which as already discussed is what is expected for a gravity decoupling 4d RFT. 

To motivate this block-diagonal structure beyond RFTs based on metric essential instantons, one can interpret it in terms of the graviphoton field. In general, the anti-self-dual piece of the graviphoton direction is given by $T_-= 2i e^{K/2} \CI_{IJ}X^J F^I_-$. Using the definition of the dual field strengths $G_I^-= \bar{\mathcal{N}}_{IJ} F^J_-$ as well as \eqref{eq:constraintscovholosections}, one can rewrite it as follows
\begin{eqn}\label{eq:graviphotondef}
    T_-= e^{K/2} \left(\mathcal{F}_I F^I_- - X^IG_I^-\right)= e^{K/2} \mathsf{F}_-^{\mathsf{T}} \eta\, {\bf \Pi} \qquad  \text{with} \qquad  \mathsf{F_-}= \begin{pmatrix}F^I_- \\ G_I^- \end{pmatrix}\, .
\end{eqn}
This means that the period vector $e^{K/2}{\bf \Pi}$ describes the graviphoton components in a democratic basis of U(1)'s, while $e^{K/2}{\bf \Pi}_{\rm pol}$ gives such a description at the endpoint of the asymptotic limit. In the neighbourhood around this point, the graviphoton will continuously rotate among different U(1)'s through theta angles, and when circling around the moduli space singularity it will transform by the action of the monodromy group. In limits with metric essential instantons, $\Gamma_{\rm ess}$ will map to a sublattice of  quantised U(1) directions that do not mix with the graviphoton via theta-angle rotations. Moreover, because $\gamma_\q \to \infty$ the charge of these U(1)'s will be parametrically larger than its graviphoton component. Reciprocally, those charges $\q$ with $\gamma_\q = \CO(1)$ represent U(1) directions whose graviphoton charge is always significant, and as such they should be involved in the monodromy orbit associated to the latter, implemented by $M_{\rm grav}$. Finally, those charges $\q$ such that $\gamma_\q \to \infty$ but nevertheless belong to $\Gamma = \Gamma_{\rm ess}^\perp$ correspond to rigid U(1) directions with negligible graviphoton component; however, either they or their magnetic duals are still coupled to some gravitational U(1) via the theta angles of the singularity and, as a consequence, through kinetic mixing. Presumably, it is due to this structure that one can build infinite-towers of asymptotically massless states by looking at the orbits that $M_{\rm grav}$ implements on $\Gamma$ \cite{Grimm:2018ohb,Grimm:2018cpv,Corvilain:2018lgw}, even when these do not themselves comprise the leading SDC tower \cite{Hassfeld:2025uoy,Monnee:2025ynn}.

From these observations, we conclude that a symplectic block-diagonal structure for the monodromy matrix around a singularity seems a necessary condition for gravity decoupling. This prompts our definition of core RFT:

\vspace{0mm}
\begin{mdframed}[leftmargin=1cm, rightmargin=1cm]
\begin{center}
{\em The core RFT $\{U(1)_\lambda\}_{\lambda \in \rm cRFT}$ is specified by a subset of electro-magnetic  \\ pairs of three-cycles $\{A^\lambda, B_\lambda\}_{\lambda \in \rm cRFT}$ with diverging charge-to-mass ratios \\ and such that  $\Gamma_{\rm cRFT} \equiv \langle \{A^\lambda, B_\lambda\}_{\lambda \in \rm cRFT}\rangle$ is monodromy invariant.} 
\end{center}
\end{mdframed}
\vspace{2mm}

In general setups, in particular singularities of codimension $\leq n_V$, the presence of RFT sectors is not exclusively linked to metric essential instantons, and so typically $\Gamma_{\rm cRFT}$ will differ from $\Gamma_{\rm ess}$. Nevertheless, due to the core RFT definition above we expect that many of the features observed in the last subsection will still hold. In particular, we expect that the charges $\q \in \Gamma_{\rm cRFT}$ have BPS masses  of the form \eqref{massess}. Namely a mass that, in units of $m_*$, only depends on the the field directions transverse to the singularity $T^i$ through exponentials. Notice that the subset of charges $\q \in \Gamma$ that satisfy such a condition forms a sublattice, that we dub $\Gamma_*$.\footnote{For limits based on growth sectors (see e.g. \eqref{growth} below), the precise definition of $\Gamma_*$ should be slightly modified. Now it should contain those $\q \in \Gamma$ for which $m_\mathsf{q}$ depends on the leading field direction through exponentials.} This includes both $\Gamma_{\rm ess}$ as well as the lattice of monodromy-invariant states of the form $\Gamma_{\rm inv} = \cap_{i} \text{ker}P_i$. Physically, $\Gamma_*$ contains the states of the compactification with lighter BPS masses, many of which have charge-to-mass ratios that blow up along the limit. This is however not true for all of them since, as pointed out in \cite{Hassfeld:2025uoy,Monnee:2025ynn}, for infinite-distance limits those states $\mathsf{q} \in W_{\ell_{\rm min}}$ whose $\gamma_\mathsf{q} = \CO(1)$ contain the leading SDC tower of particles. Thus, even if $W_{\ell_{\rm min}} \subset \Gamma_{\rm inv} \subset \Gamma_*$, these are extremal BPS states which do not belong to $\Gamma_{\rm cRFT}$. Based on our definition above,  one may define the  sublattice $\Gamma_{\rm cRFT} \subset \Gamma_*$ such that:

\begin{itemize}
    \item[-] All of its elements have a diverging charge-to-mass ratio $\gamma_\mathsf{q}$. 

    \item[-] It is invariant under the monodromy group action.

    \item[-] It is symplectic.
    
\end{itemize}
As a consequence of these conditions, the monodromy group takes the block-diagonal form \eqref{blockgen} with respect to $\Gamma = \Gamma_{\rm grav} \oplus \Gamma_{\rm cRFT}$, where we have renamed $\Gamma_{\rm cRFT}^{\perp} \to \Gamma_{\rm grav}$.

While this definition provides a topological description of the core RFT, it does not guarantee that the decoupling conditions of subsection \ref{ss:decoupling} are actually met. Nevertheless, in practice it greatly simplifies checking whether they are verified. Regarding the first decoupling condition, the components of the rigid gauge kinetic matrix $N_{\lambda i}$ that mix core RFT indices and the direction of the limit are suppressed as $\CO(\eps)$ in terms of the saxions $t^i$, while the components $N_{\mu \rho}$ with pure core RFT indices depend polynomially on them. Since $N_{IJ}$ has signature $(n_V, 1)$, with the time-like direction given by $N_{IJ} X^I \bar{X}^J = -e^{-K}$, this means that for the purpose of checking if the latter is block-diagonal with respect to the core RFT, it can be treated as if it was positive definite. In other words, the absence of kinetic mixing boils down to the condition 
\begin{eqn}
    N^{\lambda \rho} N_{\rho \sigma} \simeq \delta^\lambda_\sigma\, ,
    \label{nomixing}
\end{eqn}
where all the indices belong to RFT subset $\{z^\lambda\}$, including those of the sum. To see what else the core RFT condition implies, let us write the period vector in terms of special coordinates
\begin{eqn}
  {\bf \Pi} =  X^0 \begin{pmatrix}
  1 \\ z^i \\ \CF_{00} + \CF_{0j} z^j \\ \CF_{ij} z^j + \CF_{i0}
    \end{pmatrix}\, .
\end{eqn}
The electric periods whose coordinates $z^a$ shift by a constant under the monodromy mix with the SDC tower state represented by $X^0$, and so they do not belong to the core RFT. Let $\{z^\mu\}$ be the rest of the coordinates, and let us assume that all $\{X^\mu\}$ belong to the core RFT. For their magnetic duals to belong to the core RFT as well, they cannot shift by a constant or polynomial term, so it must be that $\CF_{\mu a} = \CO(\eps)$ and $\CF_{\mu 0} ={\rm const.} + \CO(\eps)$. Hence, after imposing the core RFT condition, \eqref{nomixing} is satisfied as long as $N_{00} N_{\lambda \rho} \gg N_{\lambda 0} N_{\rho 0}$. If there are electric periods $\{X^k\}$ that neither shift by the monodromy nor belong to the core RFT, then one should also check that the mixing terms $N_{\mu k}$ with them are negligible.

Regarding the second decoupling condition of subsection \ref{ss:decoupling}, the core RFT definition does not imply any obvious simplification. It is however illustrative to revisit what are its implications. When \eqref{nomixing2} holds, the scalar curvature \eqref{IIBscalarP} can be approximated as 
\begin{eqn}
    R \simeq - 2 n_V(n_V + 1) +  \lVert \mathcal{P} \rVert_{\text{cRFT}}^2 +  \lVert \mathcal{P} \rVert_{\notin \text{cRFT}}^2\,, 
\label{IIBscalarPdec}
\end{eqn}
where
\begin{eqn}
    \lVert \mathcal{P} \rVert_{\text{cRFT}}^2 = -g^{\rho\bar{\sigma}}g^{\tau\bar{\eta}}\CI^{\Lambda \Sigma}\mathcal{P}_{\rho\tau \Lambda}\bar{\mathcal{P}}_{\bar{\sigma}\bar{\eta} \Sigma} =
   \frac{\Mpl^2}{m_*^2} R_{\rm rigid}
    \label{PaulicRFT}
\end{eqn}
is a normalised sum of pure core RFT Pauli terms. The term $\lVert\mathcal{P} \rVert_{\notin \text{cRFT}}^2$ has a similar expression but with indices that only run over fields outside of the core RFT, namely those belonging to the gravitational sector and to the rigid vector multiplets that cannot decouple from it. Given this result, one can give a more precise formulation of the Curvature Criterion:

\begin{itemize}

    \item[-]  $\lVert\mathcal{P} \rVert_{\notin \text{cRFT}}$ never generates a curvature divergence along infinite-distance limits.

    \item[-] If the decoupling conditions are not met, $\lVert\mathcal{P} \rVert_{ \text{cRFT}}^2+\lVert \CP \rVert^2_{\rm mixed}$ do not generate a divergence.
    
\end{itemize}

Notice that this statement is slightly stronger than the one made in \cite{Marchesano:2023thx}, in the sense that it does not allow $\lVert\mathcal{P} \rVert_{\notin \text{cRFT}}$ to develop a subleading divergence with respect to $\lVert\mathcal{P} \rVert_{\text{cRFT}}$. Essentially, this states that decoupled gravitational sectors cannot generate a curvature divergence, recovering the intuition behind the original proposal made in \cite{Ooguri:2006in}. Finally, if both decoupling conditions are met along an infinite distance limit, it is expected that \eqref{PaulicRFT} diverges. The only possible exception is when the core RFT sector displays $\CN=4$ supersymmetry \cite{Marchesano:2024tod}, in which case  $R_{\rm rigid}$ vanishes identically and mutual interactions are not measured by Pauli couplings.  

In the following sections, we will show how the core RFT definition applies to different kinds of limits. This will illustrate how the decoupling conditions can constrain the regions of moduli space where the core RFT decouples from gravity, and how this can be captured by a divergence of the moduli space curvature, as well as by certain quotients of UV scales.

\section{Large complex structure limits}
\label{s:LCS}

Let us illustrate how the concepts of section \ref{s:CYlimits} work in the well-known case of  infinite distance limits in the large complex structure (LCS) regime. The RFT curvature divergences in this regime have been already classified in \cite{Marchesano:2023thx,Marchesano:2024tod}, using the mirror type IIA large volume language. Even if these limits do not contain essential instantons, we will use the results in \cite{Marchesano:2023thx,Marchesano:2024tod} to show how the set of U(1)'s of the EFT split into gravitational and rigid U(1)'s, and how the core RFT definition selects a subset of the latter, which is the source of the curvature divergence. Since most of the details of these limits have been worked out in \cite{Corvilain:2018lgw,Marchesano:2023thx,Marchesano:2024tod} we will be brief in our exposition. The reader not interested in the specifics of these examples may safely skip to the next sections, where limits featuring metric essential instantons are discussed.

The prepotential in the LCS region of moduli space reads \cite{Hosono:1994av}
\be
 {\cal F} = -\frac{1}{6} \CK_{ijk}\frac{X^iX^jX^k}{X^0} + \oh K_{ij}^{(1)}X^iX^j + K_{i}^{(2)}X^i X^0 + \frac{i}{2} K^{(3)}(X^0)^2 + (2\pi i)^{-3} \sum_{\boldsymbol{k} > \boldsymbol{0}} n_{\bm{k}}^{(0)} {\rm Li}_3 \left( e^{2\pi ik_iX^i/X^0} \right) .
\label{fullF}
\ee
The integers $\CK_{ijk}$ stand for the triple intersection numbers of the mirror Calabi--Yau $X_3$, while  the $K^{(a)}$ $a=1,2,3$ also depend on its topological data, namely \cite{Coudarchet:2023mmm}
\begin{eqn}
K_{ij}^{(1)} = \frac{1}{2} {\cal K}_{iij} \, , \qquad K_{i}^{(2)} = \frac{1}{24\ell_s^6} \int_{X_3} c_2(X_6) \wedge \omega_i \, , \qquad
    K^{(3)} = \frac{\zeta(3)}{8\pi^3} \chi(X_3)\, ,
    \label{Kcurv}
\end{eqn}
with $\{\om_i\}$ a integral basis of divisors of $X_3$. Finally, Li$_3(x)$ stands for the 3rd polylogarithmic function and $n_{\bm{k}}^{(0)}$ for the genus zero Gopakumar-Vafa (GV) invariant of the curve class $k_i {\cal C}^i$ in $X_3$, with ${\cal C}^i$ such that $\ell_s^{-2} \int_{{\cal C}^i} \om_j = \delta^i_{j}$. 

Since in this regime the special coordinates are simply given by $z^i = X^i/X^0= T^i = b^i + it^i$, it is easy to separate the prepotential into its polynomial and exponentially suppressed pieces. This structure translates into a similar split of the period vector, that fits into the general Ansatz \eqref{periodv}. The classification of infinite distance limits in terms of monodromy matrices $P_i$ and the approximate periods ${\bf \Pi}_{\rm pol} = e^{T^iP_i} {\bf a}_0$ has been performed in \cite{Corvilain:2018lgw}. In the following, we will exploit this framework using the slightly different conventions of \cite{Escobar:2018rna,Marchesano:2022axe}, which express the periods as
\begin{eqn}
    {\bf \Pi}  = 
    \begin{pmatrix}
        X^0 \\ X^i \\ \CF_0 \\ \CF_i  
    \end{pmatrix} 
    =
e^{T^i\hat{P}_i} Q {\bf a}_0 + \CO(e^{2\pi i T^i})\, ,
\label{PiLCS}
\end{eqn}
where $\hat{P}_j = QP_j Q^{-1}$ are the nilpotent generators of the monodromy group, and\footnote{Note that the ${\bf a}_0$ defined in \eqref{PaQ} above follows a slightly different convention than in \eqref{periodv}. Usually, one absorbs the $T^i_0$ into the exponential, which thus disappear. Here, we separated between non-/diverging pieces.}
\begin{equation}\label{PaQ}
    P_k = \begin{pmatrix} 0 & 0 & 0 & 0\\
    \delta_{ik} &0 & 0 & 0\\
    0 & 0 &0 & - \delta_{kj}\\
    0 & -\CK_{ijk} & 0 &0
    \end{pmatrix}\, , 
    \quad
   Q = \begin{pmatrix}
     1 & 0 & 0 & 0\\
     0 & \delta_{ij} & 0 & 0\\
     0 & K_j^{(2)}  & 1  &0 \\
     K_i^{(2)} & K_{ij}^{(1)} & 0 & \delta_{ij}
    \end{pmatrix}\, , 
    \quad
    {\bf a}_0 = X^0 \begin{pmatrix}
    1 \\ T_0^i \\ \frac{1}{6} \CK_{ijk} T_0^iT_0^jT_0^k  + i  K^{(3)} \\ - \oh \CK_{ijk} T_0^jT_
    0^k 
    \end{pmatrix}\, .
\end{equation}    
The definition of core RFT in this context depends on the particular class of LCS limit. To simplify the discussion, let us focus on the 
subset of EFT string limits \cite{Lanza:2020qmt,Lanza:2021udy,Lanza:2022zyg}, which can be described by taking $\{\om_i \}$ to be a basis of Nef divisors and considering the linear trajectory
\begin{eqn}
    T^j = T^j_0 +  e^j T, \quad \text{with} \quad T = b +i t, \quad t = \phi  \to \infty ,
\label{limitEFTstring}
\end{eqn}
and $e^j \in \mathbb{Z}_{\geq 0}$. The slice of moduli space that is transverse to the singularity amounts to the coordinate $T$, and all the rest are spectator fields on which ${\bf a}_0$ and the coefficients in the instanton expansion depend.\footnote{More generally, one may consider trajectories based on growth sectors of the form
\be
t^i = e^i_0 \phi  +  e^i_1 \phi^{\a_1} +  e^i_2 \phi^{\a_2} + \dots \, ,
\label{growth}
\ee
where $e_m$, $m =0 ,1,2, \dots$, is a set of vectors with non-negative  entries such that $\sum_i e^i_m \cdot e^i_n = 0$, $\forall m \neq n$, and $1> \a_1 > \a_2 > \ldots$\ . The analysis below extends to this case by replacing $e^i$ with the leading-order vector $e_0^i$.}

In this case, the monodromy group is generated by
\begin{equation}\label{Fullmonodromy}
    e^{e^j\hat{P}_j} = \begin{pmatrix} 1 & 0 & 0 & 0\\
    e^i &\delta_{ij} & 0 & 0\\
    \frac16 \mathcal{K}_{klm} e^k e^l e^m + 2 K^{(2)}_k e^k &  K^{(1)}_{kj} e^k +\oh \CK_{jkl}e^ke^l  &1 & - e^j\\
    K^{(1)}_{ik} e^k -\oh \CK_{i kl}e^ke^l & -\CK_{ijk}e^k & 0 &\delta_{ij}
    \end{pmatrix}\, .
\end{equation}    
To see whether this matrix takes a block-diagonal form it is useful to consider those periods in \eqref{PiLCS} that are left invariant under $T \to T +1$, as well as their corresponding charge vectors $\mathsf{q}$. One finds that they are
\begin{eqn}
   \mathsf{q}_{0} = \begin{pmatrix}
       0 \\ 0 \\ 1 \\0
   \end{pmatrix} \to X^0\, , \qquad  \mathsf{q}_{c_i} = \begin{pmatrix}
       0 \\ 0 \\ 0 \\ c_i
   \end{pmatrix} \to c_i X^i\, , \qquad \mathsf{q}_{d^j} = \begin{pmatrix}
       0 \\ -d^j \\ 0 \\0
   \end{pmatrix} \to d^j  \CF_j  \, ,
    \label{invariant}
\end{eqn}
with $ c_i e^i =0$ and $d^j \in \ker {\bf K}$, where ${\bf K}_{ij} \equiv \CK_{ijk}e^k$. It is easy to see that they all fit into the Ansatz \eqref{elemetric} and \eqref{magmetric} with non-vanishing constants, and that their BPS mass is of the form shown in \eqref{massess}, where $m_* = m_{\q_0}$. We then have that (c.f. section \ref{ss:core})
\begin{eqn}
    \Gamma_{\rm inv} = \Gamma_* = \langle \q_0, \q_{c_i}, \q_{d^j} \rangle_{\mathbb{Z}}\, .
    \label{GinvLCS}
\end{eqn}
Nevertheless, let us stress that none of these charges corresponds to an actual essential instanton. Indeed, if we now build the K\"ahler potential from the period vector \eqref{PiLCS}
\begin{eqn}
    {\bf \Pi}_{\rm sl_2}  = e^{T e^i \hat{P}_i} Q {\bf a}_0\, ,
\end{eqn}
we obtain that $K = - \log \left[ (\im T)^w + \text{const.}\right]$ and $m_* \sim \phi^{w/2} \Mpl$, where $w=1,2,3$, denotes the singularity type in the language of \cite{Grimm:2018ohb,Grimm:2018cpv,Corvilain:2018lgw} and the scaling weight in \cite{Marchesano:2022axe,Marchesano:2023thx,Marchesano:2024tod}. In this approximation, the matrix $g_{T^i\bar{T}^j}$ has rank one, being independent of the directions $c_iT^i$ with $c_ie^i =0$. This only 
reflects the fact that we are studying a codimension-one singularity, but it does not imply that $c_iX^i$ correspond to metric essential instantons, since the metric that results from the polynomial terms in \eqref{fullF} (constructed from the approximate period vector ${\bf \Pi}_{\rm pol}  =  e^{T^i\hat{P}_i} Q {\bf a}_0$) is indeed non-degenerate. It then follows that the eigenvalues of $g_{T^i\bar{T}^j}$ different from the leading field direction are polynomially suppressed with respect to the latter, as checked explicitly in \cite[Appendix A]{Marchesano:2023thx}. In terms of the weight filtration \eqref{weight} of $e^i\hat{P}_i$, we have that \cite{Corvilain:2018lgw}
\begin{eqn}
    \q_0 \in W_{-w}\, ,\qquad \mathsf{q}_{c_i},\mathsf{q}_{d^i} \in W_0\, , 
\end{eqn}
where the precise position of the elements $\mathsf{q}_{c_i}$ and $\mathsf{q}_{d^i}$ within the former depends on whether or not they additionally belong to $\im (e^i\hat{P}_i)^k$ for some value of $k$.

Since all of these states are invariant under $\exp (e^i\hat{P}_i)$, to build the core RFT lattice $\Gamma_{\rm cRFT}$, one should take those electro-magnetic pairs in \eqref{invariant} whose charge-to-mass ratio $\gamma_{\mathsf{q}}$ diverges. It follows from the analysis below that these are actually given by
\begin{eqn}
    \Gamma_{\rm cRFT} = \langle \{\mathsf{q}_{c_i}, \mathsf{q}_{d^i}\}_i\rangle_{\mathbb{Z}} \quad \text{such that} \quad c_i d^i = -1\, , \quad c_i e^i = 0\, , \quad d^i \in \ker {\bf K}\, ,
    \label{coreRFTLCS}
\end{eqn}
which in terms of the weight filtration of $e^i\hat{P}_i$ always correspond to elements of $W_0 \setminus W_{-1}$. With respect to the filtration associated to the core RFT generators $\hat{P}_\mu$ they do fall into the subspaces $W_{-1}^\mu$ and $W_{1}^\mu$, respectively, and in this sense they are analogous to the electro-magnetic pairs $\{\q_{-1}, \q_1\} \in \Gamma_{\rm ess}$ that were singled out in metric essential instanton limits (see section \ref{ss:essential}).  

As we did in section \eqref{ss:core}, let us denote by $\Gamma_{\rm grav}$ the symplectic complement of $\Gamma_{\rm cRFT}$ in the full charge lattice $\Gamma$. We then have that $\eta$ leaves invariant the symplectic subspace \eqref{coreRFTLCS} which, together with $\Gamma_{\rm cRFT} \subset \Gamma_{\rm inv}$, implies that for any $\q\in \Gamma_{\rm cRFT}$ one has that
\begin{eqn}
    \mathsf{q}^{\mathsf{T}} \eta e^{e^j\hat{P}_j} = \mathsf{q}^{\mathsf{T}}  e^{- e^j\hat{P}_j^{\mathsf{T}}} \eta = \mathsf{q}^{\mathsf{T}} \eta \qquad \text{and} \qquad  e^{e^j\hat{P}_j} \mathsf{q} = \mathsf{q}\, .
    \label{blockdiagLCS}
\end{eqn}
From here one concludes that in a basis of $\Gamma = \Gamma_{\rm grav} \oplus \Gamma_{\rm cRFT}$ the monodromy generator \eqref{Fullmonodromy} corresponding to the direction of the limit takes the block-diagonal form
\begin{eqn}
    e^{e^j\hat{P}_j} = 
    \begin{pmatrix}
        M_{\rm grav} & 0 \\ 0 & \mathbb{I}
    \end{pmatrix}\, ,
\end{eqn}
with $M_{\rm grav}$ a unipotent matrix. 

In general, each of the monodromy-invariant periods \eqref{invariant} plays a non-trivial r\^ole in the description of rigid LCS limits. In the following, we will highlight their respective physical significance, and from there build a dictionary between the monodromy action and the different sectors of an RFT limit, valid beyond the LCS case. Since every value for the scaling weight $w$ corresponds to a physically different limit \cite{Corvilain:2018lgw,Lee:2019wij}, we will analyse each of them separately, mimicking the works \cite{Marchesano:2023thx,Marchesano:2024tod,Blanco:2025qom}, to which we refer for further details.

\subsubsection*{$w=3$ limits}

These occur when ${\bf k} \equiv \CK_{ijk}e^ie^je^k \neq 0$, and correspond to decompactification limits to a 5d EFT obtained from M-theory on $X_3$. The 5d description appears above the mass scale $m_*$ interpreted as a 5d KK scale, while the core RFT survives as a dynamical theory below the latter. From the mirror type IIA viewpoint, $m_* = m_{\rm D0}$ is the D0-brane mass associated to the charge vector $\mathsf{q}_0$, namely the electric unit charge under the 5d KK U(1), which we associate to the magnetic vector $\q_{\hat{0}} = \eta \q_0$.\footnote{To easily connect with previous works in the literature, in this section we choose to label the different $U(1)_{\q}$ gauge fields via its magnetic monopole charge vector, namely $U(1)_\q \leftrightarrow \eta \q$ (c.f. discussion around \eqref{U1q}).} The rest of the 4d EFT U(1)'s can be associated to an integer divisor class of $X_3$: ${\bm v}_\a = v_\a^i [\om_i]$, $v_\a^i \in \mathbb{Z}$, which represents a D4-brane wrapped on ${\bm v}_\a$ stripped out of B-field-induced electric charges. In other words, it corresponds to the magnetic BPS state \eqref{chcharges} with $p^i = v_\a^i$. In terms of these, we have:

\begin{itemize}
    \item[-] ${\bm v}_T = {\bm e}$. This gives the direction of the limit, and it mixes with the graviphoton\footnote{\label{fnote:graviphotonLV}More precisely, the (anti-selfdual component of the) graviphoton within the large volume regime is given by $T_- =-\frac{X^0}{4} e^{-K/2}\left[F^0_- (1-4 i e^{K}\mathcal{K}_i b^i)+4i e^K \mathcal{K}_j F_-^j\right]$, with $\mathcal{K}_i=\mathcal{K}_{ijk}t^j t^k$. Notice that the field direction $F^0 \cong \q_{\hat{0}}$ always provides a non-trivial contribution to $T_-$, whereas that of $F^i$ will depend on the asymptotics of $\mathcal{K}, \mathcal{K}_i$.} via the monodromy $e^i \hat{P}_i$. One has that $\gamma_{\q_{\bm e}} = \CO(1)$ and that its coupling  behaves as $g_T \sim \phi^{-1/2}$. 

    \item[-] ${\bm v_\mu}$ such that $v_\mu^i {\bf k}_i =0$, with ${\bf k}_i \equiv \CK_{ijk}e^je^k$. These are the rigid U(1)'s of the limit, under which the invariant electric states $\mathsf{q}_{c_i}$ in \eqref{invariant} are charged. If ${\bm v_\mu} \notin \ker {\bf K}$, $\langle \q_{d^\mu}, \q_{c_\mu}\rangle$ is not monodromy invariant (it mixes with $\q_0$) and the gauge coupling asymptotes to $g_\mu \sim \phi^{-1/2}$. 

    \item[-] ${\bm v}_\lambda \in \ker {\bf  K}$. This is the subset of rigid U(1)'s that corresponds to the core RFT, and it matches the invariant magnetic states $\mathsf{q}_{d^i}$ in \eqref{invariant}. Their gauge couplings tend toward a constant in the limit, becoming parametrically larger than those of the other U(1) factors. 
    
\end{itemize}

The charge-to-mass ratio for the magnetic periods of each of these sectors behaves as:
\begin{eqn}
    \gamma_{\q_{\hat{0}}}^2, \gamma_{\q_{\bm e}}^2 = \CO(1)\, , \qquad \gamma_{\q_{\bm v_{\mu\neq\lambda}}}^2 \sim \phi^2\, , \qquad \gamma_{\q_{\bm v_{\lambda}}}^2 \sim \phi^3\, ,
\end{eqn}
with the core RFT ones diverging the fastest. In this latter case, the ratio $\gamma_{\mathsf{q}}$ has some actual physical meaning, since the  magnetic periods of this sector correspond to (anti-)effective shrinkable divisors, and so to actual BPS particles. This is in general not true for the rest of the rigid components in the four-dimensional $\mathcal{N}= 2$ EFT. 

The rigid gauge kinetic matrix \eqref{NIJ} in the basis $\{\q_{\bm e}, \q_{\hat{0}},  \q_{{\bm v}_{{\mu} \neq \lambda}}, \q_{{\bm v}_\lambda}\}$ reads 
\begin{eqn}
  N_{IJ} \sim \begin{pmatrix}
        t & t & \text{const.} &  e^{-t} \\
        t &  t^3 & t &  \text{const.} \\
        \text{const.} & t & t & \text{const.} \\
         e^{-t} &  \text{const.} & \text{const.} & \text{const.}
    \end{pmatrix}\,  ,
\end{eqn}
Notice that the kinetic mixing between the core RFT and the U(1) associated to the direction of the limit $\q_{\bm e}$ is exponentially suppressed, whilst within the RFT sector $\q_{{\bm v}_{{\mu} \neq \lambda}}, \q_{{\bm v}_\lambda},$ and the graviphoton $\q_{\hat{0}}$ all couplings behave polynomially. Hence, even if the core RFT U(1)'s exhibit the largest gauge couplings, there are two different decoupling mechanisms at play. With respect to the gravitational sector, it decouples due to the exponential suppression in the mixed kinetic terms or through diverging charge-to-mass ratios. On the other hand, the decoupling from the rest of the RFT U(1)'s is due to the standard field theory mechanism related to stronger gauge couplings and canonical normalisation of fields. 

These two mechanisms also extend to the Pauli interactions. The Pauli mixing between the core RFT and the gravitational sector is either projected out (along the graviphoton direction, see footnote \ref{fnote:Pauli}) or exponentially suppressed at the holomorphic level. Similarly, the couplings within the rigid sector amount to intersection numbers and are all of order one. However, upon canonical normalisation, all of these terms will be suppressed compared to the core RFT ones, from where \eqref{nomixing2} is then satisfied. This effect is in fact measured by the curvature divergence, which can be seen to behave as 
\begin{eqn}
    R_{\rm div} \simeq \frac{\Mpl^2}{m_*^2} R_{\rm rigid} \sim \phi^3 g_\lambda^{6}\, .
\end{eqn}
If one deforms the EFT string trajectory \eqref{limitEFTstring} to a growth sector of the form \eqref{growth} with $\a_i =\a$, the core RFT gauge couplings will asymptote as $g_\lambda \sim \phi^{-\a/2}$, such that at $\a =1$ the curvature no longer diverges. It is precisely at this point that the core RFT gauge couplings become as weak as the rest of the RFT, and the kinetic and Pauli mixings with the latter become significant. 

Alternatively, following \cite{Castellano:2024gwi} one may describe the lack of decoupling in terms of the following two UV scales
\begin{eqn}
   \Lambda_{\rm wgc} \equiv g \Mpl\, , \qquad  \Lambda_g \stackrel{\text{LCS}}{\equiv} g^{-2} m_*\,  .
\label{scales}
\end{eqn}
For a U(1) at weak coupling, 
$\Lambda_{\rm wgc}$ is the UV completion estimate predicted by the WGC \cite{Arkani-Hamed:2006emk}, while $\Lambda_g$ denotes its QFT counterpart, with $m_*$ playing the r\^ole of cut-off scale. One finds that
\begin{eqn}
    \Lambda_{g,T} \simeq \phi m_* = \Lambda_{\rm sp} \, , \qquad  \Lambda_{g,\mu \neq \lambda} \simeq \phi m_* \, , \qquad  \Lambda_{g,\lambda} \simeq \phi^{\a} m_* \, ,
\end{eqn}
where in order to identify $\Lambda_{g,T} \simeq \Lambda_{\rm sp}$ we used the explicit results in \cite{Marchesano:2022axe} (see also \cite{vandeHeisteeg:2022btw,Martucci:2024trp}), and
\begin{eqn}
    \Lambda_{{\rm wgc},T}, \Lambda_{{\rm wgc},\mu \neq \lambda} \simeq  \Lambda_{\rm sp} \, , \qquad   \Lambda_{{\rm wgc},\lambda} \simeq \phi^{-\a/2} \Mpl = \phi^{\frac{3-\a}{2}} m_* \, .
\end{eqn}
We thus see that the decoupling occurs precisely when  both scales $\Lambda_{\rm wgc}$ and $\Lambda_g$ parametrically decouple from each other and from the species scale, which can only happen for the core RFT. In fact, for this sector we have that $(\Lambda_{{\rm wgc}}/\Lambda_g)^2 \sim R_{\rm div}$, and so we can translate the curvature divergence to an equivalent parametric decoupling between the scales introduced in \eqref{scales} \cite{Castellano:2024gwi}.

\subsubsection*{$w=2$ limits}

For these degenerations, ${\bf k} \equiv \CK_{ijk}e^ie^je^k = 0$ but ${\bf k}_i \equiv \CK_{ijk}e^je^k \neq 0$. They correspond to decompactification limits to a 6d theory realised by F-theory on $X_3$. The SDC tower has $m_*$ again as its mass scale, but now there are two types of electric states, namely $\mathsf{q}_{0}$ and $\mathsf{q}_E \equiv \mathsf{q}_{{\bf k}_i}$ with ${\bf k}_i \equiv \CK_{ijk}e^je^k$, that generate the infinite set. The remaining sectors correspond to:

\begin{itemize}
    \item[-] ${\bm v}_T = {\bm e}$. This defines the direction of the limit and it mixes again with the graviphoton via \eqref{Fullmonodromy}, further exhibiting a $\gamma_{\q_{\bm e}} = \CO(1)$. This time, however, the gauge coupling becomes asymptotically constant, i.e. $g_T \sim \text{const.}$

    \item[-] ${\bm v}_E$ such that $v_E^i {\bf k}_i \neq 0$. This vector corresponds to the 6d KK U(1), it provides another significant contribution to the graviphoton in addition to $\q_{\hat{0}}$ (see footnote \ref{fnote:graviphotonLV}) and thus becomes part of the gravitational sector. One also has that $\gamma_{\q_{E}} = \CO(1)$, with $g_E \sim \phi^{-1}$. 

    \item[-] ${\bm v_\mu}\neq {\bm v}_T$ and such that $v_\mu^i {\bf k}_i =0$. These provide the rigid U(1)'s of the limit under which the electric states $\mathsf{q}_{c_i}$ are charged, once we remove $c_i ={\bf k}_i$. If ${\bm v_\mu} \notin \ker {\bf K}$, the span $\langle \q_{d^\mu}, \q_{c_\mu}\rangle$ is not monodromy invariant, and their gauge couplings asymptote to $g_\mu \sim \phi^{-1/2}$. 

    \item[-] ${\bm v}_\lambda \in \ker {\bf  K}$. Both the magnetic states in \eqref{invariant} and their electrically charged counterparts are monodromy invariant, thereby describing the core RFT. Their physical gauge couplings behave as $g_\lambda \sim \text{const.}$ along the infinite distance limit. 
    
\end{itemize}

The charge-to-mass ratios for the magnetic states of each sector now read:
\begin{eqn}
    \gamma_{\q_{\hat{0}}}^2, \gamma_{\q_{E}}^2, \gamma_{\q_{\mathbf{e}}}^2 = \CO(1)\, , \qquad \gamma_{\q_{\bm v_{\mu\neq\lambda}}}^2 \sim \phi\, , \qquad \gamma_{\q_{\bm v_{\lambda}}}^2 \sim \phi^2\, ,
\end{eqn}
and so the core RFT rations diverge again the fastest.  Furthermore, the rigid gauge kinetic matrix in the basis $\{\q_{\bm e},\q_{\hat{0}}, \q_E, \q_{{\bm v}_{{\mu} \neq \lambda}}, \q_{{\bm v}_\lambda}\}$ reads 
\begin{eqn}
  N_{IJ} \sim \begin{pmatrix}
        \text{const.} & t & t &  \text{const.} & e^{-t} \\
        t & t^2  & t & t & \text{const.} \\
        t & t & t & \text{const.} & \text{const.}\\
        \text{const.} & t & \text{const.} & t& \text{const.} \\
         e^{-t} &  \text{const.} & \text{const.} & \text{const.} & \text{const.}
    \end{pmatrix}\,  .
\end{eqn}
Recall, that the physical gauge couplings are given by gauge kinetic functions $\CI_{IJ}$, which only reduce to $-\oh N_{IJ}$ along the rigid directions. Due to this, the above expression for $N_{IJ}$ is compatible with $g_{\hat{0}}, g_E \sim \phi^{-1}$, and so $g_T \sim \text{const.}$ is the only gravitational U(1) gauge coupling that becomes comparable to the ones of the core RFT. Therefore, in order to be able to decouple the latter from the U(1) associated with the direction of the limit, a suppressed mixing between the two is required, which is not available for any rigid vector boson outside of the core RFT. With respect to all other sectors, the core RFT hosts the largest gauge couplings, and so one can simply use the standard field theory mechanism of decoupling. Again, this holds both at the level of kinetic mixing and of physical Pauli interactions.

\subsubsection*{$w=1$ limits}

In this case, ${\bf k}_i =0$ and one recovers an emergent string limit dual to a heterotic compactification on $K3 \times T^2$ \cite{Lee:2019wij}. The SDC tower contains all the supersymmetric states associated to a fundamental heterotic string of tension $m_*^2$, resulting in a large number of BPS particles beyond the KK replica corresponding to $\q_0$. In terms of divisor classes of $X_3$, one divides the sectors as:

\begin{itemize}
    \item[-] ${\bm v}_T = {\bm e}$. Since ${\bm e} \in \ker {\bf K}$, here this sector corresponds to an invariant magnetic period, which also belongs to the monodromy orbit of the graviphoton (see discussion after \eqref{eq:graviphotondef}). In the type IIA mirror description, this divisor is a K3 surface fibered over a $\mathbb{P}^1$ base, wrapped by a D4-brane that becomes a particle state of the SDC tower, as well as by an NS5-brane that describes the 4d worldsheet of the heterotic dual string. In this case, this sector becomes strongly coupled, since $g_T \sim g_{T\bar{T}}^{-1/2} \sim \phi^{1/2}$. 

    \item[-] ${\bm v_I}\notin\ker {\bf K}$. They  correspond to gravitational U(1)'s. There are invariant electric periods charged under them, but in the type IIA frame they correspond to curves inside the $K3$, and as such they belong to the SDC tower. Their gauge couplings behave as $g_I \sim \phi^{-1/2}$. 

    \item[-] ${\bm e} \neq{\bm v}_\lambda \in \ker {\bf  K}$. They describe the core RFT via electro-magnetic monodromy-invariant pairs. Their gauge couplings asymptote to a constant along the EFT string limit. 
    
\end{itemize}

The charge-to-mass ratios for the magnetic periods of these sectors are:
\begin{eqn}
    \gamma_{\q_{\hat{0}}}^2, \gamma_{\q_{\bm v_{I}}}^2, \gamma_{\q_{\mathbf{e}}}^2 = \CO(1)\, , \qquad  \gamma_{\q_{\bm v_{\lambda}}}^2 \sim \phi\, ,
\end{eqn}
whilst the rigid gauge kinetic matrix in the 
$\{\q_{\bm e}, \q_{\hat{0}},  \q_{I}, \q_{{\bm v}_\lambda}\}$ basis reads
\begin{eqn}
  N_{IJ} \sim \begin{pmatrix}
        e^{-t} & \text{const.} & \text{const.} &  e^{-t} \\
        \text{const.} &  t & t &  \text{const.} \\
        \text{const.} & t & t & \text{const.} \\
         e^{-t} &  \text{const.} & \text{const.} & \text{const.}
    \end{pmatrix}\,  ,
\end{eqn}
Again, the mixed terms between $\q_{\bm e}$ and $\q_{\lambda}$ are exponentially suppressed, which explains why one can decouple the core RFT from the U(1)$_{\q_\mathbf{e}}$ even when the latter exhibits parametrically larger gauge interactions. Note that, with respect to any other vector multiplet sector, the core RFT displays a larger coupling and thus one can apply the standard field theory mechanism. 

\subsubsection*{Summary}

Even if LCS limits do not contain essential instantons, they do display monodromy-invariant charges, each of them with a specific r\^ole in RFT limits. In particular, core RFT vector multiplets correspond to pairs of electro-magnetic charges $(\q_{d^\lambda}, \q_{c_\lambda})$ that are monodromy invariant and moreover disconnected from the SDC tower of states. To sum up, we have found that:

\begin{itemize}
    \item[-] Each monodromy-invariant charge in \eqref{GinvLCS} unrelated to SDC tower states, has a diverging charge-to-mass ratio, signaling the presence of an RFT subsector of vector multiplets. 

    \item[-] Within the RFT U(1)'s, the core RFT ones correspond to electro-magnetic pairs $(\q_{d^\lambda}, \q_{c_\lambda})$ that are both monodromy invariant and unrelated---via \eqref{Fullmonodromy}---to the SDC tower.

    \item[-] The decoupling of the core RFT from other rigid vector multiplets occurs because their gauge and Pauli interactions are parametrically stronger, with the latter measured by the curvature divergence. Indeed, let us extend the analysis made for $w=3$ limits to the other cases. The behaviour of the core RFT couplings can be modified to $g_\lambda \sim \phi^{-\a/2}$ by considering trajectories of the form \eqref{growth}, with $\a_i=\a$. The asymptotic dependence of the SDC scale is also deformed to $\Mpl/m_* \sim \phi^{w/2} g_\lambda^{w-3}$, when taking into account the structure of triple intersection numbers of each limit \cite{Marchesano:2024tod}. It follows that the divergence in the curvature behaves as
    \begin{eqn}
    R_{\rm div} \simeq \frac{\Mpl^2}{m_*^2} R_{\rm rigid} \sim \phi^w g_\lambda^{2w}\, ,
    \label{RdivLCS}
    \end{eqn}
    an thus disappears whenever $g_\lambda \sim \phi^{-1/2}$, which happens precisely when the core RFT couplings become as weak as other couplings within the 4d EFT. 

    \item[-] The decoupling of the core RFT from the gravitational sector follows a different mechanism. In particular, the prepotential couplings of the core RFT to the leading field direction $T$ only appear through exponentially suppressed terms (see section \ref{sss:specialcoords&instantons}). This kind of suppression wins over the effect of transplanckian vevs and canonical normalisation, with the latter being polynomial in the fields $T^i$. Hence, one can decouple the core RFT from  gravity even when the U(1) associated to the field direction $T$ exhibits a parametrically larger gauge coupling, as happens e.g. for emergent string limits. 

    \item[-] One can translate the decoupling conditions as relations between the scales \eqref{scales}, as well as the species cut-off. Using $\Lambda_{\rm sp} \simeq \phi^{-1/2}\Mpl$ and considering a trajectory like \eqref{growth} with $\a_i=\a$, one finds that for a core RFT U(1) with gauge coupling $g_\lambda$, the quotients read
    \begin{eqn}
       \frac{\Lambda_{\rm wgc}}{\Lambda_{\rm sp}} \simeq g_\lambda \phi^{1/2}\, , \qquad 
       \frac{\Lambda_{\rm sp}}{\Lambda_{g}} \simeq  (g_\lambda \phi^{1/2})^{w-1} \, ,\qquad \frac{\Lambda_{\rm wgc}}{\Lambda_{g}} \simeq (g_\lambda \phi^{1/2})^{w} \, .
        \label{decouplingLCS}
    \end{eqn}
    More precisely, the behaviour of these scale ratios indicates whether the RFT decouples from gravity or not. A divergence will signal that this is achieved, while for $g_\lambda \sim \phi^{-1/2}$ there is a lack of decoupling even for core RFT U(1)'s. Notice that $\Lambda_g/\Lambda_{\rm sp}$ fails to encode this information for $w=1$ limits. Finally, applying these criteria to gravitational U(1)'s---which are significantly coupled to gravity---does not seem to give any useful information.

\end{itemize}

To great extent, these lessons are independent of the specific structure behind LCS limits, in the sense that they only rely on the definition of core RFT. In the upcoming sections, we would like to argue that they also apply to finite- and infinite-distance points beyond the LCS regime. As shown in \cite{Bastian:2021eom}, for such limits essential instantons are indeed generic, even when we do not freeze any moduli and we consider singularities of complex codimension $n_V$.  While many details will differ, part of the intuition gained in the present chapter will extend to these cases as well. 


\section{Conifold-like singularities}
\label{s:conifoldlikesingus}

In this section, we apply the picture obtained in section \ref{s:CYlimits} to one of the simplest types of singularities, namely the conifold. As it is well-known, finite-distance singularities in the moduli space---like the conifold---can also lead to curvature divergences \cite{Candelas:1990rm,Strominger:1995cz}. These tend to be milder than their infinite distance counterparts, but nevertheless share some properties with them \cite{Castellano:2024gwi}. In fact, the origin of the singularity itself is purely field-theoretical, which hints to the presence of an underlying RFT. In the following, we will confirm this expectation by applying our general RFT and gravity-decoupling criteria to this case. Finally, we extend this analysis to the class of conifold-like infinite-distance singularities dubbed coni-LCS \cite{Demirtas:2020ffz,Alvarez-Garcia:2020pxd}.

\subsection{The conifold}\label{ss:conifold}

Let us thus consider a conifold degeneration in a one-dimensional vector multiplet moduli space. The associated leading prepotential in some convenient integral symplectic basis reads \cite{Strominger:1995cz}
\begin{eqn}\label{eq:prepotentialconifold}
	\mathcal{F} (X^0, X^1) =  \frac{\alpha}{2} (X^0)^2 + \left( \frac{\delta}{2} - \frac{k}{8\pi  i}\right) (X^1)^2 + \beta X^0 X^1 + \frac{k}{4 \pi i} (X^1)^2 \log \left( \frac{X^1}{X^0}\right) + \ldots\, ,
\end{eqn}
with $k\in \mathbb{N}$ denoting the order of a certain finite subgroup of $SU(2)$ (see below), and $\lbrace \alpha, \beta, \delta \rbrace$ being some model-dependent complex coefficients satisfying the constraint \cite{Bastian:2023shf}
\begin{eqn}
     \label{eq:constraintcoeffsconi}
	\delta+\tfrac{k}{2\pi i} (1+\log A_1)=\tfrac{\beta^2-|\b|^2}{\alpha-\bar\a}\, .
\end{eqn}
Here $A_1$ is a non-vanishing complex number associated to the single metric essential instanton in the conifold geometry, c.f. section \ref{ss:essential}. From \eqref{eq:prepotentialconifold} one deduces the period vector
\begin{equation}\label{eq:periodvectorconifoldgeneral}
    \mathbf{\Pi}_{\rm con} = 
    \begin{pmatrix}
    X^0 \\
    X^1 \\
    \CF_0 \\
    \CF_1
\end{pmatrix}
    =
    X^0 \begin{pmatrix}
    1\\
    z \\
    \alpha + \beta z\\
    \beta + \left(\frac{k}{2\pi i} \log z + \delta\right) z 
\end{pmatrix}\,  +\, \mathcal{O}(z^2) \, ,
\end{equation}
with $z = X^1/X^0 = e^{2\pi i (b + i t)}$ the special coordinate.  We only display the leading order terms in an expansion around the conifold locus $z=0$. The corresponding K\"ahler potential \eqref{Kahlerpot} is
\begin{eqn} \label{eq:kahlerpotconifoldgeneral}
    e^{-K}  = \left| X^0\right|^{2}\, \left[ -2\text{Im}\, \alpha  - 2\text{Im}\, \beta \left( z+\bar{z}\right) + \frac{k}{2\pi} |z|^2 \log |z|^2 - 2\text{Im}\, \delta |z|^2 -\frac{k}{4\pi}(z^2+\bar{z}^2)+\ldots\right]\, .
\end{eqn}
We then find 
\begin{align}
    \left| X^0\right|^{-2}\, e^{-K}\big\rvert_{z =0} = -2\text{Im}\, \alpha\, ,
    \label{conimstar}
\end{align}
requiring $\alpha$ to have some strictly negative imaginary part. Note that this is nothing but $\Mpl^2/m_*^2$ at the conifold point, which is therefore finite---as expected for any finite-distance singularity. The rest of the EFT couplings are:
\begin{eqn}
    g_{z\bar{z}} = \frac{k}{2\pi \im \alpha}\, \log \left|\frac{z}{A_1}\right| +\CO(z)  \, ,
\end{eqn}
and
\begin{align}
 {\cal F}_{IJ} =  
    \begin{pmatrix}
        \alpha & \beta \\ \beta & \frac{k}{2\pi i} ( \log z+1) + \delta
     \end{pmatrix} + \CO(z) \, ,
        \qquad
    {\cal N}_{IJ} =\begin{pmatrix}
         \alpha & \beta \\  \beta &  -\frac{k}{2\pi i}\left(\log \frac{\bar{z}}{|A_1|^2}-1\right)  + \delta 
     \end{pmatrix} + \CO(z)\, .
     \label{FNconi}
\end{align}

In this case, one may connect the supergravity EFT with a rigid $\mathcal{N}=2$ Abelian gauge theory by simply expanding the K\"ahler potential \eqref{eq:kahlerpotconifoldgeneral} as follows\footnote{Note that the purely (anti-)holomorphic terms inside $K_{\rm rig}$ can be removed by a K\"ahler transformation \cite{Bastian:2021eom}. Still, they capture through $\beta$ relevant physical information such as the monopole mass or the graviphoton projector.}
\begin{align}
    K = - \log \left( -2\text{Im}\, \alpha\right) - \frac{1}{2 \text{Im}\, \alpha} \underbrace{\left[  2\text{Im}\, \beta \left( z+\bar{z}\right) - \frac{k}{2\pi} |z|^2 \log |z|^2 + 2\text{Im}\, \delta |z|^2 +\frac{k}{4\pi}(z^2+\bar{z}^2)\right]}_{K_{\rm rig}} + \ldots\, ,
\end{align}
where we denote by $K_{\rm rig}$ the rigid K\"ahler potential measured in units of $m_*$, with metric
\begin{eqn}\label{eq:rigidmetricconi}
    g_{z\bar{z}}^{\rm rig} = - \frac{k}{\pi}\log \left|\frac{z}{A_1}\right| + \frac{2(\im \beta)^2}{\im \alpha} +\CO(z)\, .
\end{eqn}
Such a K\"ahler potential can, in turn, be obtained by integrating out $k$ identical hypermultiplets of mass $|z|m_*$ in the context of 4d $\CN=2$ QFT \cite{Strominger:1995cz,Gopakumar:1997dv}.

Let us now illustrate how this system is analysed from the viewpoint of section \ref{s:CYlimits}. Clearly, this is an example with metric essential instantons, since upon truncating \eqref{eq:periodvectorconifoldgeneral} at polynomial level in the periodic variable $T = \frac{1}{2\pi i}\log z$, we obtain a period vector with constant entries, and thus a vanishing moduli space metric. More precisely, using that the monodromy matrix, which is defined in terms of $T$ via the relation ${\bf \Pi}_{\rm con}(T + 1) = {\cal T} \cdot {\bf \Pi}_{\rm con}(T)$, has the form
\begin{eqn}
{\cal T} = e^{T P} =
\begin{pmatrix}
    1&0&0&0 \\
     0& 1&0&0 \\
    0& 0 & 1&0\\
    0&k&0& 1
\end{pmatrix}\, ,\qquad  \text{with}\quad P =
\begin{pmatrix}
    0&0&0&0 \\
     0& 0&0&0 \\
    0& 0 & 0&0\\
    0&k&0& 0
\end{pmatrix}\, ,
\label{monocon}
\end{eqn}
one may rewrite the period vector \eqref{eq:periodvectorconifoldgeneral} as (c.f. \eqref{periodv})
\begin{eqn}
    {\bf \Pi}_{\rm con} 
    =
e^{T P} \left( {\bf a}_0 + \sum_{n \in \mathbb{N}} {\bf a}_{n}\, e^{2\pi i n T} \right)\, ,
\label{eq:NilpotentApproxConi}
\end{eqn}
where the first two terms in the instanton expansion above are given by
\begin{eqn}
    {\bf a}_0  = (1,0,\a, \b)^{\mathsf{T}}\, ,\qquad {\bf a}_1  = (0,1,\b, \delta)^{\mathsf{T}}\, .
\label{eq:instantonvectorsconi}
\end{eqn}
In addition, we have that 
\begin{eqn}
    P\cdot {\bf a}_0  =0\, ,\quad P \cdot {\bf a}_1  = (0,0,0, k)^{\mathsf{T}}\, ,\quad P^2=0\, ,
\end{eqn}
from where we deduce that ${\bf a}_1$ indeed corresponds to the unique metric essential instanton.\footnote{\label{fnote:2ndessentialinstanton}Note that one still needs an additional (essential) instanton correction in \eqref{eq:NilpotentApproxConi} of the form ${\bf a}_2 = -\frac{ik}{4\pi}(0,0,1, \lambda)^{\mathsf{T}}$, for some $\lambda \in \mathbb{C}$, so as to fully reproduce the entire complex three-form cohomology, see section \ref{ss:essential} for details.} The details of the construction of the weight filtration associated to the log-monodromy matrix defined in \eqref{monocon} have already been worked out in section \ref{sss:EssentialFiniteDistance}. Here, we will focus instead on identifying which charge vectors belong to what subspaces so as to be able to pinpoint the core RFT states purely from a topological point of view. Let us start from those elements which are invariant under the monodromy group generated by $\mathcal{T}$. One readily sees that they are given by
\begin{eqn}
   \mathsf{q}_{0} = \begin{pmatrix}
       0 \\ 0 \\ 1 \\0
   \end{pmatrix}\, , \qquad  \mathsf{q}_{\hat{0}} = \begin{pmatrix}
       1 \\ 0 \\ 0 \\ 0
   \end{pmatrix}\, , \qquad \mathsf{q}_{-1} = \begin{pmatrix}
       0 \\ 0 \\ 0 \\1
   \end{pmatrix}\, ,
    \label{eq:invariantstatesconifold}
\end{eqn}
and linear combinations thereof. Out of these, the first two cannot be obtained upon acting with the operator $P$ on any quantised charge vector, as opposed to the last one, which thus belongs to $W_{-1}$ (hence the notation). In fact, its associated BPS mass can be seen to be of the form \eqref{massessIel}, thereby providing as candidate for core RFT U(1) the field strength to which it couples electrically, namely $F^1$ (c.f. \eqref{U1q}). Following the discussion in section \ref{sss:EssentialFiniteDistance}, one may select the element $\q_1=(0,1,0,0)^{\mathsf{T}} \in W_1\setminus W_0$ as magnetic dual to $\q_{-1}$ by minimising the BPS mass \eqref{massessI}. Therefore, noting that $P\cdot\q_1= k \q_{-1}$, we conclude that the integer span of these pair of states generates the lattice $\G_{\rm ess}$ which is, by construction, monodromy invariant. Similarly, its symplectic complement $\G_{\rm ess}^{\perp}$ can be seen to be generated by the first two vectors in \eqref{eq:invariantstatesconifold}, such that one has that $\Gamma = \Gamma_{\rm ess}^\perp \oplus \Gamma_{\rm ess}$, obtaining a block diagonal structure for the monodromy.

Nevertheless, to be able to claim that $\G_{\rm ess}$ corresponds to an actual core RFT, we need to show that the charge-to-mass ratio of all its elements diverges at the limit endpoint. By studying this quantity for the basis of the full charge lattice introduced above, we easily see that
\begin{eqn}
    \gamma^2_{\q_0}, \gamma^2_{\q_{\hat{0}}} \stackrel{z\to 0}{\longrightarrow} 1  \, , \qquad  \gamma_{\q_{-1}}^2 \simeq \frac{2\pi \im \alpha}{k |z|^2 \log |z|} \stackrel{z\to 0}{\longrightarrow} \infty \, , \quad  \gamma^2_{\q_1} \simeq \frac{k \im \alpha}{2\pi |\beta|^2}\log |z| \stackrel{z\to 0}{\longrightarrow} \infty \, .
    \label{conigammas}
\end{eqn}
This signals that the U(1)'s associated to $\q_0, \q_{\hat0},$ comprise the gravitational sector of the theory, whilst the electric-magnetic pair given by U(1)$_{\q_{-1}}$ and U(1)$_{\q_{1}}$ should indeed correspond to a rigid vector multiplet verifying the condition \eqref{cond}.

\noindent Furthermore, note that upon identifying $\CF_1/X^0$ and $\partial_z \mathscr{F}_{\rm rig} (z)$, one obtains the rigid prepotential 
\begin{eqn}
    \label{eq:rigidprepotconifold}
	\mathscr{F}_{\rm rig} (z) = \beta z + \oh \left(\frac{k}{2\pi i} \left(\log z -\oh\right) + \delta\right) z^2 + \dots \, ,
\end{eqn}
which can also be retrieved by truncating \eqref{eq:prepotentialconifold} to its dependence in $z = X^1/X^0$. From \eqref{eq:rigidprepotconifold} one also finds the rigid K\"ahler potential by using the standard relation $K_{\rm rig} =  2\im \left[ \bar{z}^{\bar{\mu}} \p_{z^\mu} \mathscr{F}_{\rm rig} \right]$, and from there the rigid metric shown in eq. \eqref{eq:rigidmetricconi}. 

One interesting aspect of this setup is that the divergence rates for $\gamma_{\q_{-1}}$ and $\gamma_{\q_1}$ are quite different, despite the electro-magnetic symmetries of $\gamma_{\mathsf{q}}$ described around \eqref{gammar}. This difference comes from the fact that $\CF_{IJ}$ in \eqref{FNconi} is not diagonal, or equivalently due to the presence of the constant $\beta$ in the magnetic period $\CF_1$, which translates into a magnetic monopole charged under the rigid U(1) whose mass tends to $m_{\rm mon} = |\beta|m_*$ as $z \to 0$. When $\beta= r+ s\a$ with $r, s\in \mathbb{Q}$, one can find an integral symplectic trasformation that effectively removes it from the magnetic period or, equivalently, an appropriate charge vector $\mathsf{q} \in \mathbb{Z}^4$ where it does not appear. In practice, this happens when a finite order monodromy is present \cite{Bastian:2023shf}, in agreement with our discussion in section \ref{s:CYlimits}.\footnote{Indeed, notice that because of \eqref{conimstar}, the pair $(X^0, \CF_0)$ cannot transform under a $\mathbb{Z}_N$ monodromy acting on $z$, whilst $(X^1, \CF_1)$ can if $\beta=0$. Moreover, in such periods one may replace $z \to z^{1/N}$ (see \eqref{monodec} and below).} With $\beta=0$ one has that $\gamma_{\q_1} = \gamma_{\q_{-1}} + \CO(z)$ as $z\to 0$, reflecting the fact that the mass of this state behaves analogously to the 't Hooft-Polyakov monopole \cite{tHooft:1974kcl,Polyakov:1974ek}. 

However, in almost every conifold degeneration known in the literature (see e.g. \cite{Candelas:1990qd}) the parameters $r,s$ are trascendental numbers, which means that $\beta$ can sometimes be made small by an integral symplectic trasformation, but not entirely removed. From the viewpoint of the full EFT one can see that $\beta$ corresponds to a mixing between the rigid U(1) and the graviphoton.\footnote{More precisely, the anti-self dual component of the graviphoton field strength, defined as $T_-=T_I F^I_-\,$, reduces to $T_I\to \sqrt{2 \im \a} \left(1, \im \b/\im \a\right)$ along the limit $z\to 0$.} One may thus wonder if its presence prevents the RFT to decouple from the gravitational sector. To answer this question, let us apply our decoupling criteria of subsection \ref{ss:decoupling} for a core RFT. Regarding the moduli space metric, there is no possible kinetic mixing with another vector multiplet; however notice that
\begin{eqn}\label{rigcondconifold}
    g^{z\bar{z}} e^K|X^0|^2 g_{z\bar{z}}^{\rm rig} = 1 - \frac{2\pi}{k\log |z/A_1|} \frac{(\im \beta)^2}{\im \alpha} +\CO(z/\log|z|)\, .
\end{eqn}
In other words, the first decoupling criterion is satisfied at a polynomial rate in the variable $t$ if $\im \beta \neq 0$, while it is achieved exponentially fast when $\im \beta = 0$. A similar statement can be obtained from the matrix $\CN_{IJ}$, which is diagonal up to exponential terms in $T$ when $\beta$ is absent.

Before coming back to the r\^ole of this $\b$-parameter and its link to the coupling with gravity, let us analyze the second decoupling criterion. As described in section \ref{ss:decoupling}, this condition aims at checking that the gaugino-gauge boson interactions do not mix the core RFT with other U(1)'s. To do this, we should have a look at the normalised Pauli coupling involving the subsector that we previously identified as the rigid one. In this extremely simple case, where the rigid direction corresponds to the only field-space variable that is available---i.e. $z$, we find
\begin{equation}\label{rigidcon}
    \lVert \CP \rVert^2_{\in\text{cRFT}} =- \CI^{11} g^{z\bar{z}} g^{z\bar{z}} \CP_{111} \bar{\CP}_{1\bar{1}\bar{1}} \simeq \frac{\pi \im \alpha}{k|z|^2 (\log |z/A_1|)^3}\, ,
\end{equation}
which must be compared with the other normalised sum of Pauli couplings that can be written, namely
\begin{equation}\label{mixedcon}
    \lVert \CP \rVert^2_{\rm mixed}= - 2\CI^{10} g^{z\bar{z}} g^{z\bar{z}} \text{Re}\left(\CP_{111} \bar{\CP}_{0\bar{1}\bar{1}}\right) - \CI^{00} g^{z\bar{z}} g^{z\bar{z}} \CP_{011} \bar{\CP}_{0\bar{1}\bar{1}} \simeq \frac{ \pi  \im\b }{k  (\log (|z/A_1|))^3}\frac{z+\bar{z}}{|z|^2}\,.
\end{equation}
In fact, since the rigid direction is the only one that exists, we do not have any further $\lVert \CP \rVert^2_{\rm grav}$. Hence, for us to be able to claim the absence of Pauli mixing, we just need to show that \eqref{nomixing2} holds, as may be readily confirmed by direct inspection. Indeed, the ratio $\lVert \CP \rVert^2_{\in\text{cRFT}}/\lVert \CP \rVert^2_{\rm mixed}$ vanishes at a rate $|z|$, signaling a very fast (exponential) decoupling. Notice, though, how this mixing term is again controlled by $\im\b$. Moreover, when the latter is absent one observes that the decoupling is achieved even faster, with the quotient $\lVert \CP \rVert^2_{\in\text{cRFT}}/\lVert \CP \rVert^2_{\rm mixed}$ vanishing like $|z|^2$.

\noindent To conclude with this example, we compute the scalar curvature. At leading order it reads
\begin{equation}\label{eq:conifoldcurvature}
    R\, \simeq\, \frac{\pi \im \alpha}{k|z|^2 (\log |z/A_1|)^3}\, ,
\end{equation}
and hence diverges when $z\to0$, as it is well known.

As a final check, we can also compare the sectional curvature on the rigid slice with the Riemann tensor of the rigid theory itself. This corresponds to comparing $ R_{z\bar z z\bar z} \simeq \oh \CP_{M zz} \bar{\CP}_{N \bar{z}{\bar{z}}} \CI^{M N}$ with $R_{z\bar zz\bar z}^{\rm rigid}\simeq \oh \frac{\Mpl^2}{m_*^2} |\CP_{zzz}|^2 \CI^{zz}$, thereby verifying (see discussion around \eqref{curvRFT})
\begin{equation}
    R_{z\bar zz\bar z}\, \simeq\,  \frac{m_*^2}{\Mpl^2} R_{z\bar zz\bar z}^{\rm rigid}=-\frac{k }{8\pi\im \alpha}\frac{1}{|z|^2 \log |z/A_1|}\;.
\end{equation}

\subsection{Coni-LCS singularities}\label{ss:coniLCS}

Coni-LCS singularities have been recently considered in the literature of string compactifications in order to engineer type IIB flux vacua with small superpotentials \cite{Demirtas:2019sip,Demirtas:2020ffz,Alvarez-Garcia:2020pxd} (see also \cite{Marchesano:2021gyv}). In the following, we will analyse them using the results of \cite{Bastian:2021eom}, where a general description of two-moduli coni-LCS singularities was given. Within this setting, the integral period vector expanded around the locus $x^1 = x^2 = 0$ up to $\mathcal{O}(x^3)$ reads (we fix $k=1$ as compared to \eqref{eq:prepotentialconifold})
\begin{equation}\label{periodconiLCS}
    \mathbf{\Pi}_{\rm cLCS}= \begin{pmatrix}
        X^0 \\ X^1 \\ X^2\\ \CF_0  \\ \CF_1 \\ \CF_2
    \end{pmatrix}= X^0 \left(\begin{array}{c}
         1  \\
         A_1x^1+A_2(x^1)^2 \\
         \frac{\log x^2}{2\pi i}\\
         -\frac{(\log x^2)^3}{48\pi^3 i}- \varepsilon_2 \frac{\log x^2}{2\pi i} -i \delta_2 - i
          \delta_1 A_1 x^1-\frac{(A_1 x^1)^2}{4\pi i}\left(1+n\log x^2-\frac{4\pi\delta_1 A_2}{A_1^2}\right)\\
         -i\delta_1+\frac{A_1 x^1 }{2\pi i}\left(\log x^1+n\log x^2\right)+\frac{A_2 (x^1)^2 }{2\pi i}\left(\log x^1+n\log x^2+\tfrac12\right)\\
                  \frac{(\log x^2)^2}{8\pi^2}  - \varepsilon_1 \frac{\log x^2}{2\pi i} - \varepsilon_2 +\frac{A_1^2 n}{2}(x^1)^2
    \end{array}\right)\, ,
\end{equation}
where $A_1,A_2\in \mathbb{C}$, $\delta_1,\delta_2, \varepsilon_1, \varepsilon_2 \in \mathbb{R}$ and $n \in \mathbb{Z}_{\geq 0}$ denote some model-dependent parameters.\footnote{We have added $\varepsilon_1, \varepsilon_2$ to the expressions in \cite[Appendix B]{Bastian:2021eom} in order to obtain a vector of quantised periods. By mirror symmetry considerations \cite{Hosono:1994av,Mayr:2000as}, these parameters should satisfy $\varepsilon_1 = \oh \mod \mathbb{Z}$ and $\varepsilon_2 = \frac{1}{12} \mod \mathbb{Z}$.} They correspond to the prepotential
\begin{eqnarray}\nonumber
    \mathcal{F}& =& -\frac16\frac{(X^2) ^3}{X^0} +\frac{n}{2}  \frac{X^2(X^1)^2}{X^0} - \oh \varepsilon_1 (X^2)^2 - \varepsilon_2 X^2X^0 - \frac{i  \delta_2}{2}(X^0)^2-i  \delta_1 X^0X^1\\&& -\frac{1}{4 \pi i}  (X^1)^2\log\left(\frac{X^1}{X^0}\right) +\frac{i }{8 \pi }  \left(1+2\log A_1\right)(X^1)^2 + \dots\, ,
   \label{prepconiLCS}
\end{eqnarray}
with the special coordinates $z^i=\frac{X^i}{X^0}$ being to leading order given by $z^1=A_1x^1$, $z^2=\frac{\log x^2}{2\pi i}$.

To fully appreciate the similarities with the conifold case, we shall compute the relevant physical couplings for this class of theories as well. Let us start from the K\"ahler potential:
\begin{align}\label{coniLCSkahlerpot}
    e^{-K} |X^0|^{-2}
    =&\,\frac{4}{3}(\im z^2)^3+2\delta_2\\\notag
    &-\frac{|z^1|^2}{2\pi}(4n\pi\im z^2-\log\left|\frac{z^1}{A_1}\right|^2)+\frac1{4\pi}(4\pi n\im z^2-1)((z^1)^2+(\bar z^1)^2)+2\delta_1(z^1+\bar z^1)\, ,
\end{align}
where we have included terms up to $\mathcal{O}(|z^1|^2)$, thereby neglecting higher powers that cannot be trusted in the expansion provided in \eqref{periodconiLCS}. Notice how, upon setting $z^1=$ const., we retrieve a typical LCS K\"ahler potential with cubic and constant terms at the polynomial level, whilst fixing $z^2$ instead allows us to identify \eqref{coniLCSkahlerpot} with that of a conifold singularity located at $z^1=0$. The rest of the EFT couplings (within the vector multiplet sector) read as follows
\begin{subequations}\label{eq:couplingsconiLCS}
\begin{equation}\label{metricconiLCS}
    g_{i\bar j }=\frac{1}{\left(\im z^2\right)^4}\left(\begin{array}{cc}
        \frac{3}{2}\left(n \im z^2-\frac{1}{2\pi }\log\left|z^1{/ A_1}\right|\right)\im z^2& \frac{9}{2}i\delta_1 \\
        -\frac{9}{2}i\delta_1 & \frac{3}{4}\left(\im z^2\right)^2
    \end{array}\right)\,,
\end{equation}
\begin{equation}\label{FABconiLCS}
  \CF_{IJ}= -\left(
\begin{array}{ccc}
 \frac{1}{3 }(z^2)^3+i \delta _2 & i \delta _1 & -\frac{1}{2 }(z^2)^2+\epsilon _2 \\
 i \delta _1 & -nz^2+\frac{i}{2\pi}\left(\log\left(\frac{z^1}{A_1}\right)+1\right) & 0 \\
 -\frac{1}{2 }(z^2)^2+\epsilon _2& 0 & z^2+\epsilon _1 \\
\end{array}
\right)\,,
\end{equation}
\begin{equation}\label{NIJconiLCS}
  N_{IJ}= \left(
\begin{array}{ccc}
  \frac{2}{3}(\im z^2)^3 &  -2\delta _1 & 2 \re z^2 \im z^2 \\
 -2\delta _1 &  2n \im z^2-\frac{1}{\pi }\log\left|z^1{/ A_1}\right| & 0 \\
 - \re z^2 \im z^2 & 0 & -2\im z^2 \\
\end{array}
\right)\,,
\end{equation}
\begin{equation}\label{gaugekinconiLCS}
  \CI_{IJ}= \left(
\begin{array}{ccc}
 -\frac{1}{6}(\im z^2)^3& \frac{\delta _1}{2} & \frac{1}{2} \re z^2 \im z^2 \\
 \frac{\delta _1}{2} &  -n \im z^2+\frac{1}{2\pi }\log\left|z^1\right|& 0 \\
 \frac{1}{2} \re z^2 \im z^2 & 0 &-\frac{1}{2}\im z^2 \\
\end{array}
\right)\,,
\end{equation}
\end{subequations}
where we have only included the first few non-vanishing polynomial contributions in $\im z^2$ and $\log|z^1|$ within $g_{i\bar j }$, $N_{IJ}$, and $\CI_{IJ}$, whereas $\CF_{IJ}$ also displays some of the exponential corrections. 

From those, and following our discussion in section \ref{ss:rigid}, one may already identify the rigid U(1)'s in this example by looking at whether \eqref{cond} is satisfied for some scalar direction. Indeed, comparing \eqref{metricconiLCS} with \eqref{NIJconiLCS}, it is easy to see that the vector multiplet associated to the special coordinate $z^1$ verifies this criterion, thus defining an RFT subsector of the theory. However, to ensure that the latter actually decouples from the gravitational fields, several additional conditions must be met. Before checking these, and mimicking the analysis of previous sections, it becomes useful to show how to identify the core RFT sector from a topological perspective.  

The monodromy matrices that act on $\mathbf{\Pi}_{\rm cLCS}(x^i)$ upon circling the singular point read as
\begin{eqn}\label{eq:monodromiesconiLCS}
    {\cal T}_1 = \begin{pmatrix}
    1&0&0&0& 0 & 0 \\
     0& 1&0&0& 0& 0 \\
    0& 0 & 1&0& 0 & 0\\
    0&0&0&1&0& 0 \\
    0&1&0&0& 1 &0\\0&0&0& 0& 0 & 1 \\
\end{pmatrix} \, , \qquad {\cal T}_2 = \begin{pmatrix}
  1&0&0&0& 0 & 0 \\
     0& 1&0&0& 0& 0 \\
    1 & 0 & 1&0& 0 & 0\\
    
    -2\varepsilon_2+\frac{1}{6}&0&-\varepsilon_1+\oh&1&0& -1 \\0& n&0&0& 1 &0\\
    -\varepsilon_1 -\oh &0&-1& 0& 0 & 1 \\
\end{pmatrix} \, ,
\end{eqn}
which are unipotent of order 2 and 4, respectively. Moreover, their associated log-monodromy operators $P_1,P_2,$ are easily shown to coincide with those displayed in eqs. \eqref{monocon} and \eqref{PaQ} above, upon appropriately identifying the constant coefficients. The latter also allow us to rewrite the period vector \eqref{periodconiLCS} as
\begin{eqn}
    {\bf \Pi}_{\rm cLCS} 
    =
e^{T^i P_i} \left( {\bf a}_0\ + \sum_{(r_1,r_2) \in \mathbb{Z}^2_{\geq0}\setminus (0,0)} {\bf a}_{r_1, r_2} e^{2\pi i \left(r_1 T^1+r_2 T^2\right)} \right)\, ,
\label{eq:NilpotentApproxConiLCS}
\end{eqn}
where we have introduced the periodic coordinates $T^1 = \frac{1}{2\pi i}\log z^1$ and $T^2=z^2$. In particular, for the first two terms in the instanton expansion one finds the following complex vectors
\begin{eqn}
    {\bf a}_0  = (1,0,0,-i\delta_2,-i\delta_1,-\varepsilon_2)^{\mathsf{T}}\, ,\qquad {\bf a}_{1,0}  = (0,1,0, -i\delta_1,-\log (A_1)/2\pi i,0)^{\mathsf{T}}\, ,
\label{eq:instantonvectorsconiLCS}
\end{eqn}
which are again in agreement with our previous results. Note that the leading order contribution to ${\bf \Pi}_{\rm cLCS}$ in \eqref{eq:NilpotentApproxConiLCS} is such that $P_1\cdot {\bf a}_0=0$, which means that there exists a (single) metric essential instanton for this geometry (c.f. \eqref{metriker}).\footnote{Contrary to the conifold case, since $P_2\cdot {\bf a}_{1,0}=n P_1\cdot {\bf a}_{1,0}=(0,0,0,0,1,0)^\mathsf{T}$, one does not need any additional essential instanton here to reconstruct the entire $H^3(Y_3, \mathbb{C})$ for this example, see footnote \ref{fnote:2ndessentialinstanton}.} This may also be confirmed upon computing the moduli space metric $g_{i \bar{j}}$ that results from truncating \eqref{periodconiLCS} to polynomial order in the $T^i$, which ends up being degenerate---as opposed to the actual one shown in \eqref{metricconiLCS}.

Next, we should construct the weight filtration associated to the full log-monodromy matrix $P= P_1+P_2$. Using that $P^3 {\bf a}_0 =(0,0,0,1,0,0)^\mathsf{T}$ and $P^4=0$, we realise that this example coincides with the abstract case studied already in section \ref{sss:EssentialInfiniteDistance}, from which we may borrow some of its most immediate results. Hence, following the discussion therein, we first isolate the chain of quantised vectors $\q$ that can be obtained upon acting with $P$ on some element $\q \in W_3\setminus W_2$ up to three times. A convenient basis for these is given by
\begin{eqn}
   \mathsf{q}_{p^0} = \begin{pmatrix}
       1 \\ 0 \\ 0 \\ 0 \\ 0 \\ 0
   \end{pmatrix}\, , \qquad  \mathsf{q}_{p^2} = \begin{pmatrix}
       0 \\ 0 \\ 1 \\ 0 \\ 0 \\ 0
   \end{pmatrix}\, , \qquad \mathsf{q}_{q_2} = \begin{pmatrix}
       0 \\ 0 \\ 0 \\ 0 \\ 0 \\ 1
   \end{pmatrix}\, ,\qquad \mathsf{q}_{q_0} = \begin{pmatrix}
       0 \\ 0 \\ 0 \\ 1 \\ 0 \\ 0
   \end{pmatrix}\, ,
    \label{eq:SCDTowerOrbit}
\end{eqn}
whose charge-to-mass ratio may be readily determined to be
\begin{equation}\label{eq:chargetomassSDCTower}
    \gamma^2_{\mathsf{q}_{p^0}} \simeq\gamma^2_{\mathsf{q}_{q_0}}\simeq4\,,\qquad  \gamma^2_{\mathsf{q}_{p^2}}\simeq\gamma^2_{\mathsf{q}_{q_2}}\simeq\frac43\,,
\end{equation}
and are thus finite. Note that \eqref{eq:SCDTowerOrbit} includes, in particular, the SDC tower of states---whose mass scale coincides with $m_*$ defined in \eqref{mstar}, as well as the U(1)$_{\q_{q_2}}$ associated to the direction of the infinite distance limit, namely $T^2\to i\infty$. Hence, these comprise the gravitational sector.\footnote{The graviphoton direction, on the other hand, reads as $T_I= ie^{K/2} X^0 \left(-\frac14 e^{-K} |X^0|^{-2}, \delta_1-2n z^1\text{Im}\,z^2,-i (\text{Im}\, z^2)^2\right)$.}

The remaining set of states are those who belong to the subspaces $\text{ker}\, P^2\setminus W_0$ and $W_{-1} \setminus W_{-2}$. These elements may be spanned by the pair of vectors
\begin{eqn}
   \mathsf{q}_{p^1} = (0,1,0,0,0,0)^\mathsf{T}\, , \qquad \mathsf{q}_{q_1} = (0,0,0,0,1,0)^\mathsf{T}\, ,
    \label{eq:cRFTOrbit}
\end{eqn}
which are related to each other via the log-monodromy matrix $P$ (as well as $P_1$), and moreover exhibit diverging charge-to-mass ratios of the form
\begin{equation}\label{eq:chargetomasscRFT}
    \gamma^2_{\mathsf{q}_{p^1}}\simeq \frac{1}{6\pi^2|A_1|^2}\frac{(\log |x^2|)^3}{|x^1|^2\log\left[|x^1||x^2|^n\right]}\,,\qquad  \gamma^2_{\mathsf{q}_{p^1}}\simeq \frac{1}{6\pi^4\delta_1^2}(\log |x^2|)^3\log\left[|x^1||x^2|^n\right]\,.
\end{equation}
Consequently, we find the core RFT lattice to be given by $\G_{\rm cRFT}= \langle \mathsf{q}_{p^1}, \mathsf{q}_{q_1} \rangle_{\mathbb{Z}}$, which, together with its symplectic complement $\G_{\rm cRFT}^\perp = \G_{\rm grav} =\langle \mathsf{q}_{p^0}, \mathsf{q}_{p^2}, \mathsf{q}_{q_2}, \mathsf{q}_{q_0} \rangle_{\mathbb{Z}}$, span the full lattice of quantised charge vectors in a way that makes manifest the block-diagonal structure of the monodromy group generated by \eqref{eq:monodromiesconiLCS}.

Let us note here again the difference between the divergence rates of $\gamma_{\mathsf{q}_{p^1}}$ and $\gamma_{\mathsf{q}_{q_1}}$ in \eqref{eq:chargetomasscRFT}. The latter comes from the presence of the complex constant $\delta_1$, that here plays the same r\^ole $\beta$ had for the conifold. As before, this difference translates into a magnetic monopole charged under the rigid U(1)$_{\mathsf{q}_{q_1}}$ whose mass tends to $m_{\rm mon}=|\delta_1|m_*$ as we approach the singularity. The same discussion thus applies in here, namely when $\delta_1$ and $\delta_2$ are related by a pair of rational numbers one can find an integral symplectic transformation that effectively removes the former from $\CF_1$ or, equivalently, an appropriate charge vector $\mathsf{q}\in\mathbb{Z}^6$ that verifies \eqref{massessIVmag} with $a_{\mathsf{q}}=|\mathsf{q}^\mathsf{T} \eta {\bf a}_0 |=0$. Indeed, whenever this happens, one can effectively set $\delta_1=0$, such that the electric and magnetic charge-to-mass ratios coincide
\begin{equation}
\gamma^2_{\mathsf{q}_{p^1}}\stackrel{\delta_1=0}{\simeq}\gamma^2_{\mathsf{q}_{q_1}}\simeq\frac{1}{6\pi^2|A_1|^2}\frac{(\log |x^2|)^3}{|x^1|^2\log\left[|x^1||x^2|^n\right]}\, ,
\end{equation}
in perfect analogy with section \ref{ss:conifold} and with the behavior of the 't Hooft-Polyakov monopole. 

Once we have identified the core RFT and gravitational sectors within our EFT, we are finally ready to check whether the former can indeed fully decouple from the latter. To this end, let us consider the first decoupling condition, which requires the absence of kinetic mixing in the rigid gauge kinetic functions $N_{IJ}$. Indeed, similarly to what happened for the conifold when computing $g^{z\bar{z}} e^K|X^0|^2 g_{z\bar{z}}^{\rm rig}$ in \eqref{rigcondconifold}, here we find
\begin{equation}
    ({N^{-1}})^{11}N_{11} = 1+\frac{3\delta_1^2}{\left( n \im z^2-\frac{1}{2\pi }\log\left|z^1\right|\right)(\im z^2)^3}+\ldots\, ,
\end{equation}
and thus a (kinematic) decoupling that happens polynomially fast in $\im z^2$ and $\log\left|z^1\right|$ if $\delta_1\neq0$, or exponentially fast in $\log\left|z^1\right|$ when $\delta_1=0$. Note that this parameter also controls (some of) the off-diagonal terms in $g_{i\bar j}$, $N_{IJ}$ and $\CI_{IJ}$, that mix the core RFT with the gravitational sector.

On the other hand, to check the second decoupling condition introduced in section \ref{ss:decoupling}, we need to calculate explicitly the normalised Pauli couplings associated to the rigid sector 
\begin{equation}
    ||\CP||^2_{\in\text{cRFT}}=-g^{1\bar{1}}g^{1\bar{1}}\CI^{1 1}\mathcal{P}_{111}\bar{\mathcal{P}}_{\bar{1}\bar{1} 1}\simeq\frac83 \frac{(\im z^2)^3}{\pi^2|z^1|^2}\left( n \im z^2-\frac{1}{2\pi }\log\left|z^1\right|\right)^{-3}\, ,
\end{equation}
and compare them with the mixed contributions 
\begin{equation}
    ||\CP||^2_{\rm mixed}\simeq \frac{ 5\delta_1}{8\pi }\frac{n\im z^2 }{ \left( n \im z^2-\frac{1}{2\pi }\log\left|z^1\right|\right)^{2}}\left(\frac1z+\frac1{\bar z}\right)\, .
\end{equation}
Once again, we find that the condition \eqref{nomixing2} is always satisfied, albeit at a faster rate when $\delta_1=0$. The scalar curvature reads
\begin{equation}
    R\simeq\frac{8}{3}\frac{(\im z^2)^3}{\pi^2|z^1|^2}\left( n \im z^2-\frac{1}{2\pi }\log\left|z^1\right|\right)^{-3}\, ,
\end{equation}
which is thus divergent and moreover controlled by the self-interactions within the rigid sector. The latter provide the dominant contribution in the normalised sum of all the Pauli interactions. 

\noindent Since all the conditions for the decoupling are met, we can now safely extract the rigid quantities from the local (supergravity) ones. These are fully specified by the rigid K\"ahler potential
\begin{align}
    K =& - \log \left( \frac{4}{3}(\im z^2)^3+2\delta_2\right) \\\notag&+ \frac{1}{\frac{4}{3}(\im z^2)^3+2\delta_2} \underbrace{\left[ 2|z^1|^2\left(n\im z^2-\frac{1}{4\pi}\log\left|\frac{z^1}{A_1}\right|^2\right)-\left( n\im z^2-\frac1{4\pi}\right)((z^1)^2+(\bar z^1)^2)\right]}_{K_{\rm rig}} + \ldots\, ,
\end{align}
its associated K\"ahler metric
\begin{equation}
    g_{z\bar z}^{\rm rig}=\partial_{z^1}\bar{\partial}_{\bar{z}^1}K_{\rm rig}=2n\im z^2-\frac{1}{2\pi}\left(\log\left|\frac{z^1}{A_1}\right|^2+2\right)\, ,
\end{equation}
as well as the rigid prepotential 
\begin{equation}
    \label{eq:rigidprepotconiLCS}
	\mathscr{F}_{\rm rig} (X^1) = -\frac{1 }{2 } (X^1)^2 \left(n  X^2 +\frac{1}{4\pi i }\left(\log\left(\frac{X^1}{ A_1}\right)^2 -1\right)\right) \, .
\end{equation}
As a final check, let us compare the sectional curvature on the rigid slice with the Riemann tensor of the rigid theory itself. This correspond to comparing $ R_{1\bar 1 1\bar 1} \simeq \oh \CP_{M 11} \bar{\CP}_{N \bar{1}{\bar{1}}} \CI^{M N}$ with $R_{z\bar zz\bar z}^{\rm rigid}\simeq \oh \frac{\Mpl^2}{m_*^2}|\CP_{zzz}|^2 \CI^{zz}$, which themselves verify
\begin{equation}
    R_{1\bar 11\bar 1} \simeq  \frac{m_*^2}{\Mpl^2} R_{z\bar zz\bar z}^{\rm rigid}=\frac{1}{3\pi^2(\im z^2)^3|z^1|^2}\left( n \im z^2-\frac{1}{2\pi }\log\left|z^1\right|\right)^{-1}\;.
\end{equation} 
%


\section{Seiberg--Witten limits}
\label{s:swpoints}

Seiberg--Witten (SW) limits are some of the best-known examples of geometric engineering of gravity-decoupled  4d $\CN=2$ gauge theories  \cite{Seiberg:1994rs,Kachru:1995wm,Kachru:1995fv,Andrianopoli:1996cm,Katz:1996fh,Billo:1998yr}. In light of the Curvature Criterion \cite{Marchesano:2023thx}, one would expect to find a divergent moduli space curvature along such limits, that is moreover related to the RFT curvature as in \eqref{PaulicRFT}, where $m_* \simeq m_{\rm h}$ stands for the string scale of the dual heterotic string construction. It turns out that the curvature divergence occurs if an only if the decoupling criteria of section \ref{ss:decoupling} are met. In the following, we illustrate this and other features of Seiberg--Witten limits by studying two familiar examples of type II Calabi--Yau manifolds. 

The strategy to analyse both constructions follows the general approach of previous sections. First, we consider a basis of quantised periods in this region of moduli space. By looking at the monodromy action in this basis, we determine the invariant sublattices. Then, utilising the charge-to-mass ratios, we single out the one corresponding to the core RFT, which coincides with the SW sector. Moreover, by a suitable choice of special coordinates, we obtain the prepotential of the 4d supergravity EFT and, upon looking at the core RFT periods, that of the rigid sector. 

Finally, it remains to verify if the SW sector is decoupled from the gravitational U(1)'s of the compactification, like those under which the SDC tower is charged. We check such a decoupling at the level of kinetic mixing and of Pauli interactions, with a similar outcome for both examples. The decoupling condition associated to the Pauli terms gives a non-trivial constraint, such that along certain limits the decoupling of the SW sector from gravity occurs, while for others it does not. This constraint can be formulated as the feature that, along the infinite distance trajectory, all the core RFT scales of the compactification decouple from those of gravitational nature, like the species cut-off (see e.g. \cite{Castellano:2024bna}). Alternatively, one can detect the decoupling by looking at those limits where the moduli space scalar curvature, when measured in Planck units, diverges.

\subsection{Mirror of $\mathbb{P}^{1,1,2,2,6}[12]$}
\label{ss:SW1}

Seiberg--Witten limits can be  constructed by considering type IIA strings compactified on a Calabi--Yau that is a K3 fibration over $\mathbb{P}^1$. An example of this that has been studied in great detail in the literature is the three-fold  $\mathbb{P}^{1,1,2,2,6}[12]$, see \cite{Candelas:1993dm,Hosono:1993qy,Kachru:1995fv,Curio:2000sc} for early work and \cite{Lee:2019wij,Bastian:2021eom,Monnee:2025msf} for more recent analyses in the context of the Swampland Programme.

\subsubsection*{Periods and monodromies}

The Calabi--Yau $X_3 = \mathbb{P}^{1,1,2,2,6}[12]$ has $h^{1,1}(X_3) =2$, and hence a complex two-dimensional vector multiplet moduli space. As standard in the literature, we will study the moduli space in the vicinity of the SW point from the viewpoint of type IIB compactified on the mirror manifold $Y_3$, in terms of two complex coordinates $(x^1,x^2)$ that locate the singularity at $x^1 = x^2 = 0$. The period vector ${\bf \Pi}$ for the three-form $\Omega$ in that region can be constructed by solving the corresponding Picard--Fuchs differential equations, and then comparing the result with the quantised basis of three-cycles within the LCS regime \cite{Curio:2000sc,Lee:2019wij}. Following the conventions in \cite{Lee:2019wij}, one recovers a period vector of the form
\begin{eqn}\label{eq:periodvectorSW1Integral}
    {\bf \Pi}_{\rm SW_1} = \begin{pmatrix}
    X^0 \\ X^1 \\ X^2 \\ \CF_0 \\ \CF_1 \\ \CF_2
         \end{pmatrix} =
    \mathsf{M}_{1}^{-1} \begin{pmatrix}
    \varpi_1 \\ \varpi_2 \\ \varpi_3 \\ {\varpi_4} \\ {\varpi_5} \\ {\varpi_6}
         \end{pmatrix} =  \mathsf{M}_{1}^{-1} {\bf \Pi}_{\rm PF}\, ,
\end{eqn}
where $\varpi_i$ solve the Picard--Fuchs equations, see Appendix \ref{ap:SWp} for their explicit expressions. Here, ${\bf \Pi}_{\rm PF}$ is a real symplectic vector, meaning that it obeys the symplectic pairing \eqref{eta} but its periods do not correspond to a quantised basis of three-cycles in $Y_3$. To obtain such a basis, one needs to apply $\mathsf{M}_{1}$, which is a symplectic  transition matrix to the LCS regime:\footnote{Besides the different period ordering, our transition matrix differs slightly from the one in \cite{Lee:2019wij}. In the period basis ${\bf \Pi}_{\rm PF}$ and ${\bf \Pi}_{\rm SW_1}^{\mathsf{T}} = (X^0, X^1, X^2, \CF_2, \CF_1, \CF_0)$ used therein, we would have that $\mathsf{M}_{1}^{\rm here} = \text{diag} (1,1,\frac{1}{2},2,1,1)\cdot \mathsf{M}_{1}^{\rm LLW}$. The choice \eqref{intquantised} leads to monodromy matrices with integer entries, c.f. eq. \eqref{eq:SWnewmonodromies} below.}
\begin{equation}\label{intquantised}
{\mathsf{M}_{1}} = \begin{pmatrix}
       0 & -i  \mathcal{X} & 0 &  0 & 0 &\frac{\mathcal{X}}{2}  \\
    0 & \frac{i}{\mathcal{X}} & 0  & 0 & 0 & \frac{1}{2 \mathcal{X}} \\
1 & 0 & 0 & 0 & 0 & -\frac{1}{2}\\
    0 & \xi_3 & -\frac{1}{\mathcal{X}} & \frac{1}{2\mathcal{X}} & \frac{i}{2 \mathcal{X}} &  i  \xi_4     \\
    0 & \xi_2 & -\mathcal{X} & \frac{\mathcal{X}}{2} & -\frac{i  \mathcal{X}}{2} &  i\xi_1 \\
    0 & 0 & 0 & 1 & 0 & 0
\end{pmatrix}\, ,
\end{equation}
with $\mathcal{X}=\frac{\Gamma \left( \frac34 \right)^4}{\sqrt{3} \pi^2}$, and $\xi_i$ some real constants satisfying the relation $\xi_1 + \xi_2/2 = (\xi_4 - \xi_3/2) \mathcal{X}^2$. From here, one obtains the following expressions for the quantised periods
\begin{equation}
\begin{aligned}\label{eq:periodsSW}
   X^0 &=\frac{1}{2 \pi \mathcal{X}} - \frac{\sqrt{x^1}}{\pi} \, ,\\
   X^1 &= \frac{i}{2\pi \mathcal{X}}\, ,\\
    X^2 & = -\frac{i}{4 \pi^2 \mathcal{X}} \left(\log \left(x^2 (x^1)^2 \right)-4\pi\mathcal{X}\xi_4+5\right) +\frac{i}{2 \pi^2 }\sqrt{x^1}\left(\log \left(x^2 \right)-6\log(2)+7\right) \, ,\\
     \CF_0 & =\frac{i}{\pi^2} \sqrt{x^1}\left( \log \left(x^2 \right)-6\log(2)+7\right) \, ,\\
    \CF_1 & = -\frac{1}{2 \pi^2 \mathcal{X}} \left(\log \left(x^2 (x^1)^2 \right)+2\pi\mathcal{X}\xi_3+5 \right)\, ,\\
  \CF_2 & = \frac{1}{\pi \mathcal{X}}\, ,
\end{aligned}
\end{equation}
where we recall that $x^1=x^2=0$ describes the Seiberg--Witten point, and the expressions are given up to order $\mathcal{O}(x^i)$. This new set of periods corresponds to a quantised basis of three-cycles, which moreover has the standard geometric interpretation in terms of the type IIA compactification on the mirror manifold $X_3$, within the large volume regime. Namely, $\CF_0$ is the period that corresponds to a D6-brane wrapping the whole three-fold, $\CF_i$ to D4-branes wrapping Nef divisors of $X_3$, $X^i$ to D2-branes wrapping their dual curves, and $X^0$ to the D0-brane. More precisely, the Nef divisors exhibit the following triple intersection numbers of $X_3$
\begin{eqn}
    \CK_{111} = 4\, , \quad \CK_{112}  =2\, , \quad \CK_{122} = 0 \, \quad \CK_{222}=0\, ,
\end{eqn}
which identify $\CF_2$ with the D4-brane wrapping the K3-fibre of the mirror Calabi--Yau \cite{Lee:2019wij}.

The monodromy matrix for ${\bf \Pi}_{\rm SW_1}$ can be obtained directly from the transformations of these periods or, equivalently, by a change of basis applied to the analogous one associated to the real symplectic period vector ${\bf \Pi}_{\rm PF}$ as ${\cal T}_i \to \mathsf{M}_1^{-1} {\cal T}_i \mathsf{M}_1$. Either way, one finds
\begin{align}\label{eq:SWnewmonodromies}
   {{\cal T}_1}=\begin{pmatrix}
        -1 & 0 & 0 & 0 & 0 & 1\\
        0 & 1 & 0 & 0 & 0 & 0\\
        0 & 0 & 1 & -1 & 0 & 1\\
        0 & 0 & 0 & -1 & 0 & 0\\
        0 & -4 & 0 & 0 & 1 & 0\\
        0 & 0 & 0 & 0 & 0 & 1
    \end{pmatrix}\, ,\qquad {{\cal T}_2}=\begin{pmatrix}
        1 & 0 & 0 & 0 & 0 & 0\\
        0 & 1 & 0 & 0 & 0 & 0\\
        1 & 0 & 1 &  0 & 0 & 0\\
        2 & 0 & 0 & 1 & 0 & -1\\
        0 &  -2 & 0 & 0 & 1 & 0\\
        0 & 0 & 0 &  0 & 0 & 1
    \end{pmatrix}\, ,
\end{align}
where ${\cal T}_j$ corresponds to the monodromy for ${\bf \Pi}_{\rm SW_1}$ around $x^j \to e^{2\pi i} x^j$. The former may be further decomposed into $\mathbb{Z}_2$-semisimple\footnote{\label{fnote:WeylgroupSU2}This coincides with the Weyl group of the underlying SU(2) gauge symmetry in the rigid Seiberg--Witten theory.} and unipotent (of order 2) parts, as follows (c.f. \eqref{monodec}) 
\begin{align}\label{eq:semisimple&unipotentT1}
   {{\cal T}_1}={\cal T}_1^{(s)} {\cal T}_1^{(u)}\, ,\qquad \text{with}\quad {\cal T}_1^{(s)}=\begin{pmatrix}
        -1 & 0 & 0 & 0 & 0 & 1\\
        0 & 1 & 0 & 0 & 0 & 0\\
        0 & 0 & 1 & -1 & 0 & 0\\
        0 & 0 & 0 & -1 & 0 & 0\\
        0 & 0 & 0 & 0 & 1 & 0\\
        0 & 0 & 0 & 0 & 0 & 1
    \end{pmatrix}\, ,\quad {\cal T}_1^{(u)}=\begin{pmatrix}
        1 & 0 & 0 & 0 & 0 & 0\\
        0 & 1 & 0 & 0 & 0 & 0\\
        0 & 0 & 1 &  0 & 0 & -1\\
        0 & 0 & 0 & 1 & 0 & 0\\
        0 & 4 & 0 & 0 & 1 & 0\\
        0 & 0 & 0 &  0 & 0 & 1
    \end{pmatrix}\, ,
\end{align}
whereas the latter is also unipotent (with the same order). To show explicitly that this example features metric essential instantons, one shall rewrite the period vector \eqref{eq:periodvectorSW1Integral} in terms of the expansion defined in \eqref{periodv}, namely
\begin{eqn}\label{eq:NilpotentPeriodsSW}
    {\bf \Pi}_{\rm{SW}_1} =
e^{T^i P_i} \left( {\bf a}_0\ + \sum_{(2r_1,r_2) \in \mathbb{Z}^2_{\geq0}\setminus (0,0)} {\bf a}_{r_1, r_2} e^{2\pi i \left(r_1 T^1+r_2 T^2\right)} \right)\, ,
\end{eqn}
where the first few non-vanishing terms are given by
\begin{eqn}
\begin{aligned}
    {\bf a}_0  &= \frac{1}{2\pi \mathcal{X}}\left(1,i,i\,\frac{-5+4\pi \mathcal{X} \xi_4}{2\pi},0,-\frac{5+2\pi \mathcal{X} \xi_3}{\pi},2\right)^{\mathsf{T}}\, ,\\
    {\bf a}_{\oh,0}  &= -\left(\frac{1}{\pi},0,i\,\frac{-7+6\log(2)}{2\pi^2}, i\,\frac{-7+6\log(2)}{\pi^2},0,0 \right)^{\mathsf{T}}\, ,
\end{aligned}
\label{eq:instantonvectorsSW}
\end{eqn}
and we introduced the periodic variables $T^i= \frac{1}{2\pi i} \log x^i$, as well as the log-monodromy matrices associated with \eqref{eq:SWnewmonodromies}, which read as
\begin{align}\label{eq:logmonodromiesSW}
   P_1=\log {\cal T}_1^{(u)}=\begin{pmatrix}
        0 & 0 & 0 & 0 & 0 & 0\\
        0 & 0 & 0 & 0 & 0 & 0\\
        0 & 0 & 0 & 0 & 0 & 1\\
        0 & 0 & 0 & 0 & 0 & 0\\
        0 & -4 & 0 & 0 & 0 & 0\\
        0 & 0 & 0 & 0 & 0 & 0
    \end{pmatrix}\, ,\qquad P_2= \log {{\cal T}_2}=\begin{pmatrix}
        0 & 0 & 0 & 0 & 0 & 0\\
        0 & 0 & 0 & 0 & 0 & 0\\
        1 & 0 & 0 &  0 & 0 & 0\\
        2 & 0 & 0 & 0 & 0 & -1\\
        0 &  -2 & 0 & 0 & 0 & 0\\
        0 & 0 & 0 &  0 & 0 & 0
    \end{pmatrix}\, .
\end{align}
From the above expressions, it is easy to see that neither $P_1$ nor $P_2$ annihilates the leading-order term $\mathbf{a}_0$ in \eqref{eq:NilpotentPeriodsSW}, whilst the combination $P_\mu=P_1-2P_2$ does. This then implies that there exists a single metric essential instanton---i.e. $\mathbf{a}_{\oh,0}$---that must be included within the period vector so as to yield a non-degenerate moduli space metric, as one may readily confirm.

With this information at hand, one may now proceed to identify the rigid sector of the theory following the prescription outlined in section \ref{ss:essential}. Hence, we start by noticing that $P \mathbf{a}_0 \neq 0$ whilst $P^2=0$, where $P=P_1+P_2$ denotes the full (log-)monodromy generator. This means that the SW singularity is of Type II$_1$ \cite{Grimm:2018cpv} and, as such, it lies at infinite distance---as we already knew. Next, we should construct the corresponding weight filtration $W_{-1} \subset W_0 \subset W_1= \G\cong H_3(Y_3,\mathbb{Z})$, for which it becomes useful to first look for those elements of the quantised charge lattice which are left invariant by the monodromy group. Indeed, a convenient basis for these is given by
\begin{eqn}\label{eq:invariantstatesSW}
    \mathsf{q}_{\rm D4_2} =\begin{pmatrix}
    0 \\ 0 \\ 1 \\ 0 \\ 0 \\ 0
    \end{pmatrix}\, , \qquad
     \mathsf{q}_{\rm D2_1} =\begin{pmatrix}
    0 \\ 0 \\ 0 \\ 0 \\ 1 \\ 0
    \end{pmatrix}\, ,
    \qquad
     \mathsf{q}_{\rm 2D0+D4_2} =\begin{pmatrix}
    0 \\ 0 \\ 1 \\ 2 \\ 0 \\ 0
    \end{pmatrix}\, ,
\end{eqn}
all of which belong to the subspace $W_{-1}= \text{Im}\, P$. Strictly speaking, though, the states above are only invariant under $({\cal T}_1)^2$ and ${\cal T}_2$. More precisely, $\mathsf{q}_{\rm D4_2}$ and $ \mathsf{q}_{\rm D2_1}$ are both left unchanged under any monodromy operation, while $\q_{\rm D4_2+2D0}$ picks up a minus sign under ${\cal T}_1 $. Consequently, this last one is charged under the discrete $\mathbb{Z}_2$ monodromy and thus, by the discussion of section \ref{ss:essential}, its charge-to-mass ratio diverges along the limit. The same applies to its preimage under $P$, which contains the element $\mathsf{q}_{\rm D6}^{\mathsf{T}} = (1,0,0,0,0,0)$, hence defining rigid U(1)'s as per \eqref{RFTlimit}. 

This expectation is in agreement with the discussion presented in \cite{Lee:2019wij}. Indeed, using the type IIA mirror perspective, one sees that the charge elements $\mathsf{q}_{\rm D4_2}$ and $\mathsf{q}_{\rm D2_1}$ correspond to a D4-brane wrapping the K3 fibre of $\mathbb{P}^{1,1,2,2,6}[12]$ and a D2-brane wrapping a curve inside the latter, respectively. Both states belong to the SDC tower, and more concretely to BPS particles associated with the emergent heterotic string \cite{Lee:2019wij}. The element $\mathsf{q}_{\rm D4_2+2D0}$ is instead identified with the W-boson of the Seiberg--Witten sector, while $\mathsf{q}_{\rm D6}$ corresponds to its monopole \cite{Curio:2000sc}. When acted upon by the monodromy generators, the latter verifies
\begin{eqn}
    {\cal T}_1 \cdot \mathsf{q}_{\rm D6} = - \mathsf{q}_{\rm D6}\, , \qquad {\cal T}_2 \cdot \mathsf{q}_{\rm D6} = \mathsf{q}_{\rm D6} +  \mathsf{q}_{\rm D4_2+2D0}\, .
    \label{monoSW}
\end{eqn}
Therefore, $\G_{\rm ess}=\langle \mathsf{q}_{\rm D4_2+2D0}, \mathsf{q}_{\rm D6}\rangle_{\mathbb{Z}}$ forms a symplectic and monodromy-invariant sublattice disconnected from the SDC tower, thus being an obvious candidate for core RFT. This picture matches the asymptotic behaviour exhibited by the charge-to-mass ratios for each of these states. We have that their masses read 
\begin{eqn}
    m_{\q_{\rm D4_2}} \simeq 2 m_{\q_{\rm D2_1}} \simeq \frac{e^{K/2}}{\pi \mathcal{X}} \Mpl\, , \quad   m_{\q_{\rm  2D0+D4_2}} \simeq 4\mathcal{X}|x^1|^{\oh} m_{\q_{\rm D2_1}}\, , \quad m_{\q_{\rm D6}} \simeq -\frac{2\mathcal{X}}{\pi}|x^1|^{\oh}\log|x^2| m_{\q_{\rm D2_1}}\, ,
    \label{massSW1}
\end{eqn}
where $e^{-K} \pi^2 \mathcal{X}^2\simeq -\log(|x^1|^2|x^2|)/\pi$, see Appendix \ref{ap:SWp} for details. Their physical charges are
\begin{eqn}
    \CQ_{\q_{\rm D4_2}}^2 \simeq 4\CQ_{\q_{\rm D2_1}}^2 \simeq  -\frac{2\pi}{\log(|x^1|^2|x^2|)} \, , \quad  
    \CQ_{\q_{\rm  2D0+D4_2}}^2 \simeq -\frac{2\pi}{\log |x^2|} \, , \quad \CQ_{\q_{\rm D6}}^2 \simeq -\frac{\log |x^2|}{2\pi } \, ,
    \label{chSW1}
\end{eqn}
from where we obtain the charge-to-mass quotients
\begin{eqn}
    \gamma^2_{\rm D4_2} \sim  \gamma_{\rm D2_1}^2 \sim  2 \, , \qquad  \gamma^2_{\rm 2D0+D4_2} \sim \gamma^2_{\rm D6} \sim \frac{\log(|x^1|^2|x^2|)}{2\mathcal{X}^2 |x^1|\log |x^2|}\, .
    \label{gammasSW1}
\end{eqn}
Notice that the last two ratios diverge at $x^1=0$. Following section \ref{ss:core}, this observation together with \eqref{monoSW}, identify the core RFT of the limit with the lattice $\G_{\rm ess}$ or, equivalently, with the U(1) under which $\mathsf{q}_{\rm D4_2+2D0}$ is charged. The rest of the U(1)'s are instead of gravitational type.

\subsubsection*{The core RFT}

To properly analyse the core RFT sector with respect to the rest of the EFT, it proves useful to first perform the following additional transformation on the period vector
\begin{equation}\label{Y}
    {\bf \Pi}_{\rm SW_1}\to \mathsf{Y} {\bf \Pi}_{\rm SW_1}\, ,\qquad\qquad \text{with}\quad \mathsf{Y} = \begin{pmatrix}
    0 & 2 & 0 & 0 & 0 & 0 \\
    0 &0 & 2 & -1 & 0 & 0 \\
      2  & 0 & 0 & 0 & 0 & -1 \\
         0 & 0 & 0 & 0 & 1 & 0  \\
          0 & 0 & 0 & 0 & 0 & 1 \\
    0 & 0 & 0 & 1 & 0 & 0 
\end{pmatrix} \, .
\end{equation}
Notice that this is not an actual symplectic transformation, since it satisfies
\begin{eqn}
 Q \equiv -\mathsf{Y}^{\mathsf{T}}\eta\mathsf{Y}=-2\eta\, .
 \label{Y2}
\end{eqn}
Nevertheless, it allows us to isolate the different rigid and gravitational U(1)'s of the limit. Indeed, after such a transformation the period vector reads
\begin{eqn}
    {\bf \Pi}_{\rm SW_1}' = \begin{pmatrix}
    X^0 \\ X^1 \\ X^2 \\ \CF_0 \\ \CF_1 \\ \CF_2
         \end{pmatrix} =
         \mathsf{Y} \mathsf{M}_{1}^{-1} {\bf \Pi}_{\rm PF}
          =
         \begin{pmatrix}
    \frac{i}{\pi \mathcal{X}} \\  -\frac{i}{2 \pi^2 \mathcal{X}} \left(\log \left(x^2 (x^1)^2 \right)-4\pi\mathcal{X}\xi_4+5\right) \\ - \frac{2}{ \pi}\sqrt{x^1}\\ -\frac{1}{2 \pi^2 \mathcal{X}} \left(\log \left(x^2 (x^1)^2 \right)+2\pi\mathcal{X}\xi_3+5 \right)  \\  \frac{1}{\pi \mathcal{X}} \\ \frac{i}{ \pi^2 }\sqrt{x^1}\left(\log \left(x^2 \right)-6\log(2)+7\right)
         \end{pmatrix}\, +\, \CO(x^i)
         \, ,
         \label{swperiodY}
\end{eqn}
and the monodromy matrices become
\begin{align}\label{eq:SWnewmonodromiesY}
    {\cal T}_1'=\begin{pmatrix}
         1 & 0 & 0 & 0 & 0 & 0\\
        0 & 1 & 0 & 0 & 2 & 0\\
        0 & 0 & -1 & 0 & 0 & 0\\
        -2 & 0 & 0 & 1 & 0 & 0\\
        0 & 0 & 0 & 0 & 1 & 0\\
        0 & 0 & 0 & 0 & 0 & -1
    \end{pmatrix}\, ,\qquad {\cal T}_2'=\begin{pmatrix}
        1 & 0 & 0 & 0 & 0 & 0\\
        0 & 1 & 0 & 0 & 1 & 0\\
        0 & 0 & 1 & 0 & 0 & 0\\
        -1 & 0 & 0 & 1 & 0 & 0\\
        0 & 0 & 0 & 0 & 1 & 0\\
        0 & 0 & 1 & 0 & 0 & 1
    \end{pmatrix}\, .
\end{align}
In this new basis, the charge-to-mass ratios of $\mathsf{p}^i = (\delta^i_j, \ldots, 0)^{\mathsf{T}}$ and $\mathsf{q}_i = (0, \ldots, \delta^j_i)^{\mathsf{T}}$ are
\begin{eqn}
    \gamma^2_{\mathsf{p}^0} \simeq  \gamma^2_{\mathsf{q}_0} \simeq 2 \, , \qquad  \gamma^2_{\mathsf{p}^1} \simeq \gamma^2_{\mathsf{q}_1} \simeq 2\, , \qquad  \gamma^2_{\mathsf{p}^2} \simeq  \gamma^2_{\mathsf{q}_2} \simeq  \frac{\log(|x^1|^2|x^2|)}{2\mathcal{X}^2 |x^1|\log |x^2|} \, .
    \label{gammasSW1Y}
\end{eqn}
Hence, now $U(1)_2$ corresponds to a rigid vector multiplet, while $U(1)_0$, $U(1)_1$ belong to the gravitational sector of the limit.\footnote{The asymptotic graviphoton direction is given instead by $T_I= ie^{K/2} X^0 \left(-\oh e^{-K} |X^0|^{-2}, -1,-\frac{1}{\pi} z^2 \log |z^2/\hat{\Lambda}_{\rm sw}|^2\right)$, with $T_-=T_I F^I_-\,$. Note that the component along the rigid direction is controlled by the mass of the monopole.} Notice, however, that in the current basis some of these choices are not physical, since properly quantised states are of the form $\mathsf{Y}\cdot\q$, with $\q \in \mathbb{Z}^6$. In particular, in this new basis the rigid states of minimal charge are
\begin{eqn}
    \q_{\rm 2D0+D4_2}^{\mathsf{T}} = (0,0,0,0,0,2)\, , \qquad  \q_{\rm D6}^{\mathsf{T}} = (0,0,2,0,0,0)\, .
    \label{newvecSW}
\end{eqn}
As discussed in Appendix \ref{ss:changebasis&pairing}, the wedge product should now be computed via $Q^{-1}=\oh\eta$, which means that the BPS masses read
\begin{eqn}
    m_\mathsf{q} = \oh e^{K/2} |{\bf \Pi}^{\mathsf{T}} \eta\, \mathsf{q}| \, \Mpl\, ,
\end{eqn}
which, from \eqref{newvecSW}, matches the previously derived masses for the SW states.

\noindent To describe the full EFT in this frame, let us consider the prepotential that reproduces \eqref{swperiodY}:
\begin{align}\label{prepotentialY}
    \mathcal{F}/(X^0)^2
    =& -i z^1 + i A_1 + \frac{i}{\pi} \left(\frac\pi2 z^1+\log (z^2/A_2)\right)(z^2)^2 + \dots\, ,
\end{align}
where the dots stand for higher order terms in the $z^i$ expansion. The special coordinates are
\begin{equation}\label{specialcoordY}
\begin{aligned}
    z^1 & =-\frac{1}{2\pi} \left( \log\left[ (x^1)^2x^2\right] +5-4\pi\mathcal{X}\xi_4\right)+\CO(x^1, x^2)\, , \\
    z^2 &= 2i\mathcal{X}\sqrt{x^1}+\CO((x^1)^{3/2}, x^2)\, , 
\end{aligned}
\end{equation}
and we have defined the following two constants
\begin{eqn}
    A_1   = \oh \left(\xi _3+2 \xi _4\right) \mathcal{X}\, , \qquad 
    A_2  =i \frac{\mathcal{X}}{\sqrt{2}} e^{1+\pi \mathcal{X}\xi_4}\, .
\end{eqn}

Let us now build the core RFT prepotential from the pair $(X^2, \CF_2)$. Following the strategy of the previous section, we identify $\CF_2/X^0$ with $\partial_{z^2} \mathscr{F}_{\rm rig} (z^2)$ which leads us to
\begin{align}\label{eq:weakcouplingprepotential}
   \mathscr{F}_{\rm sw} (a)= \frac{i  a^2}{2\pi} \left(\log \left( \frac{a}{\hat{\Lambda}_{\rm sw}}\right)^2 -3\right)  + \ldots\, , 
\end{align}
where
\begin{eqn}
a =  z^2 \, , \qquad
    \hat{\Lambda}_{\rm sw} = A_2\,    e^{- \frac32-\frac{\pi}{2} z^1} \, , \qquad \Lambda_{\rm sw} = \hat{\Lambda}_{\rm sw} m_*\, .
\end{eqn}
Note that the natural scale for the rigid prepotential is $m_* = 2 m_{\q_{\rm D2_1}}   \simeq m_{\q_{\rm D4_2}}$, since otherwise $a$ and $a_D = \p_a \mathscr{F}_{\rm sw}$ do not reproduce the masses for the Seiberg--Witten W-boson and monopole, respectively. Also, the dynamically-generated strong coupling scale $\Lambda_{\rm sw}$  arises from absorbing the dependence of the SW monopole mass on the special coordinate $z^1$ into the SW prepotential. 

The reference scale $m_*$ appears as well when embedding the SW metric into the full K\"ahler potential associated to \eqref{prepotentialY}. Indeed, one can write such a quantity as
\begin{eqn}
 \kappa_4^2\,   K = - \log \left(2 \re z^1 - 2A_1 -  K_{\rm sw} + \re F \right) - 2\log |X^0| \, ,
 \label{KahlerSW1}
\end{eqn}
where 
\begin{eqn}
    	K_{\rm sw} = \text{Im}\, \left( \frac{\partial \mathcal{F}_{\rm sw}}{\partial a}\, \bar{a}\right) = \frac{|a|^2}{\pi} \left( -2 + \log \left| \frac{a}{\hat{\Lambda}_{\rm sw}}\right|^2 \right) + \ldots\, ,
        \label{eq:rigidlimitkahlerpot}
\end{eqn}
has been computed taking into account that, in our conventions, the Dirac pairing between the SW monopole and W-boson is 2, and
\begin{eqn}
    F = \left(\frac1\pi + z^1 - \bar{z}^1\right)(z^2)^2  + \ldots\, . 
\end{eqn}
Near the SW point, $x^1 = x^2 = 0$ (equivalently $z^2 =0, z^1 = \infty$), one finds 
\begin{eqn}
    m_* = e^{\kappa_4^2 K/2} |X^0| \Mpl \simeq \frac{\Mpl}{ \sqrt{2(\re z^1 - A_1)}}\, ,
\end{eqn}
and so expanding \eqref{KahlerSW1} in powers of $|z^2/z^1|$ we obtain
\begin{eqn}
\label{KSW1exp}
K  = -  \kappa_4^{-2} \log \left(2 |X^0|^2 (\re z^1 - A_1)\right) + \frac{m_*^2}{8\pi} K_{\rm sw} + \frac{m_*^2}{16\pi} \left(F + \bar{F}  \right) + \ldots\, .
\end{eqn}
From here, it is already easy to see why along the special coordinate $z^2$ one recovers a rigid metric $g_{2\bar{2}}$, in the sense of \eqref{cond}, that matches the Seiberg--Witten one derived from \eqref{eq:rigidlimitkahlerpot}. The first term in \eqref{KSW1exp} is independent of $z^2$ and thus can be treated as a constant when computing $g_{2\bar{2}}$, while the third one is the real part of a holomorphic function in the SW variable. Hence, from the viewpoint of $z^2$, it can be seen as a K\"ahler transformation which does not contribute to the element $g_{2\bar{2}}$ either.

As expected, the 4d EFT along the special coordinate displays the metric and gauge kinetic function of a rigid SU(2) Seiberg--Witten theory, namely
\begin{equation}
\begin{aligned}
    &\tau_{\rm sw} = \frac{\theta_{\rm sw}}{2\pi} + \frac{4\pi i}{g_{\rm sw}^2} = \oh\frac{\partial^2 \mathcal{F}}{\partial^2 a} = \frac{i}{2\pi} \log \left(\frac{a}{\hat{\Lambda}_{\rm sw}}\right)^2 + \ldots \, , \\
      &K_{a \bar a} = \frac{1}{ \pi} \log \left| \frac{a}{\hat{\Lambda}_{\rm sw}}\right|^2 + \ldots = \frac{8\pi}{g_{\rm sw}^2}\, ,
\end{aligned}
\end{equation}
and the following moduli space curvature
\begin{align}\label{eq:formalrigidcurvature}
   R_{\rm sw} = - 2K^{a \bar a} \frac{\partial^2}{\partial a \partial \bar{a}} \log K_{a \bar a} = \frac{8}{\pi^2 |a|^2 K_{a \bar a}^3} + \ldots\, .
\end{align}
The semi-classical monodromy ${\cal T}_{\rm sw}$ of the Seiberg--Witten model is the one that acts as $z^2 \to - z^2$ while leaving $\Lambda_{\rm sw}$, and hence $z^1$, invariant. In terms of the monodromy generators provided above (see \eqref{eq:SWnewmonodromiesY}), this is given by
\begin{eqn}
    {\cal T}_{\rm sw}' =  {\cal T}_1'( {\cal T}_2')^{-2} = \begin{pmatrix}
        1 & 0 & 0 & 0 & 0 & 0\\
        0 & 1 & 0 & 0 & 0 & 0\\
        0 & 0 & -1 & 0 & 0 & 0\\
        0 & 0 & 0 & 1 & 0 & 0\\
        0 & 0 & 0 & 0 & 1 & 0\\
        0 & 0 & 2 & 0 & 0 & -1
    \end{pmatrix}\, .
\end{eqn}
Notice the appearance of a factor of 2 in the monodromy action, as opposed to the 4 in \cite{Kachru:1995fv,Curio:2000sc,Lee:2019wij}, which is due to the choice of periods after performing the transformation \eqref{Y}.

To see how  the core RFT sector is embedded in the full supergravity EFT, one may derive the metric and gauge kinetic mixing from the prepotential \eqref{prepotentialY}:  
\begin{equation}\nonumber
     g_{i\bar j}= \left(\begin{array}{cc}
        \frac{1}{4(\text{Re} z^1-A_1)^2}  & 0 \\
          0 & \frac{\log |z^2/\Lambda_{\rm sw}|^2}{2\pi(\text{Re} z^1-A_1)}
     \end{array}\right) + \CO(|z^2|)\, , \quad
       N_{IJ} = 
    \begin{pmatrix}
        2A_1 & -1  & 0\\ -1 & 0 & 0\\ 0 & 0 &  \frac{1}{\pi}\log \left|\frac{z^2}{\Lambda_{\rm sw}}\right|^2 
    \end{pmatrix} + \CO(|z^2|)\, ,
\end{equation}
 \begin{equation}
     {\CI_{IJ}} = 
     \begin{pmatrix}
         -\oh\text{Re}\, z^1  &\frac{A_1}{2\text{Re} z^1} & 0\\ \frac{A_1}{2\text{Re} z^1} & -\frac{1}{2\text{Re} z^1} & 0\\
         0 & 0 & - \frac{1}{2\pi}\log \left|\frac{z^2}{\Lambda_{\rm sw}}\right|^2
     \end{pmatrix}
     + \CO(|z^2|)\, ,
     \label{gNISW1}
\end{equation}
where we have taken into account the non-canonical intersection matrix $Q = 2 \eta$. For instance, the rigid gauge kinetic matrix is computed as $N_{IJ} = \im \CF_{IJ}$ instead of the usual expression shown in \eqref{NIJ}, as can be deduced from 
\begin{eqn}
    N_{IJ} = \int_{Y_3} \p_{X^I} \Omega \wedge \p_{\bar{X}^J} \bar{\Omega} =  -\p_{X_I}{\bf \Pi }^\mathsf{T} Q^{-1} \p_{\bar{X}^J} \bar{\bf \Pi }\, .
    \label{NIJFP}
\end{eqn}
Notice that one reproduces the charges previously computed in \eqref{chSW1} upon taking into account that the charge vectors are given now by \eqref{newvecSW}.

As emphasised in \cite{Lee:2019wij}, the Seiberg--Witten point corresponds to an emergent string limit with a dual $E_8 \times E_8$ heterotic string compactified on $K3 \times T^2$ \cite{Kachru:1995fv}. One may easily compute the heterotic string tension in the mirror type IIA setup, by identifying it with an NS5-brane wrapped on the K3 fibre of $X_3 =\mathbb{P}^{1,1,2,2,6}[12]$. Since this emergent string is also a BPS axionic string---or EFT string in the sense of \cite{Lanza:2020qmt,Lanza:2021udy,Lanza:2022zyg}, one may compute its tension directly from the K\"ahler potential. Indeed, by using that $z^1$ implements the 4d axionic monodromy of the NS5-brane wrapped on K3  (i.e., it can be identified with the heterotic 4d dilaton), one finds
\begin{eqn}
   2\pi T_{\rm NS5} = - 2\pi \frac{\p K}{\p \re z^1} = \frac{\Mpl^2}{8\re z^1} = \frac{1}{4} m_*^2 =  m_{\rm D0}\cdot m_{\rm D4_2+D0}\, , 
\end{eqn}
where we have adapted the conventions of \cite{Lanza:2020qmt,Lanza:2021udy,Lanza:2022zyg} to ours. Notice that here the NS5 and the D4$_2$-brane are both wrapping the K3-fibre of the type IIA CY $X_3$, and that the bound state  D4$_2+$D0 with the D0-brane precisely removes the curvature-induced contribution to the D4-brane action. One then finds that $2\pi T_{\rm NS5}/m_{\rm D4_2+D0} = m_{\rm D0}$ reproduces the familiar relation from the type IIA compactifications at large volume, even if we are far away from that regime. Finally,  using that $T_{\rm NS5} = 2\pi \ell_{\rm h}^{-2}$, one deduces that $m_* = 4 \pi \ell_{\rm h}^{-1}$, and so in practice one can identify $m_*$ with the heterotic dual string  scale $m_{\rm h} = \ell_{\rm h}^{-1}$, as expected.

\subsection{Mirror of $\mathbb{P}^{1,1,2,2,2}[8]$}

Let us now consider the mirror of type IIA compactified on the Calabi--Yau $X_3 = \mathbb{P}^{1,1,2,2,2}[8]$, which is also a K3 fibration over $\mathbb{P}^1$. This case has been studied in detail in \cite{Candelas:1993dm,Kachru:1995fv,Curio:2000sc,Billo:1998yr,Eguchi:2007iw}. In the following, we will mostly follow \cite{Billo:1998yr} and the notion of rigid limit developed therein, in order to compare it with the general framework developed in the previous sections. We will adopt a similar strategy as for $X_3 = \mathbb{P}^{1,1,2,2,6}[12]$, being brief in those steps that are identical.

\subsubsection*{Periods and monodromies}

The Calabi--Yau $X_3 = \mathbb{P}^{1,1,2,2,2}[8]$  has a complex two-dimensional vector multiplet moduli space parametrised by the coordinates $(x^1,x^2)$, with the SW point again located at $x^1 = x^2 = 0$. In \cite{Billo:1998yr}, they use the alternative coordinates
\begin{equation}\label{map11222}
    \tilde{\epsilon}=\frac{1}{2}\frac{x_1\sqrt{x_2}}{1-x_1}\, ,\qquad \tilde{u}=\frac{1}{\sqrt{x_2}}\, ,
\end{equation}
and take $\tilde{\epsilon}$ as the natural expansion parameter for the entries of the period vector ${\bf \Pi}$, in the sense that $\tilde{\eps} \to 0$ describes the rigid limit. Such a period vector is computed by direct integration of the holomorphic three-form $\Omega$ on a basis of integrally quantised three-cycles. This basis, however, has a non-canonical intersection matrix 
\begin{eqn}\label{noncanintmat11222}
    q' = \begin{pmatrix}
        0 & 2 & 0 & 0 & 0 & 0 \\
        -2 & 0 & 0 & 0 & 0 & 0 \\
         0 & 0 & 0 & 4 & 0 & 0 \\
         0 & 0 & -4 & 0 & 0 & -4 \\
         0 & 0 & 0 & 0 & 0 & 2 \\
         0 & 0 & 0 & 4 & -2 & 0
    \end{pmatrix}\, ,
\end{eqn}
c.f. \cite[eq.(7.24)]{Billo:1998yr}. To take this system to a basis with a simple special coordinates and superpotential description, we consider the following change of basis
\begin{equation}\label{X}
   \mathsf{X}=\begin{pmatrix}
        0 & 0 & 1 & 0 & 0 & 0 \\
        0 & 0 & 0 & 0 & 0 &2\\
        0 & 1 & 0 & 0 & 0 & 0\\
        0 & 0 & 0 & 1 & 2 & 0 \\
        0 & 0 & 0 & 0 & -1 & 0 \\
        -2 & 0 & 0 & 0 & 0 & 0
    \end{pmatrix}\, ,
\end{equation}
such that $\mathsf{X}\,q'\mathsf{X}^{\mathsf{T}}=-4\eta$ and 
\begin{equation}\label{periovec11222can}
    \mathbf{\Pi}_{\rm SW_2}= \mathsf{X} {\bf \Pi} =\mathbf{v}_0+\mathbf{v}_1\tilde\epsilon^{1/2}+\mathbf{v}_2\tilde\epsilon+ \mathcal{O}(\tilde\epsilon^{3/2})\, ,
\end{equation}
where
\begin{equation}\label{periovec11222can}
    \mathbf{v}_0=\mathcal{Y}\left(\begin{array}{c}
        -\frac{i}{\sqrt2}    \\
        \frac{1}{\pi i}\log\tilde\epsilon+2i\frac{k_2}{\mathcal{Y}}\\
        0 \\
        -
        \frac{1}{\sqrt{2}\pi}\log\tilde\epsilon+\frac{k_1}{\mathcal{Y}}+1   \\
        -\frac{1}{2}   \\
        0     
    \end{array}\right)\,,\quad\mathbf{v}_1=\mathcal{Y}\left(\begin{array}{c}
        0    \\
        0  \\
        I_1(\tilde u) \\
        0 \\
        0 \\
        -2I_2(\tilde u)    
    \end{array}\right)\,,\nonumber
    \end{equation}
    \begin{equation}\quad\mathbf{v}_2=\mathcal{Y}\left(\begin{array}{c}
        i\frac{13}{8\sqrt2}   \\
        \frac{i}{8\pi}\left(\frac{13 \pi  \sqrt{2} k_1}{\mathcal{Y} }+\frac{38 \pi  k_2}{\mathcal{Y} }+\frac{8 \pi  \sqrt{2} l_1}{\mathcal{Y} }-19 \log \tilde\epsilon +32 \pi  \sqrt{2}+64\right)  \\
        0 \\
        \frac{1}{16\pi}(\frac{16 \pi  l_1}{\mathcal{Y} }+13 \sqrt{2} \log (\tilde\epsilon )+38 \pi)\,  \\
        -\frac{19}{16}    \\
        0   
    \end{array}\right)\tilde u \, ,
\end{equation}
Above, $\mathcal{Y}=-\frac{1}{2\pi^3}\Gamma^2\left(\frac18\right)\Gamma^2\left(\frac38\right)$ and $k_1,k_2,l_1,$ are certain numerical constants. The functions $I_1,I_2$, on the other hand, only depend on the coordinate $\tilde u$, and are defined through the integrals
\begin{eqn}
     I_1=\frac{\sqrt{2}}{\pi}\int_{-1}^{1}dt\,\sqrt{\frac{\tilde{u}-t}{1-t^2}}\, ,
     \qquad 
     I_2=\frac{\sqrt{2}}{\pi}\int_1^{\tilde u}dt\,\sqrt{\frac{\tilde{u}-t}{1-t^2}}\, ,
\end{eqn}
reproducing the definitions in \cite{Seiberg:1994rs} for the Seiberg--Witten coordinates $a$ and $a_D$. In this new basis, the monodromy matrices around $\tilde u =0$ and $\tilde \eps =0$ are, respectively, given by
\begin{equation}\label{mon11222}
        \mathcal{T}_{\tilde u}=\left(
\begin{array}{cccccc}
 1 & 0 & 0 & 0 & 0 & 0 \\
 0 & 1 & 0 & 0 & 0 & 0 \\
 0 & 0 & -1 & 0 & 0 & 0 \\
 0 & 0 & 0 & 1 & 0 & 0 \\
 0 & 0 & 0 & 0 & 1 & 0 \\
 0 & 0 & 4 & 0 & 0 & -1 \\
\end{array}
\right)\;,\quad\mathcal{T}_{\tilde \epsilon}=\left(
\begin{array}{cccccc}
 1 & 0 & 0 & 0 & 0 & 0 \\
 0 & 1 & 0 & 0 & 4 & 0 \\
 0 & 0 & -1 & 0 & 0 & 0 \\
 -2 & 0 & 0 & 1 & 0 & 0 \\
 0 & 0 & 0 & 0 & 1 & 0 \\
 0 & 0 & 0 & 0 & 0 & -1 \\
\end{array}
\right)\, ,
\end{equation}
while the charge-to-mass ratios for $\mathsf{p}^i = (\delta^i_j,\ldots, 0)^{\mathsf{T}}$ and $\mathsf{q}_i = (0, \ldots, \delta^j_i)^{\mathsf{T}}$ read
\begin{gather}
    \gamma^2_{\mathsf{p}^0} \simeq  \gamma^2_{\mathsf{q}_0} \simeq 2 \, , \qquad  \gamma^2_{\mathsf{p}^1} \simeq \gamma^2_{\mathsf{q}_1} \simeq 2\, , \qquad \gamma^2_{\mathsf{p}^2} \simeq  \gamma^2_{\mathsf{q}_2} \simeq -\frac{1}{2}\frac{ 1}{|\tilde u|\log |\tilde u|}  \frac{\log |\tilde \epsilon|}{|\tilde \epsilon|} \, .
\end{gather}
We thus see that in the limit $\tilde{\eps} \to 0$ only the charge-to-mass ratios of the electro-magnetic pair $(\mathsf{p}^2,\mathsf{q}_2)$ blow up, which moreover span a monodromy-invariant symplectic lattice. We thus find again that, in the selected basis, the core RFT is made of the vector multiplet that corresponds to U(1)$_2$, while U(1)$_0$ and U(1)$_1$ are part of the gravitational sector of the limit. 

\subsubsection*{The core RFT}

Just like in our previous example, from the period vector one can compute the special coordinates and from there deduce the form of the prepotential. We find that
\begin{align}\label{prepotential11222}
    \mathcal{F}/(X^0)^2
    =& -\frac{i}{\sqrt2} z^1 + i A_1 + \frac{2i}{\pi} \left(\frac\pi{2\sqrt2} z^1+\log (z^2/A_2)\right)(z^2)^2 + \CO(\eps)\, ,
\end{align}
where
\begin{equation}
\begin{aligned}
     z^1&=-\frac{\sqrt{2} }{\pi  }\log \tilde\epsilon +\frac{ 2\sqrt{2}   k_2}{ \mathcal{Y} }+\mathcal{O}(\tilde\epsilon)\, ,\\
    z^2&=2i\sqrt{\tilde u\tilde\epsilon}+\mathcal{O}(\tilde\epsilon^{3/2})\, ,
\end{aligned}
\end{equation}
and
\begin{equation}
    A_1=\frac{k_2}{\mathcal{Y}}-\frac{k_1}{\sqrt2\mathcal{Y}}-\frac{1}{\sqrt2}\,,\quad A_2=i\frac1{\sqrt2}e^{\tfrac32+\frac{\pi k_2}{\mathcal{Y}}}\, .
\end{equation}
Notice that the prepotential looks exactly like the one derived in our previous example \eqref{prepotentialY}, except for the substitution $z^1 \to z^1/\sqrt{2}$. This implies that most of the analysis carried out above will also apply to the present case. The  core RFT prepotential therefore reads 
\begin{align}
   \mathscr{F}_{\rm sw} (a)= \frac{i  a^2}{\pi} \left(\log \left( \frac{a}{\hat{\Lambda}_{\rm sw}}\right)^2 -3\right)  + \ldots\, , 
\end{align}
where now
\begin{eqn}
a =   z^2 \, , \qquad
    \hat{\Lambda}_{\rm sw} =  A_2\,    e^{- \frac32-\frac{\pi}{2\sqrt2} z^1} \, , \qquad \Lambda_{\rm sw} = \hat{\Lambda}_{\rm sw} m_*\, .
\end{eqn}
Note that $\mathscr{F}_{\rm sw}(a)$ is twice as \eqref{eq:weakcouplingprepotential}, but if we take into account that in this case the Dirac pairing between electric and magnetic SW periods is 4, we again recover \eqref{eq:rigidlimitkahlerpot}. The K\"ahler potential for the full EFT can be rather computed  as $\kappa_4^2 K= - \log \left[\frac{i}{4}(\mathsf{X}\mathbf{\Pi})^{\mathsf{T}}\eta(\mathsf{X}\bar{\mathbf{\Pi}})\right]$. One finds
\begin{eqn}
 \kappa_4^2\,   K = - \log \left(2^{-1/2} \re z^1 - A_1 -  K_{\rm sw} + \re F \right) -2\log |X^0| \, ,
 \label{KahlerSW2}
\end{eqn}
where
\begin{eqn}
    F = \left(\frac1\pi +\frac1{2\sqrt2}( z^1 - \bar{z}^1)\right)(z^2)^2  + \ldots\, . 
\end{eqn}
The reference scale $m_*$ at the Seiberg--Witten point $z^1 =\infty$, $z^2=0$ is
\begin{eqn}
    m_* = e^{\kappa_4^2 K/2} |X^0| \Mpl \simeq \frac{\Mpl}{ \sqrt{2^{-1/2} \re z^1 -  A_1}}\, ,
\end{eqn}
from where one finds a similar expansion as in \eqref{KSW1exp}. The metric and gauge kinetic matrices now take the form (up to $\CO(|z^2|)$)
\begin{equation}\nonumber
     g_{i\bar j}= \left(\begin{array}{cc}
        \frac{1}{4(\text{Re} z^1-(A_1/\sqrt2))^2}  & 0 \\
          0 & \frac{\sqrt2\log |z^2/\hat{\Lambda}_{\rm sw}|^2}{\pi(\text{Re} z^1-(A_1/\sqrt2))}
     \end{array}\right)\, , \quad
       N_{IJ} = 
    \begin{pmatrix}
        A_1 & -{2^{-3/2}}  & 0\\ -{2^{-3/2}} & 0 & 0\\ 0 & 0 &  \frac{{1}}{\pi}\log \left|\frac{z^2}{\hat{\Lambda}_{\rm sw}}\right|^2 
    \end{pmatrix}\, ,
\end{equation}
 \begin{equation}
     {\CI_{IJ}} = 
     \begin{pmatrix}
         -\frac{\text{Re}\, z^1 }{{4}\sqrt2} &\frac{A_1}{{4}\text{Re} z^1} & 0\\ \frac{A_1}{{4}\text{Re} z^1} & -\frac{1}{{4}\sqrt2\text{Re} z^1} & 0\\
         0 & 0 & - \frac{1}{{2}\pi}\log \left|\frac{z^2}{\hat{\Lambda}_{\rm sw}}\right|^2
     \end{pmatrix}
     \, ,
\end{equation}
being thus very similar to the ones obtained in the previous example.

\subsection{Decoupling conditions and curvature divergences}
\label{ss:decouplingSW}

As pointed out in section \ref{ss:decoupling}, the fact that along the special coordinate $z^2$ we find a rigid metric and gauge kinetic function $g_{2\bar{2}}\frac{\Mpl^2}{m_*^2}= N_{22} = - 2\CI_{22}$ that correspond to a Seiberg--Witten Lagrangian,  does not mean that the associated vector multiplet is decoupled from the rest of the EFT. To see if this is the case, one needs to analyse the kinetic and Pauli terms that mix the Seiberg--Witten sector with the rest of the compactification, and verify whether \eqref{nomixing} and \eqref{nomixing2} are satisfied. In the following, we will address this question for the SW CY limits described above. Given that their EFTs are very similar to each other, the analysis is identical in both cases modulo numerical $\CO(1)$ factors. For concreteness, we will frame our discussion in terms of the first example, constructed from the mirror of $\mathbb{P}^{1,1,2,2,6}[12]$, although the conclusions can be translated verbatim to the second case as well.

By looking at \eqref{gNISW1}, it is clear that there is no significant kinetic mixing between the Seiberg--Witten sector and the rest of the EFT, neither at the moduli nor the gauge sector. Equivalently, one can see that \eqref{nomixing} is satisfied up to exponential corrections, and so the first decoupling condition is verified. Notice, nevertheless, that something very similar to $w=1$ LCS limits occurs. The largest gauge couplings and smallest kinetic terms are those along the field theory direction $z^1$ that dominates the K\"ahler potential, and whose U(1) belongs to the gravitational sector of the limit. A priori, this could result in large kinetic mixings with other EFT sectors, and in particular with the SW one we are interested in. However, this is not the case due to the structure of the prepotential \eqref{prepotentialY}, which is such that $\CF_{11I} \simeq \CO(\eps), \forall I=1,2$. In particular, we note that $\CF_{112} = \CO(\eps)$ reflects the definition of core RFT (see section \ref{s:CYlimits}).

Let us now turn to the second decoupling condition \eqref{nomixing2}. To analyse the different Pauli terms around the Seiberg--Witten point, one may look at the asymptotic behaviour of the scalar curvature, and more precisely at that exhibited by $\lVert \CP \rVert_{\rm EFT}^2$ in  \eqref{IIBscalarP}. As shown in Appendix \ref{ap:SWp}, one of the most relevant terms contributing to this quantity is 
\begin{eqn}
    \lVert \CP \rVert^2_{\in \text{cRFT}} = - (g^{2\bar{2}})^2 \CI^{22} |\CP_{222}|^2= \frac{{16}\pi\left( \re z^1-A_1\right)}{\left(\log \left|\frac{z^2}{\hat{\Lambda}_{\rm sw}}\right|^2\right)^3|z^2|^2}\simeq  \frac{\Mpl^2}{m_*^2} R_{\rm sw} \, ,
    \label{SW1Pauli222}
\end{eqn}
whereas the Pauli couplings mixing the core RFT and gravitational U(1)'s give
\begin{eqn}
\label{PmixedSW1}
  \rVert \CP\lVert_{\rm mixed}^2\, \simeq\, {24} \left( \re z^1 \right)^2 \left(\log \left|\frac{z^2}{\Lambda_{\rm sw}}\right|^2\right)^{-2} \, .
\end{eqn}
Applying the decoupling criteria of section \ref{ss:decoupling}, we find that the SW sector decouples from the gravitational sector as long as
\begin{eqn}
    \frac{ \rVert \CP\lVert_{\rm mixed}^2}{ \lVert \CP \rVert^2_{\in \text{cRFT}}} \simeq \frac{3}{2\pi} \re z^1  |z^2|^2 \log \left|\frac{z^2}{\Lambda_{\rm sw}}\right|  \to 0 \, .
    \label{SW1decoup}
\end{eqn}
Recall that the SW point is located at $\re z^1 = \infty, |z^2| =0$, and so depending on which trajectory one takes gravity will be decoupled from the SW sector or not. By direct inspection, one finds that the scalar curvature diverges like \eqref{SW1Pauli222} whenever \eqref{SW1decoup} is met. In the opposite regime, in which the quotient in \eqref{SW1decoup} blows up, the curvature  asymptotes the positive constant $R \to {12}$, which means that $\lVert \CP \rVert_{\rm EFT}^2 \to {24}$. 

To understand the physics behind this decoupling condition, let us rewrite it in terms of the relevant scales of this problem. In particular, we have that the SW monopole mass is given by
\begin{eqn}
    \frac{m_{\rm mon}}{ m_*} = |\p_{a} \mathscr{F}_{\rm sw} (a)| = \left|\frac{z^2}{\pi} \log \left(\frac{z^2}{\hat{\Lambda}_{\rm sw}}\right)^2 - \frac{2 z^2}{\pi} + \ldots\, \right|\, .
    \label{mmonSW1}
\end{eqn}
In addition, we have the two scales
\begin{eqn}
    \Lambda_{\rm wgc} \equiv \frac{g_{\rm sw}}{\sqrt{8\pi}} \Mpl \, , \qquad 
    \Lambda_g \equiv \frac{8\pi}{g_{\rm sw}^2} \Lambda_{\rm RFT} = m_{\rm mon} \, ,
    \label{scalesSW}
\end{eqn}
used in the literature in order to parametrise gravity-decoupling limits \cite{FierroCota:2023bsp,Castellano:2024gwi}, which we already encountered in section \ref{s:LCS}, see \eqref{scales}. In this case, the rigid cut-off is given by $\Lambda_{\rm RFT} = m_{\rm W} = |a|m_*$, namely the scale below which the gauge couplings stop running, and it differs from the one in LCS limits, which is given by $m_*$. The fact that $\Lambda_g$ coincides with the SW monopole mass is explained by recalling that the D6-brane state can be actually regarded as a 't Hooft-Polyakov monopole in the underlying $SU(2)$ gauge theory \cite{tHooft:1974kcl,Polyakov:1974ek}. 

In terms of these scales, we have that the Pauli mixing term reads
\begin{eqn}
    \rVert \CP\lVert_{\rm mixed}\, \simeq\, \sqrt{\frac32}\,\frac{1}{(2\pi)^3} \left(\frac{\Lambda_{\rm wgc}}{\Lambda_{\rm sp}}\right)^2 > \frac{2\sqrt{6}}{\pi}\, ,
    \label{Qwgcsp}
\end{eqn}
where we used that $\Lambda_{\rm sp} = m_{\rm h} = m_*/4\pi$. The inequality can be seen by translating this quotient into the original coordinates $x^i$, since $2\pi \re z^1 \simeq - \log |x^1|^2 - \log |x^2| > -\log |x^2| \simeq 4\log |z^2/\hat\Lambda_{\rm sw}|$, which implies that $\Lambda_{\rm wgc}/m_* = \Lambda_{\rm wgc}/4\pi \Lambda_{\rm sp} >  \sqrt{2}$.\footnote{For the SW limit based on the mirror of $\mathbb{P}^{1,1,2,2,2}[8]$, we have the analogous relation $\Lambda_{\rm wgc}/\Lambda_{\rm sp} >  (4\pi)^{3/2}$, which follows as well from $\log |z^2/\Lambda_{\rm sw}|^2 \simeq \oh \log |\tilde{u}| = - \frac{1}{4} \log |x^2| \leq - \oh  \log |x^1| - \frac{1}{4} \log |x^2| \simeq - \oh \log |\tilde{\eps}| = (\pi/2\sqrt{2}) \re z^1$.} The core RFT Pauli term reads instead
\begin{eqn}
    \lVert \CP \rVert_{\in \text{cRFT}}\,  \simeq\, \frac{2\sqrt{2}}{\pi} \frac{\Lambda_{\rm wgc}}{\Lambda_g} \, ,
\end{eqn}
and the decoupling criterion becomes
\begin{eqn}
    \frac{\sqrt{3}}{32\pi^2} \left(\frac{\Lambda_g}{\Lambda_{\rm wgc}}\right)  \left(\frac{\Lambda_{\rm wgc}}{\Lambda_{\rm sp}}\right)^2 \to 0\, .
   \label{SW1decoup2}
\end{eqn}
Since $\Lambda_{\rm wgc}/\Lambda_{\rm sp}$ is bounded from below, the only way to satisfy the decoupling condition is that the Pauli term of the SW sector diverges in Planck units, together with the moduli space curvature. Reciprocally, if the scalar curvature diverges, the decoupling criterion is automatically satisfied whenever $\Lambda_{\rm wgc}/\Lambda_{\rm sp}$ asymptotes to a constant. If, on the other hand, we have that $\Lambda_{\rm wgc}/\Lambda_{\rm sp} \to \infty$, then an explicit computation reveals that this forces $-\log |x^1| \gg -\log |x^2|$, such that both $\lVert \CP \rVert_{\in \text{cRFT}}\to\infty$ and \eqref{SW1decoup} trivially hold. Notice that $\Lambda_{\rm wgc}/\Lambda_{\rm sp} \to \infty$  corresponds to the gravity-decoupling criterion proposed in \cite{FierroCota:2023bsp}. In the case at hand, we find that it does not capture the full set of limits where the SW sector decouples, since for some of those $\Lambda_{\rm wgc}/\Lambda_{\rm sp}$ asymptotes to a constant. The only quotient whose divergences seem to encode every time the second decoupling condition is satisfied is $\Lambda_{\rm wgc}/\Lambda_{g}$ or, equivalently, the moduli space curvature.

Remarkably, another way to rewrite the decoupling condition is 
\begin{eqn}
  \frac{\sqrt{3}}{32\pi^2} \left(\frac{\Lambda_g}{\Lambda_{\rm sp}}\right) \left(\frac{\Lambda_{\rm wgc}}{\Lambda_{\rm sp}}\right) \to 0 \, .
   \label{SW1decoup3}
\end{eqn}
Together with the lower bound on $\Lambda_{\rm wgc}/\Lambda_{\rm sp}$, this implies the following hierarchies among scales 
\begin{eqn}
\frac{\sqrt{6}}{8\pi}\, \Lambda_g \ll 4\pi\sqrt{2} \Lambda_{\rm sp} = \sqrt{2} m_* <  \Lambda_{\rm wgc}\, ,
\label{SW1hierarchies}
\end{eqn}
with the rhs inequality being true in all limits, whilst the lhs one is rather imposed by decoupling. Thus, condition \eqref{SW1decoup} implies that the 't Hooft-Polyakov monopole mass is below the species scale, and more precisely that $m_{\rm mon} \ll m_*$. On the contrary, if we have that $m_{\rm mon} \simeq m_*$, then the decoupling condition is never met. By the same reasoning as before, one can see that the quotient $m_{\rm mon}/m_{\rm h}$ controls whether or not there is a decoupling of the SW sector from gravity, as illustrated in figure \ref{fig:scales} below.

\begin{figure}[htb]
\begin{center}
\includegraphics[width=0.6\textwidth]{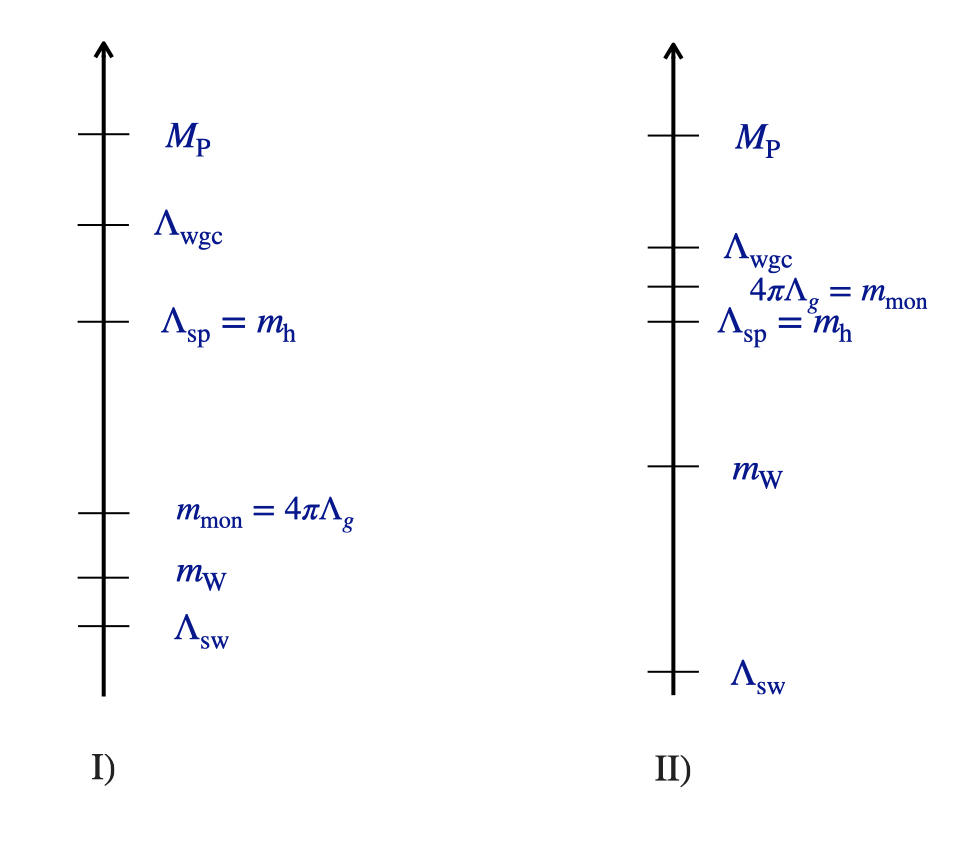}
\caption{\small Mass scales along two different classes of limits approaching the SW point. In scenario \textbf{I)} the field theory scales $(\Lambda_{\rm sw}, m_{\rm W}, m_{\rm mon})$ lie parametrically below the species scale, the decoupling occurs and the moduli space curvature diverges. In scenario  \textbf{II)} $m_{\rm mon} \gtrsim \Lambda_{\rm sp}$, there is no decoupling from the gravitational sector as well as no curvature divergence.}
\label{fig:scales}
\end{center}
\end{figure}

This observation is quite suggestive if we interpret $m_{\rm mon} = \Lambda_g$ as a field theoretic UV completion of the SW sector, as well as the scale of a QFT tower of states. Indeed, notice that for Seiberg--Witten theory at  weak coupling, $m_{\rm mon}$ represents the scale where the tower of dyonic states appears. The mass of these SW-charged BPS particles is given by $m_{\rm dyon} = |q_{\rm e} a + q_{\rm m} a_D| m_*$, where $a_D = \p_{a} \mathscr{F}_{\rm sw}$.\footnote{The rigid charge-to-mass ratio for any of these states is 
\begin{eqn}\nonumber
    \gamma^2_{\rm rigid} \equiv \frac{m_*^2}{\Mpl^2} \gamma^2_{\rm sw} \simeq \frac{m_*^2}{m_{\rm W}m_{\rm mon}}\, ,
    \label{gammasw}
\end{eqn}
in terms of which, one may rewrite the decoupling condition as $\gamma^{-2}_{\rm rigid} \frac{\Mpl^2}{m_*^2} \to 0$. Since along the infinite distance limit $\Mpl/m_* \to \infty$, the decoupling implies that $\gamma_{\rm rigid} \to \infty$. Notice that this is stronger than simply requiring $\gamma_{\rm sw} \to \infty$, which recovers a rigid SW Lagrangian but does not guarantee full decoupling from gravity.} 
In other words, for the decoupling to take effect, the UV completion of the RFT must see the whole tower of RFT-charged states---namely the dyons---before reaching the scale where quantum gravity effects become important. Infinite towers of RFT-charged states are a common feature of rigid limits, and they have been considered in \cite{Marchesano:2024tod} in the context of LCS limits. It was observed that the typical scale of such a tower, dubbed $\Lambda_{\rm ch}$ therein, lies at or below the species scale in gravity-decoupling limits, and that its position with respect to the SDC towers determines the nature of the rigid UV completion of the decoupling RFT, dubbed UVRT. If $\Lambda_{\rm ch}$ lies above some KK tower, we have a higher-dimensional UVRT, like a 5d or 6d SCFT or a Little String Theory (LST). In this sense, the fact that for the present limits  $\Lambda_{\rm ch} = m_{\rm mon}$ lies below any KK tower can be interpreted as the Seiberg--Witten sector being genuinely a 4d RFT decoupled from gravity, as opposed to an RFT that arises from dimensionally reducing a higher-dimensional rigid theory. It also reflects that, in Seiberg--Witten limits, the KK and species scales cannot be separated, unlike what happens in LCS emergent string limits \cite{Marchesano:2024tod}.

\newpage

Based on this SW example, it is tempting to speculate that divergences of either $\Lambda_{\rm sp}/\Lambda_g$ or $\Lambda_{\rm sp}/\Lambda_{\rm ch}$ could measure the decoupling of rigid sectors. However, our previous findings show that this cannot be true in general. Indeed, as pointed out around \eqref{decouplingLCS} for certain LCS $w=1$ limits one has that the core RFT decouples from gravity, while at the same time $\Lambda_{\rm sp}/\Lambda_g$ remains constant. In fact, it follows from the analysis in \cite{Marchesano:2024tod} that the same happens for $\Lambda_{\rm sp}/\Lambda_{\rm ch}$ along those limits, wherein the UVRT is a Little String theory. We then conclude that $\Lambda_{\rm sp}/\Lambda_g$ is a good gravity-decoupling indicator whenever the rigid UV completion of the core RFT is a field theory itself. If instead the UVRT is an LST, the decoupling of gravitational and rigid sectors is not based on field theory mechanisms, and so the condition that gravitational scales must decouple from rigid ones no longer needs to hold. 

For the limits that we have analysed in this work, the quotient that better captures when a rigid sector decouples from gravity seems to be  $\Lambda_{\rm wgc}/\Lambda_{g}$. However, as pointed out in \cite{Castellano:2024gwi}, unless one restricts to certain LCS trajectories, this quotient only sets an upper bound for the scalar curvature, whose divergence is the ultimate criterion for decoupling in any kind of $\CN=2$ RFT limit. A divergence in  $\Lambda_{\rm wgc}/\Lambda_{g}$ can thus only be seen as a necessary condition for decoupling, and not a sufficient one. Following \cite{Castellano:2024gwi}, one interprets $\Lambda_{\rm wgc}$ as the gravitational estimate for the UV completion scale of the core RFT, while $\Lambda_{g}$ corresponds to its field theory version. Even in those cases where the UVRT in an LST, our results indicate that $\Lambda_{g}$ must lie parametrically below $\Lambda_{\rm wgc}$ for the decoupling to take place, since otherwise quantum gravity effects could be as relevant as e.g. QFT non-perturbative effects, and so the RFT does not decouple from the gravitational sector. 


\section{Conclusions}
\label{s:conclu}

In this work, we have studied the physics of gravity decoupling along finite- and infinite-distance limits in 4d $\CN=2$ vector multiplet moduli spaces. We have done so in the context of Calabi--Yau compactifications of type II string theory, bearing in mind limits within the interior of the Coulomb branch which develop axionic shift symmetries at their endpoint. The emerging picture is that one can distinguish between three different kinds of U(1)'s:

\begin{itemize}
    \item[-] \underline{Gravitational U(1)'s}: These are U(1)$_{\eta\q}$ that remain inherently gravitational along the limit, with their mutual coupling being comparable to their coupling to the graviphoton. Via \eqref{U1q}, they map to states $\q$ with charge-to-mass ratio $\gamma_{\q} = \CO(1)$ at the limit endpoint, as expected for extremal BPS states.

    \item[-] \underline{Rigid U(1)'s}: They exhibit mutual couplings that are parametrically larger than the analogous one to the graviphoton. As a result, the latter reduce to the self-interactions of a rigid $\CN=2$ theory. The associated charge $\q$ has a ratio $\gamma_\q$ that diverges at the singularity. 

    \item[-] \underline{Core RFT U(1)'s}: These are rigid U(1)'s such that, in addition, the orbits implemented by the monodromy action do not mix them with any gravitational field. Equivalently, the shift symmetries at the limit endpoint do not rotate them with gravitational U(1)'s via theta angles. This must be valid for their electro-magnetic dual as well, see section \ref{ss:core} for the precise definition.
    
\end{itemize}

Having a rigid U(1) only implies the existence of a piece of the Lagrangian with rigid-like couplings, but it does not guarantee its decoupling from gravity. In fact, rigid U(1)'s outside of the core RFT cannot decoupled due to their theta angle mixing, which relates either them or their magnetic duals with the gravitational U(1) sector. This enforces a non-negligible mixing at the level of the rigid gauge kinetic matrix, and that the monodromy group acts on them in the same block where the graviphoton transforms. 

Regarding the core RFT U(1)'s, they still need to meet certain conditions to actually decouple from gravity, dictated by field theory standard wisdom. First, they need to have negligible mixing with the U(1)'s outside of the core RFT which, as discussed in section \ref{ss:decoupling}, boils down to requiring absence of significant mixing in the rigid gauge kinetic function. For singularities of maximal codimension, we typically find that this condition is automatically satisfied after imposing the core RFT criterion, as well as that all rigid U(1)'s are within the core RFT.

Second, there must be no significant mixing at the level of the interactions. In this case, this amounts to the Pauli couplings \eqref{Pauli}, which encode the cubic derivatives of the prepotential. We require that the physical Pauli couplings within the core RFT are parametrically stronger than those that mix the latter with the vector multiplets outside of it. Remarkably, this hierarchy in physical Pauli couplings implies a hierarchy among the terms that contribute to the moduli space curvature. In particular, except in those cases where the core RFT displays $\CN=4$ extended supersymmetry, satisfying this condition implies a scalar curvature divergence. Moreover, we have not been able to find any example where a curvature divergence is generated and no decoupling occurs, in full agreement with the Curvature Criterion \cite{Marchesano:2023thx}. Profiting from our more refined picture, we have formulated a slightly stronger version of this conjecture, which states that physical Pauli couplings fully outside of the core RFT, which correspond to the actual gravitational sector, never diverge nor induce a divergence in the curvature. 

In order to arrive at this picture, we have explored in some detail the physics of metric essential instantons. By adding a reasonable assumption like \eqref{metriker} to their definition, we have been able to identify them as specific BPS states of the theory. In fact, using the machinery of weight filtrations associated to nilpotent orbits, we have been able to translate the monodromy data into  general results on both the charge and mass of these modes. In particular, we have identified the electric states as BPS particles that lie parametrically below the maximal UV cut-off of the limit $m_*$, which in the case of infinite-distance points coincides with the scale of the leading SDC tower, and in emergent string limits with the species scale. The magnetic duals to these electric states may or may not lie below $m_*$, but we have been able to derive that in any event both electric and magnetic particles have a diverging charge-to-mass ratio. Furthermore, the associated U(1)'s satisfy all the conditions of a core RFT sector. We have illustrated such features in well-known examples of finite- and infinite-distance limits away from the LCS regime, like conifold and coni-LCS singularities, as well as explicit instances of Seiberg--Witten points. 

Out of these examples, Seiberg--Witten limits are perhaps the most interesting ones. While the core RFT can be identified with the Seiberg--Witten sector,  the conditions to decouple it from the rest of the EFT are non-trivial.  Indeed, while in an appropriate basis it is straightforward to see that gauge and kinetic mixing with the gravitational  sector is negligible, Pauli mixing terms need not be so. In certain limits, these terms are of the same order as the Pauli couplings intrinsic to Seiberg-–Witten theory and therefore constitute an obstruction to decoupling that, to our knowledge, has not been previously identified in the literature. The limits approaching the Seiberg--Witten point such that the decoupling occurs can be precisely identified with those in which the moduli space curvature diverges, again in agreement with the Curvature Criterion. Alternatively, one can interpret the decoupling (or lack thereof) in terms of the infinite towers of states present in this setting. On the one hand, there is the field-theoretic tower of dyons familiar from Seiberg--Witten theory, whose scale is set by the 't Hooft-Polyakov monopole mass, or equivalently by the rigid cut-off $\Lambda_g$. On the other hand, there is the SDC tower represented by an emergent heterotic string, and whose scale $m_*$ coincides with the species cut-off $\Lambda_{\rm sp}$. When $\Lambda_{\rm sp} \lesssim \Lambda_g$, there is an obstruction to the decoupling of the SW sector from gravity, which in the IR can only be detected through physical Pauli couplings. In the UV completion of the theory, one may interpret this lack of decoupling in terms of the massive tower spectrum, which suggests that quantum gravity effects are as significant as field theory ones. To achieve decoupling from gravity, the rigid field-theoretic scale $\Lambda_g$ must hence lie parametrically below any gravitational scale, like the species scale or the WGC UV completion estimate $\Lambda_{\rm wgc}$, as defined in \eqref{scalesSW}. 

Notice that this last interpretation goes along the lines of the philosophy of \cite{Castellano:2024gwi}, which proposed to measure gravity decoupling along asymptotic limit in terms of the behaviour of the quotient of scales $\Lambda_{\rm wgc}/\Lambda_g$. Since several such quotients have been proposed to measure this kind of decoupling, let us summarise here how our results reflect on each of these proposals:

\begin{itemize}
    \item[-] \underline{$\Lambda_{\rm sp}/\Lambda_g$}: Its divergence along a limit seems to properly measure that gravity decouples, except in those cases where the core RFT comes from the dimensional reduction of a Little String Theory, as it happens in certain LCS limits \cite{Marchesano:2024tod}. Such a divergence thus seems to correspond to a sufficient but not a necessary condition for decoupling. 

    \item[-] \underline{$\Lambda_{\rm wgc}/\Lambda_{\rm sp}$}: Divergences in this quotient measure gravity decoupling when applied to rigid U(1)'s in LCS limits. However, in Seiberg--Witten/coni-LCS points it fails to capture such a decoupling, as there are limits where it remains constant yet the decoupling still occurs.

    \item[-] \underline{$\Lambda_{\rm wgc}/\Lambda_g$}: As pointed out in \cite{Castellano:2024gwi}, this quotient has a similar behaviour to the (square-root of the) moduli space scalar curvature in a large number of asymptotic limits, while it only bounds it from above for others. In this sense, its divergence seems to represent a necessary condition, but not a sufficient one.  
    
\end{itemize}

In general, our results indicate that it is the asymptotic behaviour of the scalar curvature measured in Planck units---or, in certain cases, of specific sectional curvatures of the moduli space---that gives the most reliable indicator of whether a gauge sector decouples from gravity. As noted above, this conclusion holds provided one excludes core RFT sectors with an extended $\CN=4$ supergravity. Hence, it would be highly interesting to develop precise criteria to parametrise gravity decoupling in such (more restricted) instances. 

Besides this line of action, our results can be extended in a number of directions. For instance, it would be interesting to analyse if they can be generalised to moduli-space singularities that do not exhibit axionic shift symmetries at their endpoint. This should in particular include strong coupling singularities with monodromy generators that do not commute among themselves \cite{Akhond:2021xio}. Nevertheless, deep in their Coulomb branch one may still be able to apply many of the concepts developed here, as well as to test the implications of the monodromy group for  the EFT along the lines of \cite{Marchesano:2022avb,Delgado:2024skw}. Additionally, one could extend the distinction between gravitational and rigid sectors to other settings, like for instance the hypermultiplet sector of 4d $\CN=2$ EFTs. While in this case the moduli space structure is more involved, there is still a well-defined distinction between gravitational and rigid Lagrangians, which suggests that one may translate this structure in terms of the physics of strings and instantons. Presumably, this link could shed some light on the results obtained in \cite{Marchesano:2019ifh,Baume:2019sry,Grimm:2019wtx} when exploring infinite-distance limits in such moduli spaces. Finally, it would be interesting to apply the same kind of strategy to develop a general macroscopic understanding of rigid limits for EFTs in diverse dimensions and with different amounts of supersymmetry, as already done for 5d $\CN=1$ EFTs in \cite{Blanco:2025qom}. As usual, of particular importance are 4d $\CN=1$ EFTs, featuring chiral matter and non-trivial scalar potentials \cite{Ibanez:2012zz,Blumenhagen:2013fgp,Baumann:2014nda,McAllister:2023vgy,Marchesano:2024gul}. Understanding how gravity decoupling works in this case might give us a new perspective on the structure and limitations of low-energy effective field theories in quantum gravity. We hope to return to all these questions in the  future.

\bigskip

\bigskip

\centerline{\bf  Acknowledgments}

\vspace*{.4cm}

We would like to thank Christian Aoufia, Alexander Bernal, Gonzalo F. Casas, Matilda Delgado, Damian van de Heisteeg, \'Alvaro Herraez, Luis Ib\'añez, S\"oren Kotlewski, Luca Martucci, Luca Melotti, Gonzalo Morr\'as, Sav Sethi, Michelangelo Tartaglia, \'Angel Uranga, Max Wiesner, and Matteo Zatti for discussions.  This work is supported through the grants {\small CEX2020-001007-S, PID2021-123017NB-I00 and PID2024-156043NB-I00}, funded by {\small MCIN/AEI/10.13039/501100011033}, and ERDF, EU. L.P. is supported by the fellowship {\small LCF/BQ/DI22/11940039} from ``La Caixa" Foundation (ID 100010434). The work of A.C. is supported by a Kadanoff and an Associate KICP fellowships, as well as through the NSF grants {\small PHY-2014195 and PHY-2412985}. L.P. would like to thank  the II. Institute for Theoretical Physics at the University of Hamburg and the DESY Theory Group for hospitality during late stages of this work. A.C. would like to thank IFT-Madrid for hospitality during different stages of this work, and Teresa Lobo for her continuous encouragement and support.


\appendix

\section{Tools from Special K\"ahler Geometry}
\label{ap:special}

In this appendix, we collect the special geometry background that is needed for the main text, and also derive afresh several technical results that are used throughout the work. We first provide a summary of the generic symplectic structure of four-dimensional $\mathcal{N}=2$ supergravity theories, and then specialise to the case where a formulation in terms of a prepotential is viable. After having introduced all the necessary material, we discuss some of the requirements---mentioned in the main body of the paper---that are needed for obtaining a rigid theory. Lastly, we consider the case where we do not have a canonical symplectic pairing for the basis of $H_3(Y_3)$. 

\subsection{The symplectic structure of 4d $\mathcal{N}=2$ supergravity}
Let us denote by $\CM$ the space whose coordinates $\{z^i\}$ are given by the $n_V$ vevs of 4d complex scalar fields belonging to vector multiplets in $\mathcal{N}=2$ supergravity theories. This space can be seen as the base of a tensor bundle $\mathcal{H}=\mathcal{SV}\bigotimes\mathcal{L}$, where $\mathcal{SV}\to\mathcal{M}$ is a holomorphic flat vector bundle of rank $2n_V+2$ and structural group $\mathrm{Sp}(2n_V+2,\mathbb{R})$, and  $\mathcal{L}\to\mathcal{M}$ is a holomorphic line bundle. Additionally, $\mathcal{M}$ is also a Hodge-K\"ahler manifold, meaning that its metric can be derived from a K\"ahler potential $K$ as in
\begin{equation}\label{kahlermetric}
    g_{i\bar j}=\partial_i\bar\partial_j K\, .
\end{equation}
In turn, the above K\"ahler potential can be written as\footnote{In order to lighten the notation, we set $\kappa_4=1$ throughout this section.}
\begin{equation}
    K=-\log(i\langle\Omega,\bar \Omega\rangle)\, ,
\end{equation}
for some section $\Omega(z)$ of $\mathcal{H}$ and compatible hermitian metric $i\langle\cdot,\cdot\rangle$.

When applied to type IIB string theory compactified on a Calabi--Yau three-fold $Y_3$, this picture inherits a nice geometrical interpretation. $\mathcal{M}$ becomes the moduli space of the complex structure deformations of $Y_3$, whilst the sections of the $\mathcal{H}$ bundle belong to its complex cohomology $H^3(Y_3,\mathbb{C})$. This geometric picture also provides a natural hermitian metric
\begin{equation}\label{wedgeprod}
    i\langle A,\bar B\rangle=i\int_Y d^6y \;A\wedge\bar B\, ,
\end{equation}
as well as an interpretation for $\Omega$ as the unique, nowhere-vanishing, holomorphic, harmonic, and covariantly constant (3,0)-form on $Y_3$. From now on, we will stick to this geometric perspective.

There are two different basis for the third cohomology $H^3(Y_3,\mathbb{C})$ that are used consistently in this work. The first one is the basis of $2n_V+2$ real three-forms $\{\alpha_I,\beta^K\}$, which satisfy
\begin{equation}\label{realformsbasis}
    \langle \alpha_I,\alpha_J\rangle=\langle \beta^K,\beta^L\rangle=0\, ,\quad\langle\alpha_I ,\beta^K\rangle=-\langle \beta^K,\alpha_I\rangle=\delta_I^K\, ,
\end{equation}
whereas the second one is linked to the Hodge-theoretic decomposition 
\begin{equation}\label{hodgedec}
    H^3(Y_3,\mathbb{C})=H^{(3,0)}\bigoplus H^{(2,1)}\bigoplus H^{(1,2)}\bigoplus H^{(0,3)}\, .
\end{equation}
Because of its uniqueness, $\Omega$ and its complex conjugate $\bar\Omega$ are chosen as representatives of $H^{(3,0)}$ and $H^{(0,3)}$, respectively. The $n_V$ (2,1)- and (1,2)-forms spanning $H^{(2,1)}$ and $H^{(1,2)}$ will be denoted by $G_i$ and $\bar G_{j}$. The change of basis between $\{\Omega,G_i,\bar G_j,\bar\Omega\}$ and $\{\alpha_I,\beta^K\}$ is encoded in
\begin{subequations}\label{changeofbasis}
  \begin{equation}
    \Omega=X^I\alpha_I-\CF_I\beta^I\, ,
 \end{equation}
  \begin{equation}
   G_i=\chi^J_i\alpha_J-\tilde\chi_{iJ}\beta^J\, .
 \end{equation} 
\end{subequations}
This second basis, the one linked to the Hodge decomposition, has the advantage of satisfying the so-called orthogonality relations, namely that a $(p,q)$- and a $(p',q')$-form are orthogonal with respect to \eqref{wedgeprod} unless $p+p'=q+q'=3$. This means that 
\begin{equation}
    \langle\Omega,G_i\rangle=\langle\Omega,\bar G_{\bar j}\rangle=\langle\bar\Omega, G_{i}\rangle=\langle\bar\Omega,\bar G_{\bar j}\rangle=0\, .
\end{equation}
Exploiting now the fact that, under an infinitesimal motion on $\CM$, a closed $(p,q)$-form only mixes with closed $(p\pm1,q\mp1)$-forms, we can define $G_i$ as the part of $\partial_i\Omega$ parallel to $H^{(2,1)}$, i.e.
\begin{equation}\label{derofOmega}
    \partial_i\Omega=G_i-k_i\Omega\, .
\end{equation}
Therefore, it follows that 
\begin{equation}
k_i=-\frac{\langle\partial_i\Omega|\bar\Omega\rangle}{\langle\Omega|\bar\Omega\rangle}=\partial_iK\, ,
\end{equation}
and, from \eqref{kahlermetric}, that the only non-trivial inner products are
\begin{equation}
    i\langle\Omega|\bar\Omega\rangle=e^{-K}\, ,\qquad i\langle G_i|\bar G_{\bar j}\rangle=-e^{-K}g_{i\bar j}\, .
\end{equation}

Before proceeding, notice how the change of basis \eqref{changeofbasis} can be reinterpreted in terms of periods and three-cycles. In fact, one can define a dual basis of three-cycles $\{A^I,B_K\}$ within $Y_3$, spanning the third-homology group $H_3(Y_3, \mathbb R)$, and such that 
\begin{subequations}
  \begin{equation}\label{cyclesbasis}
    A^I\cap A^J=B_K\cap B_L=0\, , \qquad A^I\cap B_K=\delta^I_K\, ,
 \end{equation}
  \begin{equation}
   \int_{A^I} \a_J=\delta^I_J\, , \qquad \int_{B_K}\b^L=-\delta^L_K\, .
 \end{equation} 
\end{subequations}
This allows us to expand any three-form $ v\in H^3(Y_3, \mathbb C)$ as follows
\begin{equation}
     \mathbf{v}=\left(\begin{array}{c}
         \int_{A^I} v\\
         \int_{B_K} v 
    \end{array}\right)=\left(\begin{array}{c}
         \langle v,\b^I\rangle\\
         \langle v,\a_J\rangle 
    \end{array}\right)\, ,
\end{equation}
where $\mathbf{v}$ is the vector of coefficients in the corresponding basis of differential forms $\{\a_I,\b^K\}$ (equivalently the vector of periods in the basis $\{ A^I, B_K\}$). When applied to the holomorphic three-form $\Omega$, this expansion provides us the well-known period vector
\begin{equation}
     \mathbf{\Pi}=\left(\begin{array}{c}
         X^I\\
         \mathcal{F}_K
    \end{array}\right)\, .
\end{equation}

From \eqref{derofOmega}, we can define the usual K\"ahler covariant derivative as the linear differential operator $\mathcal{D}_i:\;H^{(p,q)}\to H^{(p-1,q+1)}$ such that 
\begin{equation}
    \mathcal{D}_i\Omega=(\partial_i+k_i)\Omega=G_i\,.
\end{equation}
In analogy with the holomorphic sections $\{\Omega,G_i\}$, we can also define \textit{covariantly} holomorphic ones $\{V,U_i\}$ simply by
\begin{equation}\label{covholsections}
    V=e^{K/2}\Omega\, ,\qquad U_i=e^{K/2}G_i\,.
\end{equation}
The latter satisfy the orthogonality relations 
\begin{equation}
    i\langle V,\bar V\rangle=1\,,\qquad i\langle U_i,\bar U_{\bar j}\rangle=-g_{i\bar j}\,,
\end{equation}
and the covariant derivative $D_i$ will now act on $V$ as\footnote{Here we use different symbols for the covariant derivative acting on $\Omega$ or $V$. One can alternatively say that these sections have different weight $p$, namely $p=1$ and $p=\oh$.}
\begin{equation}
    D_iV=(\partial_i+\tfrac12K_i)V=U_i\,.
\end{equation}
Notice how, similarly to the fact that $\{\Omega,G_i\}$ were such that
\begin{equation}
    \bar\partial_{\bar j}\Omega=\bar\partial_{\bar j}G_i=0\, ,
\end{equation}
the sections $\{V,U_i\}$ satisfy
\begin{equation}
    \bar D_{\bar j}V=\bar D_{\bar j}U_i=0\, .
\end{equation}
Exploiting the orthogonality relations together with the definition of K\"ahler covariant derivative, one can introduce a $(0,3)$-tensor on $\CM$ whose components are defined through
\begin{equation}\label{Cijkdef}
    C_{ijk}=\langle D_iU_j,U_k\rangle=\langle V,D_iD_jD_kV\rangle\, ,\quad D_iU_j=iC_{ijk}g^{k\bar k}\bar U_{\bar k}\, .
\end{equation}
This object is particularly important because one can express both the Riemann tensor and the Ricci scalar of $\CM$ in terms of it, namely
\begin{equation}
    R_{i\bar j k\bar l}=g_{i\bar j}g_{k\bar l}+g_{i\bar l}g_{k\bar j}- C_{ikm}\bar C_{\bar j\bar l\bar n}g^{m\bar n }\,,
\end{equation}
and
\begin{equation}
    R=-2n(n+1)+2 C_{ikm}\bar C_{\bar j\bar l\bar n}g^{i\bar j }g^{k\bar l }g^{m\bar n }\, .
\end{equation}
Notice that the change of basis between $\{V,U_i,\bar U_j,\bar V\}$ and $\{\alpha_I,\beta^K\}$ now reads
\begin{subequations}\label{changeofbasis2}
  \begin{equation}
    V=L^I\alpha_I-M_I\beta^I\, ,
 \end{equation}
  \begin{equation}
  U_i=f^J_i\alpha_J-h_{iJ}\beta^J\, .
 \end{equation} 
\end{subequations}
where, because of \eqref{covholsections}, we have
\begin{equation}
    L^I=e^{K/2}X^I\, ,\quad M_I=e^{K/2}\CF_I\;,\quad f^J_i=e^{K/2}\chi^J_i\, ,\quad h_{iJ}=e^{K/2}\tilde{\chi}_{iJ}\, .
\end{equation}

These quantities are related to the gauge kinetic function $\CI_{IJ}$ and the theta-terms $\CR_{IJ}$ of the 4d $\CN=2$ supergravity theory. In fact, the complex matrix 
$\CN_{IJ}=\CR_{IJ}+i\mathcal{I}_{IJ}$ must satisfy
\begin{equation}\label{eq:constraintscovholosections}
    M_A=\CN_{AB}L^B\, ,\qquad h_{iA}=\bar{\CN}_{AB}f^B_i\, ,
\end{equation}
and then, if we assemble the $(n+1)\times (n+1)$ matrices
\begin{equation}\label{supportmatrices}
    {\mathrm{f}^J}_I:=\left(\bar{L}^J\quad f_i^J\right)\,,\qquad {\mathrm{h}}_{JI}:=\left(\bar{M}_J\quad h_{iJ}\right)\,,
\end{equation}
we get
\begin{equation}
    \mathrm{h}_{IJ}=\CN_{IK}\,{\mathrm{f}^K}_I\,\implies\,\CN_{IJ}=\mathrm{h}_{IK}{f^{-1|K}}_J\;.
\end{equation}
By carefully changing basis from $\{V,U_i,\bar U_j,\bar V\}$ to $\{\alpha_I,\beta^K\}$, one can also recover the following useful relations
\begin{equation}\label{usefulrelations}
    \CI_{IJ}L^I\bar L^J=-\oh\,,\quad g_{i\bar j}=-2f_i^I\CI_{IJ}\bar f^J_{\bar j}\,,\quad f_i^Ig^{i\bar j}\bar f^J_{\bar j}=-\oh\CI^{IJ}-\bar L^IL^J\, .
\end{equation}

These functions are also related to the physical charge of a massive state belonging to the theory. The latter is linked to the so-called Hodge-$\star$ operator, which acts on a $(p,q)$-form $v\in H^{(p,q)}$ as follows
\begin{equation}
    \star v=i^{p-q}v\, .
\end{equation}
On the forms we already introduced, the Hodge-$\star$ operator then acts as 
\begin{equation}\label{hodgedef}
    \star \Omega=-i\Omega\,,\qquad \star \chi_i=i\chi_i\, ,
\end{equation}
and, more generally, it can be used to define the \textit{Hodge norm}
\begin{equation}
    ||v||^2:=\int_{Y_3}v\wedge\star\bar v= \langle v,\star\bar v\rangle:=\langle C v,\bar v\rangle\, ,
\end{equation}
where now $v\in H^3(Y_3, \mathbb{C})$ and we have also introduced the Weil operator $C$. Using this formalism, we shall rewrite
\begin{equation}
    e^{-K}=i\langle \Omega,\bar\Omega\rangle= i\int_{Y_3}\Omega\wedge \bar \Omega=\int_{Y_3}\Omega\wedge\star\bar \Omega=||\Omega||^2\, .
\end{equation}
Since the Hodge-$\star$ operator maps three-forms into themselves, we may obtain its action on the basis $\{\a_K,\b^K\}$ by expanding
\begin{subequations}
  \begin{equation}
    \star \a_K=C_K^L\a_L-A_{KL}\b^L\, ,
 \end{equation}
  \begin{equation}
  \star \b^K=B^{KL}\a_L-C_L^K\b^L\, ,
 \end{equation} 
\end{subequations}
and using both \eqref{changeofbasis} and \eqref{hodgedef}. What one finds is 
\begin{equation}
    C_K^L=\mathcal{R}_{KM}\mathcal{I}^{ML}\;,\quad  B^{KL}=\mathcal{I}^{KL}\, ,\quad A_{KL}=\mathcal{I}_{KL}+\mathcal{R}_{KM}\mathcal{I}^{MN}\mathcal{R}_{NL}\,.
\end{equation}
Hence, if we consider any three-form with constant integer coefficients
\begin{equation}
    \mathsf{q}=p^K\a_K-q_L\b^L\, ,
\end{equation}
we have that
\begin{equation}
\begin{aligned}
    ||\mathsf{q}||^2&=\int_{Y_3}\mathsf{q}\wedge\star \mathsf{q}=-p^I\mathcal{I}_{IJ}p^J-p^I(\mathcal{R}\mathcal{I}^{-1}\mathcal{R})_{IJ}p^J+2p^I(\mathcal{R}\mathcal{I}^{-1})_{\ I}^Jq_J-q_I(\mathcal{I}^{-1})^{IJ}q_J\label{hodgenorm}\\\nonumber
    &=-(p^I\quad q_I)\left(\begin{array}{cc}
        \mathcal{I}+\mathcal{R}\mathcal{I}^{-1}\mathcal{R} & -\mathcal{R}\mathcal{I}^{-1} \\
        -\mathcal{I}^{-1}\mathcal{R} & \mathcal{I}^{-1}
    \end{array}\right)\left(\begin{array}{c}
        p^J \\
        q_J
    \end{array}\right)=-\mathsf{q}^{\mathrm{T}}\mathcal{M}(\mathcal{N})\mathsf{q}=2{\mathcal{Q}}_\q^2\,\,,
\end{aligned}
\end{equation}
where $\mathcal{Q}_\q$ is the physical charge of a particle with charge vector given (with a slight abuse of notation) by $\q=(p^I,q_J)^{\mathsf{T}}$. The quantity $\mathcal{M}(\mathcal{N})$ is then the matrix representation of the Weil operator $C$ in the real basis $\{\a_K,\b^K\}$, and we notice that its determinant is identically one.

This physical charge ${\mathcal{Q}}_\q^2$ satisfies another important relation, namely
\begin{equation}\label{charges}
    {\mathcal{Q}}_\q^2=|Z_{\q}|^2+|Z_{\q,i}|^2\, ,
\end{equation}
where 
\begin{equation}
\begin{aligned}
        Z_{\q}&=q_IL^I-p^I\CF_I\, ,\\
        Z_{\q,i}&=D_iZ_{\q}=q_If^I_i-p^Ih_{iI}\,,
\end{aligned}
\end{equation}
and $|Z_{\q,i}|^2=g^{i\bar j}Z_{\q,i}\bar Z_{\q,\bar j}$. These give the charges of the field strength of the graviphoton and the other U(1)'s orthogonal to it, which are in one-to-one correspondence with the original quantised charges $\q=(p^I, q_I)^\mathsf{T}$, but crucially depend on the scalar fields since they refer to supermultiplet eigenstates. Eq. \eqref{charges} thus allows us to connect once again the real basis $\{\a_I,\b^K\}$ and the complex one $\{V,U_i,\bar U_j,\bar V\}$, this time from the perspective of the electric and magnetic charges associated to the U(1)${}^{n_V+1}$ gauge group.

\subsection{The prepotential formulation}

Let us introduce now the $n_V$ functions
\begin{equation}
    Z^i(z):=\frac{X^i(z)}{X^0(z)}\,.
\end{equation}
Whenever these functions are invertible with respect to $z^i$, meaning that
\begin{equation}\label{invspec}
    \det\left[\frac{\partial Z^i}{\partial z^j}\right]\neq 0\, ,
\end{equation}
we can construct a prepotential, which is a holomorphic function defined as
\begin{equation}\label{prep}
    \CF=\oh X^I\CF_I\, ,
\end{equation}
that crucially satisfies
\begin{equation}
    \CF_I=\frac{\partial\CF}{\partial X^I}\,,\qquad \CF(X^I)=(X^0)^2\CF(Z^i)\,.
\end{equation}
Using the following notation
\begin{equation}
    \CF_{IJ}=\partial_I\partial_J\CF\,,\qquad \CF_{IJK}=\partial_I\partial_J\partial_K\CF\, ,
\end{equation}
one can also prove that 
\begin{equation}\label{prepprop}
    \CF_{I}=\CF_{IJ}X^J\,,\qquad \CF_{IJK}X^K=0\,.
\end{equation}
This way, one can almost identically repeat the discussion outlined above, with $\CF_{IJ}$ taking the r\^ole of $\CN_{AB}$. One then gets 
\begin{equation}
    M_I=\CF_{IJ}L^J\,,\qquad h_{iJ}={\CF}_{JK}f^K_i\, ,
\end{equation}
as well as 
\begin{equation}\label{projectionprep}
    N_{IJ}L^I\bar L^J=-1\,,\quad g_{i\bar j}=f_i^IN_{IJ}\bar f^J_{\bar j}\,,\quad f_i^Ig^{i\bar j}\bar f^J_{\bar j}=N^{IJ}-L^I\bar L^J\,,
\end{equation}
with $N_{IJ}=2\im\CF_{IJ}$. In this case, the derivation of these equations proves to be more useful, since it allows us to recover also the relation between the determinant of the metric $g_{i\bar j}$ and that of $N_{IJ}$, i.e. \eqref{detgvsdetN}. To show this, it is sufficient to express using both the basis $\{V,U_i,\bar U_j,\bar V\}$ and $\{\a_I,\b^K\}$, the inner product $i\langle(\bar V\quad \bar U_j)^{\mathsf T},(V\quad U_i)\rangle$, which reads
\begin{equation}
    \left(\begin{array}{cc}
        -1 &  0\\
        0 & g_{i\bar{j}}
    \end{array}\right)=i\langle(\bar V\quad \bar U_j)^{\mathsf T},(V\quad U_i)\rangle=\left(\begin{array}{c}
        \bar L^K \\
        \bar f^K_j
    \end{array}\right)N_{KL}\,(L^L\quad f_i^L)\,.
\end{equation}
Next, recalling that $f_i^L=e^{K/2}\partial_iX^L+k_iL^L$, one readily notices that the $i$-th column of the matrix $(L^L\quad f_i^L)$ is just the sum of $e^{K/2}\partial_iX^L$ and a multiple of the 0-th column, namely $L^L$. By the properties of the determinant, this implies that 
\begin{equation}
    \det\left[(L^L\quad f_i^L)\right]=e^{(n_V+1)K/2}(X^0)^{n_V+1}\det\left[\frac{\partial Z^i}{\partial z^j}\right]\,,
\end{equation}
from which \eqref{detgvsdetN} automatically follows. Notice how this was not possible without a prepotential, since the matrix ${\mathrm{f}^J}_I$ in \eqref{supportmatrices}---used to derive the counterparts \eqref{usefulrelations}---does not share the same property for its columns.

We can now rewrite all the couplings we introduced before in terms of the prepotential. Expressing $\CN_{IJ}$ in terms of $\CF_{IJ}$ one gets
\begin{equation}
    \CN_{IJ}=\bar \CF_{IJ}+i\frac{N_{IK}X^KN_{JL}X^L}{N_{KL}X^KX^L}\, ,
\end{equation}
whereas for $C_{ijk}$ we have
\begin{equation}\label{CintermsofF}
   C_{ijk}=e^{-K/2}f_i^If_j^Jf_k^K\CF_{IJK}\, .
\end{equation}
This last equation is particularly useful because it allows us to rewrite the Ricci scalar as follows
\begin{equation}
    R=-2n(n+1)+2e^{-K} \CF_{IJK}\bar \CF_{\bar L\bar M\bar N}N^{IL }N^{JM }N^{KN}\, ,
\end{equation}
where we have used (together with \eqref{CintermsofF}) also the third equation in \eqref{projectionprep} and the second one in \eqref{prepprop}. Finally, note how we can also obtain \eqref{IIBscalar} just by first inserting \eqref{prepprop} into \eqref{CintermsofF}, and then using the chain rule for the special coordinates, namely 
\begin{equation}    C_{ijk}=e^{K}\partial_{Z^i}X^I\partial_{Z^j}X^J\partial_{Z^k}X^K\CF_{IJK}=(X^0)^3 e^{K} \CF_{ijk}\, .
\end{equation}
In general, \eqref{invspec} is not granted, and when it is not satisfied the function \eqref{prep} cannot be used. However, it always exists a symplectic frame where a prepotential formulation is viable, meaning that there always exists a $\mathsf{S}\in {\rm Sp}(2n_V+2)$ such that for $\mathsf{S}\cdot\mathbf{\Pi}$  \eqref{invspec} holds. Assuming that we rotate to such a frame, we can identify the generic variables $\{z^i\}$ on $\CM$ with the special coordinates, i.e. $\frac{\partial Z^i}{\partial z^j}=\delta^i_j$. In the main text, we always assume to be within a symplectic frame, and thus use $z^i$ to refer to the special coordinates directly. Here we have used $Z^i$ instead to denote them, in order to stress their properties and importance.

Using the special coordinates not only allows us to simplify drastically many equations, but it also maintains the connection with the quantised U(1) directions indicated by the periods $X^I$. It is, in fact, in terms of these that one can obtain simple expressions for the Pauli couplings introduced in \eqref{Pauli}:
\begin{equation}
    \CP_{Ijk}=2\CI_{IJ}D_jf^J_k=2ie^{-K/2}f^J_jf^K_k\CF_{JKL}N^{LM}\CI_{MI}=2ie^{K/2}(X^0)^2\CF_{jkL}N^{LM}\CI_{MI}\,.
\end{equation}
Notice, though, how in this equation only the last equality comes from exploiting the fact that the indices $i,j$ refer to special coordinates, whereas the second combines the definition \eqref{Cijkdef} with \eqref{CintermsofF}. With this expression at hand, and comparing with previous ones for the curvature and the (0,3)-tensor $C_{ijk}$, it is easy to verify that  
\begin{equation}
  \CP_{Ikm} \bar{\CP}_{J\bar{l}{\bar{n}}} \CI^{IJ}=-2e^{2K}|X^0|^6 \CF_{ikm}\bar \CF_{\bar j\bar l\bar n}g^{i\bar j}\, ,
\end{equation}
as well as
\begin{equation}
    \lVert \CP \rVert^2_{\rm EFT} =  - \CI^{IJ} g^{k\bar{l}} g^{m\bar{n}} \CP_{Ikm} \bar{\CP}_{J\bar{l}\bar{n}}=2 |X^0|^6 e^{2K} g^{i\bar{j}}g^{k\bar{l}}g^{m\bar{n}}  \CF_{ikm} \bar{\CF}_{\bar{j}\bar{l}\bar{n}}\,,
\end{equation}
from where \eqref{curvPauli} and \eqref{IIBscalarP} follow.

\noindent We conclude this subsection by giving the explicit expressions for the charges with respect to the field strengths of the graviphoton and the $n_V$ U(1)'s orthogonal to it, corresponding to the states $\q^{\mathsf T}_{\mathsf{mag}}=(p^\mu,\, \re\CF_{A\mu}\,p^\mu)$ and $\q^{\mathsf T}_{\mathsf{el}}=(0,\, q_j)$, used extensively in section \ref{s:CYlimits}. These are, respectively,
\begin{subequations}
    \begin{equation}
    Z_\mathsf{mag}=-\frac{i}{2}e^{K/2}X^IN_{I\mu}p^\mu=-\frac{i}{2}\frac{e^{-K/2}}{\bar X^0}\bar k_{\mu}p^\mu\,,
    \end{equation}
    \begin{equation}
    Z_{\mathsf{mag},i}=D_iZ_\mathsf{mag}=-\frac{i}{2}f^I_iN_{I\mu}p^\mu=-\frac{i}{2}\frac{e^{-K/2}}{\bar X^0}g_{i\mu}p^\mu\,,
    \end{equation}
    \begin{equation}
    Z_\mathsf{el}=e^{K/2}X^jq_j\,,
    \end{equation}
    \begin{equation}
    Z_{\mathsf{el},i}=D_iZ_\mathsf{el}=f^j_iq_j\,.
    \end{equation}
\end{subequations}
To derive these, one should take into account that, in special coordinates, we have
\begin{equation}
    k_i=\partial_{Z^i}K=-X^0\partial_{X^i}\log(-(XN\bar X))=e^KX^0N_{iJ}\bar X^J\,.
\end{equation}
%


\subsubsection*{Considerations on rigid requirements}\label{sss:rigidconsiderations}

In this subsection, we study the behavior of some of the physical quantities introduced before whenever a subsector of the theory is effectively decoupling from gravity.

Let us start with the formal manipulation
\begin{equation}
\begin{aligned}
\mathcal{N}_{IJ} &\equiv \CR_{IJ} + i \CI_{IJ} = \bar{\CF}_{IJ} + 2i\, \frac{\im \CF_{IK}\im \CF_{JL} X^K X^L}{\im \CF_{MN} X^M X^N } \\ & = \CF_{IJ}-i N_{IJ}+i \frac{(NX)_I (NX)_J}{XNX}\\ &= \CF_{IJ}-i \left(N_{IJ}+(\bar{X}^0)^{-2}e^{-K} \frac{XN\bar{X}}{XNX} k_{\bar{I}} k_{\bar{J}}\right)\, ,
\label{ap:calNIJ}
\end{aligned}
\end{equation}
where we used that $e^{-K}=-XN\bar{X}$ and $(NX)_I=- (\bar{X}^0)^{-1} XN\bar{X}\, k_{\bar{I}}$. In this work, we are interested in limits---both at finite and infinite field space distance---which feature axionic shift symmetries at their endpoints. More precisely, this implies that the K\"ahler potential either depends just on $\re Z^i$ ($\zeta = 1$) or on $\im Z^i$ ($\zeta = -1$), where $Z^i$ denotes the special coordinate. Hence, in the following we assume that
\begin{equation}\label{eq:shiftsymm}
   \frac{\partial K}{\partial\,\mathrm{Re}\,Z^i} \simeq 0\quad \text{or}\quad  \frac{\p K}{{\p{\rm Im}\,Z^i}} \simeq 0\implies k_{\bar{i}} \simeq \zeta_{(i)} k_i\, ,
\end{equation}
and we have made explicit that the $\zeta=\pm1$ factor may depend on the special coordinate we decide to differentiate with respect to. 

Next, we restrict \eqref{ap:calNIJ} to the moduli space directions $I,J=i,j\neq0,$ and we assume that all such fields satisfy \eqref{eq:shiftsymm} with the same $\zeta=\zeta_{(i)}$ (as happens in all our examples). This yields 
\begin{equation}\label{eq:approximateN}
\begin{aligned}
    {\CN}_{ij} \simeq \CF_{ij}-i \left(N_{ij}+\zeta\frac{X^0}{\bar X^0}\frac{XN\bar{X}}{XNX}|X^0|^{-2}e^{-K} k_{i} k_{\bar{j}}\right)\, .
\end{aligned}
\end{equation}
From here, and exploiting the $(-,+,+,\dots)$ signature of $N_{IJ}$, one can easily show that
\begin{equation}
    |XN\bar X|\leq|XNX|
\end{equation}
such that upon imposing \eqref{cond} for the rigid directions $i,j=\mu, \nu,$ one recovers that $\CN_{\mu \nu} \simeq \bar{\CF}_{\mu \nu}$.

For completeness, let us specialise to the cases of interest analysed herein. In general, after imposing the existence of the (common) shift symmetry \eqref{eq:shiftsymm}, one can rewrite the quotient appearing in the last term of \eqref{eq:approximateN} as follows
\begin{equation}
    \frac{X^0 (XN\bar{X})}{\bar{X}^0(XNX)} = \frac{Z N\bar{Z}}{ZNZ} =\left[ \zeta-|X^0|^2e^{K} ((NZ)_0-\zeta (N \bar{Z})_0)\right]^{-1}=\frac{1}{\zeta+(\zeta k_0-k_{\bar{0}})}\, .
\end{equation}
If, in addition, we have that $Z^i \simeq \zeta Z^i$ holds for the leading entries of $Z^I$, one finds
\begin{equation}
    \zeta k_0-k_{\bar{0}}=|X^0|^{2} e^K \left[ (\zeta-1) N_{00}+ N_{0i} (\zeta \bar{z}^i-z^i)\right] \simeq |X^0|^{2} e^K  (\zeta-1) N_{00}\, ,
\end{equation}
and thus
\begin{equation}
   \frac{ZN\bar{Z}}{ZNZ} \simeq\frac{1}{\zeta+|X^0|^{2} e^K  (\zeta-1) N_{00}}\, .
\end{equation}
As one may easily check, substituting this into \eqref{eq:approximateN} above correctly reproduces all the (leading-order terms of the) gauge kinetic functions $\CN_{ij}$ presented in the main text.

\subsection{Changes of basis and pairing matrices}\label{ss:changebasis&pairing}

In this subsection, we comment on the possibility of having a non-canonical intersection matrix for our basis of three-cycles, something we have to deal with in section \ref{s:swpoints}.

In fact, a good amount of the relations written above relies on the choice of real three-cycles/forms \eqref{cyclesbasis}/\eqref{realformsbasis}. Let us consider a generic basis of three-forms $\gamma_I$ (equivalently three-cycles $C^I$), and expand $H^3(Y_3, \mathbb{R})$ in terms of these. Then, the components of any generic three-form $\Sigma$ are
\begin{eqn}
    \Pi_I = \int  \Sigma   \wedge \gamma_I  \, ,
\end{eqn}
whilst (minus) the intersection matrix of the cycles Poincaré dual to $\gamma_I$ is 
\begin{eqn}\label{eq:wedgeMatrixgammaI}
    Q_{IJ} = \int \gamma_I \wedge \gamma_J\, .
\end{eqn}
In the canonical case, $\gamma_I = \{\beta^J,\alpha_I\}$ and the matrix \eqref{eq:wedgeMatrixgammaI} thus reads
\begin{equation}
    Q=\begin{pmatrix}
0  & -\mathbb{I}  \\ \mathbb{I} & 0
\end{pmatrix}\, .
\end{equation}
One can easily check that the generic three-form $\Sigma$ can be expanded as
\begin{eqn}
    \Sigma = \Pi_I Q^{IJ} \gamma_J\, ,
\end{eqn}
where $Q^{IJ}$ denotes the inverse of $Q_{IJ}$, and in the canonical frame it has the form
\begin{eqn}
    Q^{-1} = \begin{pmatrix}
0  & \mathbb{I}  \\ -\mathbb{I} & 0
\end{pmatrix}\, .
\end{eqn}
As a result, the wedge product of two such three-forms may be computed as follows
\begin{eqn}\label{genericwedge}
    \int \Sigma \wedge \Gamma = - \Pi^{\mathsf{T}}_\Sigma  Q^{-1}  \Pi_\Gamma\, .
\end{eqn}
In the symplectic basis $\{\beta^J,\alpha_I\}$ defined above, we have that $Q^{-1}= \eta = -Q$, but if we perform a change of frame that is not symplectic, the difference between $Q$ and $Q^{-1}$ must be taken into account. Let us, for instance, consider a new basis of integer three-forms given by
\begin{eqn}
 \gamma_I' = M_I^J \gamma_I  \implies  \Pi' = M \Pi\, .
\end{eqn}
Since both $\Sigma$ and $\Sigma \wedge \Gamma$ need to remain invariant under the change of basis, we obtain the following transformation of the pairing matrix
\begin{eqn}
    Q \to Q' = M Q M^{\mathsf{T}}\, .
\end{eqn}
From these considerations, it follows that every quantity arising from a wedge product should remain invariant under any such change of basis, as per \eqref{wedgeprod} and \eqref{genericwedge}. In particular, in section \ref{s:swpoints} we were dealing with transformations that satisfy
\begin{equation}\label{quasisympl}
    M \eta M^{\mathsf{T}}=n\eta\, ,
\end{equation}
with $n\in\mathbb{Z}$. This means that the new intersection matrix can be identified as $Q'=-n\eta$ whereas its inverse (needed to compute the wedge product in question) is $Q'^{-1}=\frac1n\eta$. As an example, let us consider the matrix $N_{AB}$, which can be defined via
\begin{eqn}
    N_{IJ} = \int_{Y_3} \p_{X^I} \Omega \wedge \p_{\bar{X}^J} \bar{\Omega}\, .
\end{eqn}
After a transformation of the type \eqref{quasisympl}, this quantity will be computed as
\begin{eqn}
    N_{IJ} =  -\frac1n\p_{X_I}{\bf \Pi }^\mathsf{T} \eta\,\p_{\bar{X}^J} \bar{\bf \Pi }\, .
\end{eqn}
Another relevant quantity for us, is the mass of a BPS particle, see \eqref{massD3}. Hence, after a transformation of the type \eqref{quasisympl}, this will be computed as follows
\begin{eqn}
    m_\mathsf{q} =e^{K/2}|\langle\Omega,\mathsf{q}\rangle|= \frac1n e^{K/2} |{\bf \Pi}^{\mathsf{T}} \eta\, \mathsf{q}| \, \Mpl\, .
\end{eqn}
%


\section{Details of Seiberg--Witten limits}
\label{ap:SWp}

In this appendix, we supplement technical material for section \ref{s:swpoints}. In particular, we give detailed information about how the relevant physical quantities are computed as well as the order of the expansion to which each of them should be trusted.

\subsection{Material for $\mathbb{P}^{1,1,2,2,6}[12]$}

The solution of the Picard--Fuchs system gives the following set of functions:
\begin{equation} \label{eq:periodvectorPicardFuchs}
    {\bf \Pi}_{\rm PF} = \frac{1}{\pi} \begin{pmatrix}
    1 + \frac{5}{36} x^1 + \frac{295 (2+x^2)(x^1)^2}{7776}\\
    x^1 + \frac{77}{216}(2+x^2)(x^1)^2\\
    -\sqrt{x^1} \left( 1-\frac{x^2}{16} + \frac{23}{864} (16+3x^2)x^1 - \frac{15}{1024}(x^2)^2\right)\\
    \frac{i}{2} \varpi_2 \log \left(x^2 (x^1)^2 \right) + \frac{1}{\pi} \left( x^1+\frac{787(x^1)^2}{324} + \frac{325 x^2(x^1)^2}{648}\right)\\
   \frac{i}{2} \left( \varpi_1 \log \left(x^2 (x^1)^2\right) + \frac{1}{\pi} \left(5+ \frac{25 x^1}{36} + \frac{2597 (x^1)^2}{11664} + \frac{827 x^2 (x^1)^2}{23328} \right)\right)\\
   -i \left( \varpi_3 \left(\log(x^2) - 6 \log2 + 3\right) - \frac{\sqrt{x^1}}{\pi} \left( 4-\frac{x^2}{8} - \frac{47 (x^2)^2}{1024}\right)\right)
\end{pmatrix}  =: \begin{pmatrix}
    \varpi_1 \\
           \varpi_2 \\
           \varpi_3\\
           \varpi_4 \\
           \varpi_5\\
           \varpi_6
\end{pmatrix}\, ,
\end{equation}
displayed here up to order $\CO((x^i)^2)$. The monodromies in this basis are given by the matrices
\begin{align}\label{eq:SWidptmonodromies}
    \mathcal{T}_1=\begin{pmatrix}
        1 & 0 & 0 & 0 & 0 & 0\\
        0 & 1 & 0 & 0 & 0 & 0\\
        0 & 0 & -1 & 0 & 0 & 0\\
        0 & -2 & 0 & 1 & 0 & 0\\
        -2 & 0 & 0 & 0 & 1 & 0\\
        0 & 0 & 0 & 0 & 0 & -1
    \end{pmatrix}\, ,\qquad \mathcal{T}_2=\begin{pmatrix}
        1 & 0 & 0 & 0 & 0 & 0\\
        0 & 1 & 0 & 0 & 0 & 0\\
        0 & 0 & 1 & 0 & 0 & 0\\
        0 & -1 & 0 & 1 & 0 & 0\\
        -1 & 0 & 0 & 0 & 1 & 0\\
        0 & 0 & 2 & 0 & 0 & 1
    \end{pmatrix}\, ,
\end{align}
which are obtained upon performing the rotations $x^1 \to e^{2\pi i}x^1$ and $x^2 \to e^{2\pi i}x^2$, respectively. The total monodromy around the weak coupling point is obtained by encircling $x^1$ once as well as $x^2$ twice in the opposite direction \cite{Lee:2019wij}, namely
\begin{align}
    \mathcal{T}_{\infty}= \mathcal{T}_1 (\mathcal{T}_2)^{-2} = \begin{pmatrix}
        1 & 0 & 0 & 0 & 0 & 0\\
        0 & 1 & 0 & 0 & 0 & 0\\
        0 & 0 & -1 & 0 & 0 & 0\\
        0 & 0 & 0 & 1 & 0 & 0\\
        0 & 0 & 0 & 0 & 1 & 0\\
        0 & 0 & 4 & 0 & 0 & -1
    \end{pmatrix}\, .
\end{align}
As explained in section \ref{s:swpoints}, these entries of the period correspond to a generic complex basis for the $H^3(Y_3)$ cohomology. In order to align with an integrally quantised basis one performs the transformation \eqref{intquantised}, and obtain the periods displayed in \eqref{eq:periodsSW}. The special coordinates corresponding to this basis are 
\begin{equation}
\begin{aligned}\label{speccoordsSWbeforeY}
   z^1 &=i +2 i \mathcal{X}\sqrt{x^1}+\CO(x^1) \, ,\\
   z^2 &=\frac{1}{2 \pi i} \left(\log \left[(x^1)^2x^2  \right]+5-4\pi\mathcal{X}\xi_4\right)+\frac{\mathcal{X}}{2 \pi i }\sqrt{x^1}\left(\log \left(x^1 \right)+3\log(2)-1-2\pi\mathcal{X}\xi_4\right)+\CO(x^1)\, ,
\end{aligned}
\end{equation}
and since they can be inverted as
\begin{equation}
\begin{aligned}
   x^1 &=\frac{1}{2 i \mathcal{X}}(z^1 -i)+\CO((z^1 -i)^2) \, ,\\
   \log(x^2) &= 2\pi i\, z^2-4\log(z^1 -i)-5+4\pi\mathcal{X}\xi_4+4\log(2\mathcal{X}) +2\pi i+\CO(z^1 -i)\, ,
\end{aligned}
\end{equation}
we can derive the associated prepotential
\begin{equation}
\begin{aligned}\label{prepSWbeforeY}
     F=\CF/(X^0)^2=&  2z^2-(2iz^2+A_1)(z^1-i)\\
     &- \left(z^2+\frac{2i}{\pi}\log(z^1-i)+iA_2+1\right)(z^1-i)^2+\CO((z^1 -i)^3)\,,
\end{aligned}
\end{equation}
where we have introduced
\begin{equation}
    A_1=2\mathcal{X}\xi_3+4\mathcal{X}\xi_4\;,\quad A_2=-\mathcal{X}\xi_3-4  \mathcal{X}\xi_4-2- \frac1\pi \log (\mathcal{X}^2/2)\, .
\end{equation}
Notice how the expansions are performed in powers of $x^1$ and $x^2$ or $z^1-i$ and $\frac{1}{(z^1-i)^4}e^{2\pi iz^2}$, depending on whether the function is expressed in terms of the original or the special coordinates. To ease the notation we only specified the order in $x^1$ and $z^1-i$, since this turns out to be the crucial one for the invertibility of the special coordinates \eqref{speccoordsSWbeforeY}.

The prepotential \eqref{prepSWbeforeY} can be used to compute all the couplings of the theory as functions of the special coordinates. It is written to the lowest order in $z^1-i$ that is needed to recover $\CF_0,\CF_1,\CF_2$ up to $\CO(x^1,x^2)$, as they were given in \eqref{eq:periodsSW} (one can easily check this by differentiating with respect to $z^1,z^2$ and then plugging \eqref{speccoordsSWbeforeY} in the result). To find the relevant aforementioned quantities of the 4d $\CN=2$ supergravity theory, one first computes 
\small
\begin{subequations}
\begin{equation}
\begin{aligned}
    {\CF_{IJ}} &= \partial_I\partial_J\CF\\
    &=\left(
\begin{array}{ccc}
 2 {z^2}+\frac{4 i }{\pi }\log \left({z^1}-i\right)+2 i A_1+2 i A_2+2+\frac{6 i}{\pi } & 2 i {z^2}-\frac{4 }{\pi }\log \left({z^1}-i\right)-A_1-2 A_2+2 i-\frac{6}{\pi } & 0 \\
 2 i {z^2}-\frac{4 }{\pi }\log \left({z^1}-i\right)-A_1-2 A_2+2 i-\frac{6}{\pi } & -2 {z^2}-\frac{4 i }{\pi }\log \left({z^1}-i\right)-2 i A_2-2-\frac{6 i}{\pi } & -2 i \\
 0 & -2 i & 0 \\
\end{array}
\right)\,,
\end{aligned}
\end{equation}
\begin{equation}
\begin{aligned}
  {\Theta_{IJ}}&=2\re\CF_{IJ}\\
     &= 
    \left(
\begin{array}{ccc}
 4\re {z^2}-\frac{8 }{\pi }(\arg(z^1-i)-\pi/2) & -4\im {z^2}-\frac{8 }{\pi }\log \left|{z^1}-i\right|-2A_1-4 A_2-\frac{12}{\pi } & 0 \\ -4\im {z^2}-\frac{8 }{\pi }\log \left|{z^1}-i\right|-2A_1-4 A_2-\frac{12}{\pi } & -4\re {z^2}+\frac{8 }{\pi }(\arg(z^1-i)-\pi/2) & 0 \\
 0 & 0 & 0 \\
\end{array}
\right)\,,
\end{aligned}
\end{equation}
\begin{equation}
\begin{aligned}
  {N_{IJ}}&=2\im\CF_{IJ}\\
     &= 
    \left(
\begin{array}{ccc}
 4\im {z^2}+\frac{8 }{\pi }\log \left|{z^1}-i\right|+4A_1+4 A_2+\frac{12}{\pi } & 4\re {z^2}-\frac{8}{\pi }(\arg(z^1-i)-\pi/2) & 0 \\ 4\re {z^2}-\frac{8}{\pi }(\arg(z^1-i)-\pi/2) & -4\im {z^2}-\frac{8 }{\pi }\log \left|{z^1}-i\right|-4 A_2-\frac{12}{\pi } & -4 \\
 0 & -4 & 0 \\
\end{array}
\right)\,,
\end{aligned}
\end{equation}
\end{subequations}
\normalsize
written here up to subleading $\CO(|z^1-i|)$ terms. From these, one recovers the complex matrix $\CN_{IJ}=\CR_{IJ}+i\CI_{IJ}$, even though the field dependence of its real and imaginary parts are much more complicated than the `rigid counterparts' $\Theta_{IJ}$ and $N_{IJ}$. In order to make them easier to read, in what follows we display only the leading polynomial term in $\phi$, where this parameter has been defined through the field-trajectory $(\log|z^1-i|,\im z^2)\to\phi(\log|z^1-i|,\im z^2)$ (equivalent to $(\log|x^1|,\log|x^2|)\to\phi(\log|x^1|,\log|x^2|)$). The leading order pieces of the real and imaginary parts of the complex matrix $\CN_{IJ}$ thus read 
\begin{subequations}
\begin{equation}
     {\CR_{IJ}} = 
    \left(
\begin{array}{ccc}
 \CR_{I0} & \CR_{I1}&\CR_{I2}
\end{array}
\right)\sim
    \left(
\begin{array}{ccc}
 \text{const.} & \text{const.}&\text{const.}\\
  \text{const.} & \text{const.}&1/\phi^{2}\\
   \text{const.} & 1/\phi^{2}&1/\phi^{2}
\end{array}
\right)\, ,
\end{equation}\small
\begin{equation}
\begin{aligned}
     {\CR_{I0}} =&-\frac{  1}{\left( {\im{z^2}}+\frac1\pi\log \left|z^1-i\right|\right){}^2} \times\\&\times
    \left(
\begin{array}{c}
(\arg(z^1-i)/\pi-1/2) ({\im{z^2}})^2+\frac1\pi{\re{z^2}} \left( {\im{z^2}}+\frac1\pi\log \left|z^1-i\right|\right)\log \left|z^1-i\right|^2 \\- \frac{A_1}{2}{\left( {\im{z^2}}+\frac1\pi\log \left|z^1-i\right|\right)} {\im{z^2}}  \\-{\left({\im{z^2}}+\frac1\pi\log \left|z^1-i\right|\right)}\left({\im{z^2}}+\frac2\pi\log \left|z^1-i\right|\right) 
\end{array}
\right)\, ,\\
     {\CR_{I1}} = &
    \left(
\begin{array}{c}
 \frac{A_1}{2}\frac{  1}{ {\im{z^2}}+\frac1\pi\log \left|z^1-i\right|} {\im{z^2}}  \\  -2\re {z^2}\\
 - \frac{A_1}{2\pi}\frac{  1}{\left( {\im{z^2}}+\frac1\pi\log \left|z^1-i\right|\right){}^2}(\arg(z^1-i)-\pi/2)
\end{array}
\right)\, ,\\
     {\CR_{I2}} = &
    \left(
\begin{array}{c}
 \frac{1 }{{\im{z^2}}+\frac1\pi\log \left|z^1-i\right|}\left({\im{z^2}}+\frac2\pi\log \left|z^1-i\right|\right)  \\- \frac{A_1}{2\pi}\frac{  1}{\left( {\im{z^2}}+\frac1\pi\log \left|z^1-i\right|\right){}^2}(\arg(z^1-i)-\pi/2)\\  \frac{1}{\pi}\frac{  1}{\left( {\im{z^2}}+\frac1\pi\log \left|z^1-i\right|\right){}^2}(\arg(z^1-i)-\pi/2) 
\end{array}
\right)\, ,
\end{aligned}
\end{equation}\normalsize
\begin{equation}
     {\CI_{IJ}} = 
    \left(
\begin{array}{ccc}
 \CI_{I0} & \CI_{I1}&\CI_{I2}
\end{array}
\right)\sim
    \left(
\begin{array}{ccc}
 \phi & 1/\phi&1/\phi\\
  1/\phi & \phi&1/\phi\\
   1/\phi & 1/\phi&1/\phi
\end{array}
\right)\, ,
\end{equation}\small
\begin{equation}
\begin{aligned}
     {\CI_{I0}} =& 
    \left(
\begin{array}{c}
 -\frac{1 }{{\im{z^2}}+\frac1\pi\log \left|z^1-i\right|}{\im{z^2}}\left({\im{z^2}}+\frac2\pi\log \left|z^1-i\right|\right)  \\ \frac{A_1}{2\pi}\frac{  1}{\left( {\im{z^2}}+\frac1\pi\log \left|z^1-i\right|\right){}^2} \left((\arg(z^1-i)/\pi-1/2) {\im{z^2}}-{\re{z^2}} \left( {\im{z^2}}+\frac1\pi\log \left|z^1-i\right|\right)\right)  \\\frac{1}{\left({\im{z^2}}+\frac1\pi\log \left|z^1-i\right|\right)^2}  \left({\re{z^2}} \left({\im{z^2}}+\frac1\pi\log \left|z^1-i\right|\right)-(\arg(z^1-i)/\pi-1/2) {\im{z^2}}\right) 
\end{array}
\right)\, ,\\
     {\CI_{I1}} =& 
    \left(
\begin{array}{c}
  \frac{A_1}{2\pi}\frac{  1}{\left( {\im{z^2}}+\frac1\pi\log \left|z^1-i\right|\right){}^2} \left((\arg(z^1-i)/\pi-1/2) {\im{z^2}}-{\re{z^2}} \left( {\im{z^2}}+\frac1\pi\log \left|z^1-i\right|\right)\right)  \\  -2\im {z^2}\\
  \frac{1}{{\im{z^2}}+\frac1\pi\log \left|z^1-i\right|}   \frac{A_1}2 \\
\end{array}
\right)\, ,\\
     {\CI_{I2}} = &
    \left(
\begin{array}{c}
 \frac{1}{\left({\im{z^2}}+\frac1\pi\log \left|z^1-i\right|\right)^2}  \left({\re{z^2}} \left({\im{z^2}}+\frac1\pi\log \left|z^1-i\right|\right)-(\arg(z^1-i)/\pi-1/2) {\im{z^2}}\right) \\\frac{ {A_1}}{2}\frac{1}{{\im{z^2}}+\frac1\pi\log \left|z^1-i\right|} \\ -\frac{1}{{\im{z^2}}+\frac1\pi\log \left|z^1-i\right|} \\
\end{array}
\right)\, .
\end{aligned}
\end{equation}\normalsize
\end{subequations}
The cubic derivatives of the prepotential are
\begin{equation}
\begin{aligned}
    X^0\CF_{000}&=\frac{4}{\pi(z^1-i)} -\frac3\pi(2\pi z^2+4i)+\frac{6i}\pi(2\pi z^2+2i)(z^1-i)+\frac{2}\pi(3\pi z^2+2i)(z^1-i)^2\, ,\\ X^0\CF_{001}&=\frac{4i}{\pi(z^1-i)}-4i+\frac8\pi -\frac2\pi(2\pi z^2+2i)(z^1-i)\, ,\\
    X^0\CF_{002}&=2-4i(z^1-i)-2(z^1-1)^2\, ,\\
       X^0\CF_{011}&=-\frac{4}{\pi(z^1-i)}+2z^2+\frac{4i}{\pi}\,,\quad X^0\CF_{012}=2i+2( z^1-i)\,,\quad X^0\CF_{022}=0\, ,\\
          X^0\CF_{111}&=-\frac{4i}{\pi(z^1-i)}\, ,\quad X^0\CF_{112}=-2\, ,\quad   X^0\CF_{122}=0\, ,\quad   X^0\CF_{222}=0\, ,
\end{aligned}
\end{equation}
where, once again, we emphasized the expansion in powers of $z^1-i$.

The K\"ahler potential reads
\begin{equation}
\begin{aligned}\label{KpotSW1}
    e^{-K}|X^0|^{-2}&=8\im z^2-4A_1+\CO(|z^1-i|)\\
    &=- \frac4{\pi}\log[|x^1|^2|x^2|]-4 \xi _3 \mathcal{X}+8 \xi _4 \mathcal{X}-\frac{19}{\pi }\\
    &-\frac{16\mathcal{X}}\pi\sqrt{|x^1|}\left( 2\log|x^1|+ \log|x^2|+\pi\xi _3 \mathcal{X}-2 \pi\xi _4 \mathcal{X}+5\right)\cos(\arg(\sqrt{x^1}))\\
    &+\CO(x^1, x^2)\, ,
\end{aligned}
\end{equation}
and the metric in special coordinates has the form
\begin{equation}
    g_{i\bar j}=\frac{1}{4\left(\im z^2\right)^2}\left(\begin{array}{cc}
        4\left(\im z^2+\frac1\pi\log|z^1-i|\right)^2 & \frac2\pi\log|z^1-1| \\
         \frac2\pi\log|z^1-1| & 1
    \end{array}\right)\, ,
\end{equation}
where for the latter we have only displayed the leading polynomial term in $\phi$, as before.

\subsubsection*{The transformation $\mathsf{Y}$}

In order to have a clearer picture of the physics of this compactification we performed the transformation \eqref{Y}, which satisfies \eqref{Y2}. After this transformation, the periods look like \eqref{swperiodY} and the monodromies about $x^1$ and $x^2$ are given by \eqref{eq:SWnewmonodromiesY}. As commented in section \ref{s:swpoints}, the transformation \eqref{Y} is not a symplectic transformation. This operation corresponds to a change in the basis of 3-cycles we choose for computing the period, and as such changes the intersection matrix that we use to calculate wedge products. Its net effect is not very drastic though, since it transforms the intersection matrix as $\eta\to 2\eta$ and its inverse as $\eta^{-1}\to \frac{1}{2}\eta^{-1}$. For example, the K\"ahler potential will now be computed as follows
\begin{equation}
    e^{-K}=i\langle\Omega,\bar\Omega\rangle=\frac{i}2{\bf \Pi}'^{\rm  T}_{\rm SW_1}\eta^{-1}\,\bar{{\bf \Pi}}'_{\rm SW_1}\, ,
\end{equation}
where ${\bf \Pi}'_{\rm SW_1}$ is the period vector as in \eqref{swperiodY} and the K\"ahler potential is the same as \eqref{KpotSW1} when expressed in terms of $x^1$ and $x^2$.

However, after the transformation the special coordinates change and now they look like
\begin{equation}
\begin{aligned}\label{speccoordsSWafterY}
   z^1 &= -\frac{1}{2 \pi } \left(\log \left[(x^1)^2x^2  \right]+5-4\pi\mathcal{X}\xi_4\right)\, ,\\
   &-\frac{\mathcal{X}^2}{ \pi  }\left(\log \left[(x^1)^2x^2  \right]+3+\frac\pi{\mathcal{X}}\xi_2+\pi\mathcal{X}\xi_3-4\pi\mathcal{X}\xi_4\right){x^1}+\CO((x^1)^2)\, ,\\
   z^2 &=2 i \mathcal{X}\sqrt{x^1}+\CO((x^1)^{3/2})\, .
\end{aligned}
\end{equation}
Note that they can still be easily inverted
\begin{equation}
\begin{aligned}
   x^1 &= -\frac1{4\mathcal{X}^2}(z^2)^2+\CO((z^2)^4)\, ,\\
   \log(x^2) &= -2\pi z^1-4\log z^2-5+4\pi\mathcal{X}\xi_4+4\log(2\mathcal{X})+2\pi i+\CO((z^2)^2)\, ,
\end{aligned}
\end{equation}
from which we obtain the new prepotential
\begin{equation}\label{prepSWafterY}
     F=\CF/(X^0)^2= -iz^1+iA_1+\frac{i}\pi \left(\frac\pi2z^1+\log\frac{z^2}{A_2}\right)(z^2)^2+\CO((z^2)^3)\,,
\end{equation}
where we have introduced
\begin{equation}
    A_1=\oh(\xi_3+2\xi_4)\mathcal{X}\, ,\qquad A_2  =i \frac{\mathcal{X}}{\sqrt{2}} e^{1+\pi \mathcal{X}\xi_4}\, .
\end{equation}
The expansions in terms of the special coordinates are now performed in powers of $z^2$ and $\frac{1}{(z^1-i)^4}e^{2\pi iz^2}$. Again, to lighten the notation we only write the order in $x^1$ (equivalently $z^2$). 

As before, the prepotential \eqref{prepSWafterY} can be used to compute all the couplings of the theory as a function of the special coordinates. It has been written here including all the lowest orders in $z^2$ that are needed to recover $\CF_0,\CF_1,\CF_2$ up to $\CO(x^1,x^2)$, as they were shown in \eqref{swperiodY}. To find the resulting couplings of the 4d $\CN=2$ supergravity EFT, one first computes
\small
\begin{subequations}
\begin{equation}
\begin{aligned}
    {\CF_{IJ}} = \partial_I\partial_J\CF
    =\left(
\begin{array}{ccc}
 2iA_1 & -i & 0 \\
 -i& 0 & 0 \\
 0 & 0 & \frac{i}\pi(\pi z^1+3+2\log(z^2/A_2)) \\
\end{array}
\right)\,,
\end{aligned}
\end{equation}
\begin{equation}
\begin{aligned}
  {\Theta_{IJ}}=2\re\CF_{IJ}
     = 
    \left(
\begin{array}{ccc}
 0 & 0 & 0 \\ 0 & 0 & 0 \\
 0 & 0 & -\frac{1}\pi(\pi \im z^1+2(\arg(z^2)-\arg(A_2))) \\
\end{array}
\right)\,,
\end{aligned}
\end{equation}
\begin{equation}
\begin{aligned}
  {N_{IJ}}=2\im\CF_{IJ}
     = 
    \left(
\begin{array}{ccc}
 2A_1 & -1 & 0 \\ -1 & 0 & 0 \\
 0 & 0 & \frac{1}\pi(\pi \re z^1+3+2\log(|z^2/A_2|)) \\
\end{array}
\right)\,,
\end{aligned}
\end{equation}
\end{subequations}
\normalsize
written here up to subleading $\CO(|z^2|)$ terms. From these, one recovers the complex matrix $\CN_{IJ}=\CR_{IJ}+i\CI_{IJ}$. Since the transfromation $\mathsf{Y}$ simplified a lot the field dependence, we will to display its real and imaginary parts entirely. In any event, we will introduce $\phi$, this time as $(\re z^1,\log|z^2|)\to\phi(\re z^1,\log|z^2|)$ (yet equivalent to $(\log|x^1|,\log|x^2|)\to\phi(\log|x^1|,\log|x^2|)$), to have a glimpse of the behavior of these complicated functions. The real and imaginary parts of the complex matrix $\CN_{IJ}$ thus read 
\begin{subequations}
\begin{equation}
     {\CR_{IJ}} = 
    \left(
\begin{array}{ccc}
 \CR_{I0} & \CR_{I1}&\CR_{I2}
\end{array}
\right)\sim
    \left(
\begin{array}{ccc}
 \text{const.} & 1/\phi^2 & e^{-\phi}\\
  1/\phi^2  & 1/\phi^2&e^{-\phi}\\
   e^{-\phi} & e^{-\phi}& \text{const.}
\end{array}
\right)\, ,
\end{equation}\small
\begin{equation}
\begin{aligned}
     {\CR_{I0}} =&   \oh\frac{\im z^1}{({\im z^1})^2+({\re z^1}-{A_1})^2} \left(
\begin{array}{c}
{{\re z^1} ({\re z^1}-2 {A_1})}+({\im z^1})^2\\{A_1}   \\0 
\end{array}
\right)\, ,\\
     {\CR_{I1}} = &\oh\frac{\im z^1}{({\im z^1})^2+({\re z^1}-{A_1})^2}
    \left(
\begin{array}{c}
 {A_1}  \\ -1\\0
\end{array}
\right)\, ,\\
     {\CR_{I2}} = &\oh
    \left(
\begin{array}{c}
 0  \\0\\ -\im z^1-\frac2\pi(\arg(z^2)-\arg(A_2))
\end{array}
\right)\, ,
\end{aligned}
\end{equation}\normalsize
\begin{equation}
     {\CI_{IJ}} = 
    \left(
\begin{array}{ccc}
 \CI_{I0} & \CI_{I1}&\CI_{I2}
\end{array}
\right)\sim
    \left(
\begin{array}{ccc}
 \phi & 1/\phi&e^{-\phi}\\
  1/\phi & 1/\phi&e^{-\phi}\\
  e^{-\phi} & e^{-\phi}&\phi
\end{array}
\right)\, ,
\end{equation}\small
\begin{equation}
\begin{aligned}
     {\CI_{A0}} =& \oh\frac{{\re z^1}-{A_1}}{({\re z^1}-{A_1})^2+({\im z^1})^2}
    \left(
\begin{array}{c}
 -({\re z^1})^2+2 {A_1} {\re z^1}- 2 {A_1}^2-({\im z^1})^2  \\ {A_1} \\0
\end{array}
\right)\, ,\\
     {\CI_{A1}} =& \oh\frac{({\re z^1}-{A_1})}{({\re z^1}-{A_1})^2+({\im z^1})^2}
    \left(
\begin{array}{c}
  {A_1}   \\ -1\\0
\end{array}
\right)\, ,\\
     {\CI_{A2}} = &
    \oh\left(
\begin{array}{c}
 0 \\ 0 \\-\frac{1}\pi(\pi  {\re z^1}+3+2 \log (|z^2/A_2|))
\end{array}
\right)\, .
\end{aligned}
\end{equation}\normalsize
\end{subequations}
The cubic derivatives of the prepotential are
\begin{equation}
\begin{aligned}
    X^0\CF_{000}&=-\frac{i}{\pi}(3\pi z^1+2)(z^2)^2\,,\quad  X^0\CF_{001}=i(z^2)^2\,,\quad  X^0\CF_{002}=\frac{2i}{\pi}(\pi z^1+1)z^2\, ,\\
       X^0\CF_{011}&=0\,,\quad X^0\CF_{012}=-iz^2\,,\quad X^0\CF_{022}=-\frac{i}{\pi}(2\pi z^1+2)\, ,\\
          X^0\CF_{111}&=0\,,\quad X^0\CF_{112}=0\,,\quad   X^0\CF_{122}=i\,,\quad   X^0\CF_{222}=\frac{2i}{\pi z^2}\,.
\end{aligned}
\end{equation}
On the other hand, the volume in special coordinates reads, at leading order, as 
\begin{equation}
    e^{-K}|X^0|^2=2\left(\re z^1-A_1\right)+\CO(|z^2|)\,,
\end{equation}
whereas the metric 
\begin{equation}
    g_{i\bar j}=\frac{1}{2\left(\re z^1-A_1\right)}\left(\begin{array}{cc}
       \frac{1}{2\left(\re z^1-A_1\right)}  & 0 \\
         0 & \pi\re z^1+3 +2\log|z^2/A_2|
    \end{array}\right)+\CO(|z^2|)\;.
\end{equation}
The Pauli couplings $\CP_{Ijk} $ are
\begin{equation}
\begin{aligned}
    \CP_{011}&=0\,,\quad  \CP_{111}=0\,,\quad  \CP_{211}=0 \,,\quad \CP_{012}=0\,,\quad  \CP_{112}=0\,,\quad  \CP_{212}=\frac{1}{\sqrt{\re z^1-A_1}}\, ,\\
      \CP_{022}&=-\frac{1}{\pi\sqrt{\re z^1-A_1}}\frac{(\pi(\re z^1-A_1)+1)\bar z^1+(\bar z^1-2A_1)|z^2|^2/(z^2)^2}{\bar z^1-A_1}\, ,\\
          \CP_{122}&=\frac{1}{\pi\sqrt{\re z^1-A_1}}\frac{\pi(\re z^1-A_1)+1-|z^2|^2/(z^2)^2}{\bar z^1-A_1}\, ,\\
          \CP_{222}&=\frac{2}{\pi\sqrt{\re z^1-A_1}}\frac{1}{z^2}\,,
\end{aligned}
\end{equation}
and the normalised sums of interest 
\begin{equation}
\begin{aligned}
    ||\CP||^2_{\rm cRFT}&=\frac{16\pi(\re z^1-A_1)}{\left(\pi  {\re z^1}+3+2 \log (|z^2/A_2|)\right)|z^2|^2}+\CO(|z^2|)\, ,\\
    ||\CP||^2_{\rm grav}&=\CO(|z^2|)\, ,\\
    ||\CP||^2_{\rm mixed}&\simeq\frac{96\pi^2(\re z^1)^2}{\left(\pi  {\re z^1}+2 \log (|z^2|)\right)^2}+\CO(|z^2|)\,,
\end{aligned}
\end{equation}
where in the last equation we once again only displayed the leading term in $\phi$.

The scalar curvature, at leading order, reads
\begin{equation}
\begin{aligned}
    R&=\frac{16\pi(\re z^1-A_1)}{\left(\pi  {\re z^1}+3+2 \log (|z^2/A_2|)\right)^3|z^2|^2}+\CO(|z^2|^0)\\
    &\simeq\frac{4}{\mathcal{X}^2}\frac{\log[|x^1|^2 |x^2|]}{\left(\log |x^2|\right)^3|x^1|}+\CO(|x^1|^0)\,.
\end{aligned}
\end{equation}

\subsection{Material for $\mathbb{P}^{1,1,2,2,2}[8]$}

The original period vector given in \cite{Billo:1998yr} reads
\begin{equation}\label{periovec11222}
    \mathbf{\Pi}=\mathcal{Y}\left(\begin{array}{c}
        \tilde\epsilon^{1/2}I_2 \\
        \tilde\epsilon^{1/2}I_1 \\
        -\frac{i}{\sqrt{2}}+i\frac{13}{8\sqrt{2}}\tilde\epsilon\tilde u \\
        -\frac{1}{\sqrt{2}\pi}\log\tilde\epsilon+\frac{k_1}{\mathcal{Y}}+\frac{13}{8\sqrt{2}\pi}\tilde\epsilon\log\tilde\epsilon\, \tilde u \\
        \frac{1}{2}+\frac{19}{16}\tilde\epsilon\tilde u  \\
        \frac{1}{2\pi i}\log\tilde\epsilon+\frac{k_2}{\mathcal{Y}}+\frac{19}{16\pi i}\tilde\epsilon\log\tilde\epsilon\, \tilde u  
    \end{array}\right)\, +\, \mathcal{O}(\tilde\epsilon^{3/2})\, ,
\end{equation}
where $\mathcal{Y},k_1,k_2,l_1,l_2$ are numerical constants related by
\begin{equation}
    l_2=2 \sqrt{2} \mathcal{Y} +\frac{4 \mathcal{Y} }{\pi }+\frac{13 k_1}{8 \sqrt{2}}+\frac{19 k_2}{8}+\frac{l_1}{\sqrt{2}}\,.
\end{equation}
As explained in section \ref{s:swpoints}, this period was obtained by direct integration of the holomorphic three-form $\Omega$ on a basis of integrally-quantised three-cycles. Since to this basis corresponded the non-canonical intersection matrix \eqref{noncanintmat11222}, we performed the change of basis \eqref{X} and obtained the periods \eqref{periovec11222can}. This is given in terms of the coordinates $\tilde\eps,\tilde u$, but it will closely resemble the periods of the previous example if we plug \eqref{map11222}. In fact, as described in \cite{Billo:1998yr}, this change of coordinates is quite natural for compactifications over a large class of CY's to which both $\mathbb{P}^{1,1,2,2,2}[8]$ and $\mathbb{P}^{1,1,2,2,6}[12]$ belong---namely $\mathbb{P}^{1,1,2k_3,2k_4,2k_5}[2(1+k_3+k_4+k_5)]$. The coordinates $\tilde\eps,\tilde u$ have the advantage of controlling the rigid limit just by sending $\tilde\eps\to0$, while leaving $\tilde u$ free. In practice, this means that we could choose to study the deep-quantum regime of the rigid SW theory, by expanding $I_1,I_2$ in \eqref{periovec11222can} for $\tilde u\approx1$. Despite this, we decide to keep working in the semi-classical region of the theory, where $\tilde u\gg1$, in order to make contact with the previous example. In this regime, and after the transformation $\mathsf{X}$, the monodromies about $\tilde u =0$ and $\tilde \eps =0$ are shown in \eqref{mon11222}, whereas the special coordinates read
\begin{equation}
\begin{aligned}
     z^1=&-\frac{\sqrt{2} }{\pi  }\log \tilde\epsilon +\frac{ 2\sqrt{2}   k_2}{ \mathcal{Y} }-\frac{8}{\sqrt2\pi}\tilde u\left(-  \log (\tilde\epsilon )+\frac{13 \pi  k_2}{\mathcal{Y}}+\frac{8 \pi  l_2}{\mathcal{Y}}\right)\tilde\eps+\mathcal{O}(\tilde\epsilon^2)\, ,\\
    z^2=&2i\sqrt{\tilde u\tilde\epsilon}+\mathcal{O}(\tilde\epsilon^{3/2})\, .
\end{aligned}
\end{equation}
As before, the relevant expansion is performed in powers of $\tilde\eps$, since on this depends the invertibility of the special coordinates. The expansion in $\tilde u^{-1}$ is always cut up to order $\tilde u^0$ (included), and the inverse asymptotic expansions read
\begin{equation}
\begin{aligned}
     \log\tilde\epsilon&=-\frac{\pi}{\sqrt2}\left(z^1-\frac{2\sqrt2 k_2}{\mathcal{Y}}\right)-\frac{\pi}{\sqrt2}\left(z^1-\frac{2\sqrt2 k_2}{\mathcal{Y}}+\frac{13 k_2+8l_2}{8\sqrt2\mathcal{Y}}\right)(z^2)^2+\CO((z^2)^4)\, ,\\
    \log\tilde u&=\frac{\pi}{\sqrt2}z^1+2\log z^2-\log4-\frac{2\pi k_2}{\mathcal{Y}}-i\pi+\mathcal{O}((z^2)^2)\, .
\end{aligned}
\end{equation}
Through these, we can compute the prepotential 
\begin{equation}
    F=\mathcal{F}/(X^0)^2= -\frac{i}{\sqrt2} z^1 + i A_1 + \frac{2i}{\pi} \left(\frac\pi{2\sqrt2} z^1+\log (z^2/A_2)\right)(z^2)^2 + \CO((z^2)^3)\, ,
\end{equation}
where we have defined
\begin{equation}
    A_1=\frac{k_2}{\mathcal{Y}}-\frac{k_1}{\sqrt2\mathcal{Y}}-\frac{1}{\sqrt2}\,,\qquad A_2=i\frac1{\sqrt2}e^{\tfrac32+\frac{\pi k_2}{\mathcal{Y}}}\, ,
\end{equation}
and use it to compute the couplings of the theory in terms of the special coordinates. In order to do this, we have to take into account that after the transformation $\mathsf{X}$ the intersection matrix has transformed. In this example we will have $\eta\to4\eta$ and $\eta^{-1}\to\frac14\eta^{-1}$, and thus 
\small
\begin{subequations}
\begin{equation}
    {\CF_{IJ}} = \partial_I\partial_J\CF\\
    =\left(
\begin{array}{ccc}
 2iA_1 & -i/\sqrt2 & 0 \\
 -i/\sqrt2& 0 & 0 \\
 0 & 0 & \frac{2i}\pi(\frac\pi{\sqrt2} z^1+3+2\log(z^2/A_2)) \\
\end{array}
\right)\,,
\end{equation}
\begin{equation}
  {\Theta_{IJ}}=2\re\CF_{IJ}\\
     = 
    \left(
\begin{array}{ccc}
 0 & 0 & 0 \\ 0 & 0 & 0 \\
 0 & 0 & -\frac{1}\pi(\frac\pi{\sqrt2} \im z^1+2(\arg(z^2)-\arg(A_2))) \\
\end{array}
\right)\,,
\end{equation}
\begin{equation}
  {N_{IJ}}=2\im\CF_{IJ}\\
     = 
    \left(
\begin{array}{ccc}
 A_1 & -\frac1{2\sqrt2} & 0 \\ -\frac1{2\sqrt2} & 0 & 0 \\
 0 & 0 & \frac{1}\pi(\frac\pi{\sqrt2} \re z^1+3+2\log(|z^2/A_2|)) \\
\end{array}
\right)\,,
\end{equation}
\end{subequations}
\normalsize
From these, one recovers the complex matrix $\CN_{IJ}=\CR_{IJ}+i\CI_{IJ}$, whose real and imaginary parts read
\begin{subequations}\label{eq:RandISW2}
\begin{equation}
     {\CR_{IJ}} = 
    \left(
\begin{array}{ccc}
 \CR_{I0} & \CR_{I1}&\CR_{I2}
\end{array}
\right)\sim
    \left(
\begin{array}{ccc}
 \text{const.} & 1/\phi^2 & e^{-\phi}\\
  1/\phi^2  & 1/\phi^2&e^{-\phi}\\
   e^{-\phi} & e^{-\phi}& \text{const.}
\end{array}
\right)\, ,
\end{equation}\small
\begin{equation}
\begin{aligned}
     {\CR_{I0}} &= \frac{1}{4}\frac{\im z^1}{2({\im z^1})^2+(\sqrt2{\re z^1}-2{A_1})^2} \left(
\begin{array}{c}
{{\re z^1} (\sqrt2{\re z^1}-4 {A_1})}+\sqrt2({\im z^1})^2\\2{A_1}   \\0 
\end{array}
\right)\, ,\\
     {\CR_{I1}} &= \frac{1}{4}\frac{\im z^1}{2({\im z^1})^2+(\sqrt2{\re z^1}-2{A_1})^2}
    \left(
\begin{array}{c}
 2{A_1}  \\ -\sqrt2\\0
\end{array}
\right)\, ,\\
     {\CR_{I2}} &= \oh
    \left(
\begin{array}{c}
 0  \\0\\ -\sqrt2\im z^1-\frac4\pi(\arg(z^2)-\arg(A_2))
\end{array}
\right)\, ,
\end{aligned}
\end{equation}\normalsize
\begin{equation}
     {\CI_{IJ}} = 
    \left(
\begin{array}{ccc}
 \CI_{I0} & \CI_{I1}&\CI_{I2}
\end{array}
\right)\sim
    \left(
\begin{array}{ccc}
 \phi & 1/\phi&e^{-\phi}\\
  1/\phi & 1/\phi&e^{-\phi}\\
  e^{-\phi} & e^{-\phi}&\phi
\end{array}
\right)\, ,
\end{equation}\small
\begin{equation}
\begin{aligned}
     {\CI_{I0}} &= \frac14\frac{{\re z^1}-\sqrt2{A_1}}{({\re z^1}-\sqrt2{A_1})^2+({\im z^1})^2}
    \left(
\begin{array}{c}
 -({\re z^1})^2+2 {A_1} {\re z^1}- 2 {A_1}^2-({\im z^1})^2  \\ {A_1} \\0
\end{array}
\right)\, ,\\
     {\CI_{I1}} &= \frac14\frac{{\re z^1}-\sqrt2{A_1}}{({\re z^1}-\sqrt2{A_1})^2+({\im z^1})^2}
    \left(
\begin{array}{c}
  {A_1}   \\ -\frac1{\sqrt2}\\0
\end{array}
\right)\, ,\\
     {\CI_{I2}} &=
    \oh\left(
\begin{array}{c}
 0 \\ 0 \\-\frac{2}\pi(\frac\pi{\sqrt2}  {\re z^1}+3+2 \log (|z^2/A_2|))
\end{array}
\right)\, ,
\end{aligned}
\end{equation}
\end{subequations}
where, as before, we displayed the whole polynomial dependence in $\phi$, with $\phi$ defined by the trajectory $(\re z^1,\log|z^2|)\to\phi(\re z^1,\log|z^2|)$ (equivalent to $(\log|\tilde\eps|,\log|\tilde u|)\to\phi(\log|\tilde\eps|,\log|\tilde u|)$).
The cubic derivatives of the prepotential are
\begin{equation}
\begin{aligned}
    X^0\CF_{000}&=-\frac{i}{\pi}(3\sqrt2\pi z^1+4)(z^2)^2\, ,\quad  X^0\CF_{001}=i\sqrt2(z^2)^2\, ,\quad  X^0\CF_{002}=\frac{2i}{\pi}(\sqrt2\pi z^1+2)z^2\, ,\\
       X^0\CF_{011}&=0\,,\quad X^0\CF_{012}=-i\sqrt2z^2\, ,\quad X^0\CF_{022}=-\frac{i}{\pi}(\sqrt2\pi z^1+4)\, ,\\
          X^0\CF_{111}&=0\,,\quad X^0\CF_{112}=0\, ,\quad   X^0\CF_{122}=i\sqrt2\,,\quad  X^0\CF_{222}=\frac{4i}{\pi z^2}\, .
\end{aligned}
\end{equation}
The volume instead reads
\begin{equation}
\begin{aligned}
    e^{-K}|X^0|^{-2}&=\frac1{\sqrt2}\left(\re z^1-\sqrt2A_1\right)+\CO(|z^2|^2)\\
    &= -\frac{ \log ({\tilde\epsilon})}{\pi }+\frac1{\sqrt{2}}+\frac{\sqrt{2} k_1+2 k_2}{2\mathcal{Y}}-\frac{4 }{ \pi}  \tilde u (\log (\tilde u)-2+3\log (2)){\tilde\epsilon}\\&+\frac{ 1  }{8 \pi }\tilde u\left(-6 \log ({\tilde\epsilon})+19 \pi  \sqrt{2}+32+2\frac{19 \pi  k_2+4 \pi  \sqrt{2} l_1}{\mathcal{Y} }\right){\tilde\epsilon}\cos (\arg(\tilde\epsilon)+\arg(\tilde u))\\&+\CO(\tilde\eps^2)\,,
\end{aligned}
\end{equation}
and the metric is given by
\begin{equation}
    g_{i\bar j}=\frac{\sqrt2}{\left(\re z^1-\sqrt2A_1\right)}\left(\begin{array}{cc}
       \frac{1}{4\left(\re z^1-\sqrt2A_1\right)}  & 0 \\
         0 & \frac\pi{\sqrt2}\re z^1+3 +2\log|z^2/A_2|
    \end{array}\right)+\CO(|z^2|)\;.
\end{equation}
Similarly, the Pauli couplings $\CP_{Ijk} $ are
\begin{equation}
\begin{aligned}
    \CP_{011}&=0\;,\quad  \CP_{111}=0\, ,\quad  \CP_{211}=0 \, ,\quad \CP_{012}=0\, ,\quad  \CP_{112}=0\, ,\quad  \CP_{212}=\frac{1}{\sqrt{\re z^1-A_1}}\, ,\\
      \CP_{022}&=-\frac{1}{\pi\sqrt{\re z^1-A_1}}\frac{(\pi(\re z^1-A_1)+1)\bar z^1+(\bar z^1-2A_1)|z^2|^2/(z^2)^2}{\bar z^1-A_1}\, ,\\
          \CP_{122}&=\frac{1}{\pi\sqrt{\re z^1-A_1}}\frac{\pi(\re z^1-A_1)+1-|z^2|^2/(z^2)^2}{\bar z^1-A_1}\, ,\\
          \CP_{222}&=\frac{2}{\pi\sqrt{\re z^1-A_1}}\frac{1}{z^2}\,,
\end{aligned}
\end{equation}
and the normalised sums read
\begin{equation}
\begin{aligned}
    ||\CP||^2_{\rm cRFT}&=\,\frac{16\pi(\sqrt2\re z^1-2A_1)}{\left(\frac\pi{\sqrt2}  {\re z^1}+3+2 \log (|z^2/A_2|)\right)^3|z^2|^2}+\CO(|z^2|^0)\, ,\\
    ||\CP||^2_{\rm grav}&=\,\CO(|z^2|)\, ,\\
    ||\CP||^2_{\rm mixed}&\simeq\,\frac{24\pi^2(\re z^1)^2}{\left(\pi  {\re z^1}+2 \log (|z^2|)\right)^2}+\CO(|z^2|)\,,
\end{aligned}
\end{equation}
where in the last equation we again displayed only the leading term in $\phi$.

The scalar curvature, at leading order, yields
\begin{equation}
\begin{aligned}
    R &=\,\frac{16\pi(\sqrt2\re z^1-2A_1)}{\left(\frac\pi{\sqrt2}  {\re z^1}+3+2 \log (|z^2/A_2|)\right)^3|z^2|^2}+\CO(|z^2|^0)\\
    &\simeq\, \frac{8}{\left(\log |\tilde u|\right)^3|\tilde u|}\frac{\log|\tilde\eps|}{|\tilde\eps|}+\CO(|x^1|^0)\,.
\end{aligned}
\end{equation}
%



\bibliographystyle{JHEP2015}
\bibliography{papers}

@article{McAllister:2023vgy,
    author = "McAllister, Liam and Quevedo, Fernando",
    title = "{Moduli Stabilization in String Theory}",
    eprint = "2310.20559",
    archivePrefix = "arXiv",
    primaryClass = "hep-th",
    month = "10",
    year = "2023"
}

@book{Blumenhagen:2013fgp,
    author = {Blumenhagen, Ralph and L{\"u}st, Dieter and Theisen, Stefan},
    title = "{Basic concepts of string theory}",
    doi = "10.1007/978-3-642-29497-6",
    isbn = "978-3-642-29496-9",
    publisher = "Springer",
    address = "Heidelberg, Germany",
    series = "Theoretical and Mathematical Physics",
    year = "2013"
}

@article{Marchesano:2024gul,
    author = "Marchesano, Fernando and Shiu, Gary and Weigand, Timo",
    title = "{The Standard Model from String Theory: What Have We Learned?}",
    eprint = "2401.01939",
    archivePrefix = "arXiv",
    primaryClass = "hep-th",
    reportNumber = "IFT-UAM/CSIC-24-01, ZMP-HH/24-01",
    doi = "10.1146/annurev-nucl-102622-012235",
    journal = "Ann. Rev. Nucl. Part. Sci.",
    volume = "74",
    pages = "113--140",
    year = "2024"
}

@article{Akhond:2021xio,
    author = "Akhond, Mohammad and Arias-Tamargo, Guillermo and Mininno, Alessandro and Sun, Hao-Yu and Sun, Zhengdi and Wang, Yifan and Xu, Fengjun",
    title = "{The hitchhiker's guide to 4d $\mathcal{N}=2$ superconformal field theories}",
    eprint = "2112.14764",
    archivePrefix = "arXiv",
    primaryClass = "hep-th",
    reportNumber = "IFT-UAM/CSIC-21-151, ZMP-HH/21-28",
    doi = "10.21468/SciPostPhysLectNotes.64",
    journal = "SciPost Phys. Lect. Notes",
    volume = "64",
    pages = "1",
    year = "2022"
}

@article{Grimm:2019wtx,
    author = "Grimm, Thomas W. and Van De Heisteeg, Damian",
    title = "{Infinite Distances and the Axion Weak Gravity Conjecture}",
    eprint = "1905.00901",
    archivePrefix = "arXiv",
    primaryClass = "hep-th",
    doi = "10.1007/JHEP03(2020)020",
    journal = "JHEP",
    volume = "03",
    pages = "020",
    year = "2020"
}

@article{Marchesano:2022avb,
    author = "Marchesano, Fernando and Wiesner, Max",
    title = "{4d strings at strong coupling}",
    eprint = "2202.10466",
    archivePrefix = "arXiv",
    primaryClass = "hep-th",
    reportNumber = "IFT-UAM/CSIC-22-13",
    doi = "10.1007/JHEP08(2022)004",
    journal = "JHEP",
    volume = "08",
    pages = "004",
    year = "2022"
}

@article{Delgado:2024skw,
    author = "Delgado, Matilda and van de Heisteeg, Damian and Raman, Sanjay and Torres, Ethan and Vafa, Cumrun and Xu, Kai",
    title = "{Finiteness and the emergence of dualities}",
    eprint = "2412.03640",
    archivePrefix = "arXiv",
    primaryClass = "hep-th",
    reportNumber = "MPP-2024-224, CERN-TH-2024-204",
    doi = "10.21468/SciPostPhys.19.2.047",
    journal = "SciPost Phys.",
    volume = "19",
    number = "2",
    pages = "047",
    year = "2025"
}

@article{Gunara:2013rca,
    author = "Gunara, Bobby Eka and Louis, Jan and Smyth, Paul and Tripodi, Luca and Valandro, Roberto",
    title = "{The rigid limit of $N=2$ supergravity}",
    eprint = "1305.1903",
    archivePrefix = "arXiv",
    primaryClass = "hep-th",
    doi = "10.1088/0264-9381/30/19/195014",
    journal = "Class. Quant. Grav.",
    volume = "30",
    pages = "195014",
    year = "2013"
}

@article{Hassfeld:2025uoy,
    author = "Hassfeld, Bjoern and Monnee, Jeroen and Weigand, Timo and Wiesner, Max",
    title = "{Emergent Strings in Type IIB Calabi--Yau Compactifications}",
    eprint = "2504.01066",
    archivePrefix = "arXiv",
    primaryClass = "hep-th",
    month = "4",
    year = "2025"
}

@article{Martucci:2024trp,
    author = "Martucci, Luca and Risso, Nicol{\`o} and Valenti, Alessandro and Vecchi, Luca",
    title = "{Wormholes in the axiverse, and the species scale}",
    eprint = "2404.14489",
    archivePrefix = "arXiv",
    primaryClass = "hep-th",
    doi = "10.1007/JHEP07(2024)240",
    journal = "JHEP",
    volume = "07",
    pages = "240",
    year = "2024"
}

@article{Monnee:2025msf,
    author = "Monnee, Jeroen and Weigand, Timo and Wiesner, Max",
    title = "{K-Points and Type IIB/Heterotic Duality with NS5-Branes}",
    eprint = "2510.02435",
    archivePrefix = "arXiv",
    primaryClass = "hep-th",
    reportNumber = "ZMP-HH-25/18",
    month = "10",
    year = "2025"
}

@article{Klemm:1996bj,
    author = "Klemm, Albrecht and Lerche, Wolfgang and Mayr, Peter and Vafa, Cumrun and Warner, Nicholas P.",
    title = "{Selfdual strings and N=2 supersymmetric field theory}",
    eprint = "hep-th/9604034",
    archivePrefix = "arXiv",
    reportNumber = "CERN-TH-96-95, HUTP-96-A014, USC-96-008",
    doi = "10.1016/0550-3213(96)00353-7",
    journal = "Nucl. Phys. B",
    volume = "477",
    pages = "746--766",
    year = "1996"
}

@article{Katz:1997eq,
    author = "Katz, S. and Mayr, P. and Vafa, C.",
    title = "{Mirror symmetry and exact solution of 4-D N=2 gauge theories: 1.}",
    eprint = "hep-th/9706110",
    archivePrefix = "arXiv",
    reportNumber = "HUTP-97-A025, OSU-M-97-5, IASSNS-HEP-97-65",
    doi = "10.4310/ATMP.1997.v1.n1.a2",
    journal = "Adv. Theor. Math. Phys.",
    volume = "1",
    pages = "53--114",
    year = "1998"
}

@article{Katz:1996fh,
    author = "Katz, Sheldon H. and Klemm, Albrecht and Vafa, Cumrun",
    title = "{Geometric engineering of quantum field theories}",
    eprint = "hep-th/9609239",
    archivePrefix = "arXiv",
    reportNumber = "EFI-96-37, HUTP-96-A046, OSU-M-96-24",
    doi = "10.1016/S0550-3213(97)00282-4",
    journal = "Nucl. Phys. B",
    volume = "497",
    pages = "173--195",
    year = "1997"
}

@article{Marchesano:2021gyv,
    author = "Marchesano, Fernando and Prieto, David and Wiesner, Max",
    title = "{F-theory flux vacua at large complex structure}",
    eprint = "2105.09326",
    archivePrefix = "arXiv",
    primaryClass = "hep-th",
    reportNumber = "IFT-UAM/CSIC-21-58",
    doi = "10.1007/JHEP08(2021)077",
    journal = "JHEP",
    volume = "08",
    pages = "077",
    year = "2021"
}

@article{Monnee:2025ynn,
    author = "Monnee, Jeroen and Weigand, Timo and Wiesner, Max",
    title = "{Physics and Geometry of Complex Structure Limits in Type IIB Calabi-Yau Compactifications}",
    eprint = "2509.07056",
    archivePrefix = "arXiv",
    primaryClass = "hep-th",
    month = "9",
    year = "2025"
}

@article{Blanco:2025qom,
    author = "Blanco, Alejandro and Marchesano, Fernando and Melotti, Luca",
    title = "{Curvature divergences in 5d $\mathcal{N}=1$ supergravity}",
    eprint = "2505.05558",
    archivePrefix = "arXiv",
    primaryClass = "hep-th",
    doi = "10.1007/JHEP11(2025)026",
    journal = "JHEP",
    volume = "11",
    pages = "026",
    year = "2025"
}

@article{Breitenlohner:1981sm,
    author = "Breitenlohner, Peter and Sohnius, Martin F.",
    title = "{Matter Coupling and Nonlinear $\sigma$ Models in $N=2$ Supergravity}",
    reportNumber = "ICTP/80-81/10",
    doi = "10.1016/0550-3213(81)90470-3",
    journal = "Nucl. Phys. B",
    volume = "187",
    pages = "409--428",
    year = "1981"
}

@article{Grimm:1977xp,
    author = "Grimm, R. and Sohnius, M. and Wess, J.",
    title = "{Extended Supersymmetry and Gauge Theories}",
    reportNumber = "Print-77-0884 (KARLSRUHE)",
    doi = "10.1016/0550-3213(78)90303-6",
    journal = "Nucl. Phys. B",
    volume = "133",
    pages = "275--284",
    year = "1978"
}

@book{Freedman_VanProeyen_2012,
    place={Cambridge},
    title={Supergravity},
    publisher={Cambridge University Press},
    author={Freedman, Daniel Z. and Van Proeyen, Antoine},
    year={2012}
}

@book{Lauria:2020rhc,
    author = "Lauria, Edoardo and Van Proeyen, Antoine",
    title = "{${\cal N}=2$ Supergravity in $D=4,5,6$ Dimensions}",
    eprint = "2004.11433",
    archivePrefix = "arXiv",
    primaryClass = "hep-th",
    doi = "10.1007/978-3-030-33757-5",
    isbn = "978-3-030-33755-1, 978-3-030-33757-5",
    volume = "966",
    month = "3",
    year = "2020"
}

@article{Ferrara:1988ff,
    author = "Ferrara, S. and Sabharwal, S.",
    title = "{Dimensional Reduction of Type {II} Superstrings}",
    reportNumber = "CERN-TH-5219-88, UCLA-88-TEP-35",
    doi = "10.1088/0264-9381/6/4/002",
    journal = "Class. Quant. Grav.",
    volume = "6",
    pages = "L77",
    year = "1989"
}

@article{Freed:1997dp,
    author = "Freed, Daniel S.",
    title = "{Special Kahler manifolds}",
    eprint = "hep-th/9712042",
    archivePrefix = "arXiv",
    doi = "10.1007/s002200050604",
    journal = "Commun. Math. Phys.",
    volume = "203",
    pages = "31--52",
    year = "1999"
}

@article{Billo:1998yr,
    author = "Billo, Marco and Denef, Frederik and Fre, Pietro and Pesando, Igor and Troost, Walter and Van Proeyen, Antoine and Zanon, Daniela",
    title = "{The Rigid limit in special Kahler geometry: From K3 fibrations to special Riemann surfaces: A Detailed case study}",
    eprint = "hep-th/9803228",
    archivePrefix = "arXiv",
    reportNumber = "KUL-TF-98-19, IFUM-616-FT",
    doi = "10.1088/0264-9381/15/8/003",
    journal = "Class. Quant. Grav.",
    volume = "15",
    pages = "2083--2152",
    year = "1998"
}

@article{Lerche:1996ni,
    author = "Lerche, W. and Mayr, P. and Warner, N. P.",
    title = "{Noncritical strings, Del Pezzo singularities and Seiberg-Witten curves}",
    eprint = "hep-th/9612085",
    archivePrefix = "arXiv",
    reportNumber = "CERN-TH-96-326, USC-96-026",
    doi = "10.1016/S0550-3213(97)00312-X",
    journal = "Nucl. Phys. B",
    volume = "499",
    pages = "125--148",
    year = "1997"
}

@article{Candelas:1990qd,
    author = "Candelas, Philip and De la Ossa, Xenia C. and Green, Paul S. and Parkes, Linda",
    title = "{An Exactly soluble superconformal theory from a mirror pair of Calabi-Yau manifolds}",
    reportNumber = "UTTG-41-90",
    doi = "10.1016/0370-2693(91)91218-K",
    journal = "Phys. Lett. B",
    volume = "258",
    pages = "118--126",
    year = "1991"
}

@article{Andrianopoli:1996cm,
    author = "Andrianopoli, L. and Bertolini, M. and Ceresole, Anna and D'Auria, R. and Ferrara, S. and Fre, P. and Magri, T.",
    title = "{N=2 supergravity and N=2 superYang-Mills theory on general scalar manifolds: Symplectic covariance, gaugings and the momentum map}",
    eprint = "hep-th/9605032",
    archivePrefix = "arXiv",
    reportNumber = "POLFIS-TH-03-96, UCLA-96-TEP-9",
    doi = "10.1016/S0393-0440(97)00002-8",
    journal = "J. Geom. Phys.",
    volume = "23",
    pages = "111--189",
    year = "1997"
}

@article{Strominger:1990pd,
    author = "Strominger, Andrew",
    title = "{Special Geometry}",
    reportNumber = "UCSB-TH-89-61",
    doi = "10.1007/BF02096559",
    journal = "Commun. Math. Phys.",
    volume = "133",
    pages = "163--180",
    year = "1990"
}

@article{Candelas:1990pi,
    author = "Candelas, Philip and de la Ossa, Xenia",
    title = "{Moduli Space of {Calabi-Yau} Manifolds}",
    reportNumber = "UTTG-07-90",
    doi = "10.1016/0550-3213(91)90122-E",
    journal = "Nucl. Phys. B",
    volume = "355",
    pages = "455--481",
    year = "1991"
}

@article{Bastian:2021hpc,
    author = "Bastian, Brice and Grimm, Thomas W. and van de Heisteeg, Damian",
    title = "{Engineering small flux superpotentials and mass hierarchies}",
    eprint = "2108.11962",
    archivePrefix = "arXiv",
    primaryClass = "hep-th",
    doi = "10.1007/JHEP02(2023)149",
    journal = "JHEP",
    volume = "02",
    pages = "149",
    year = "2023"
}

@article{Bastian:2021eom,
    author = "Bastian, Brice and Grimm, Thomas W. and van de Heisteeg, Damian",
    title = "{Modeling General Asymptotic Calabi{\textendash}Yau Periods}",
    eprint = "2105.02232",
    archivePrefix = "arXiv",
    primaryClass = "hep-th",
    doi = "10.1002/prop.70010",
    journal = "Fortsch. Phys.",
    volume = "73",
    number = "7",
    pages = "e70010",
    year = "2025"
}

@article{Craps:1997gp,
    author = "Craps, Ben and Roose, Frederik and Troost, Walter and Van Proeyen, Antoine",
    title = "{What is special Kahler geometry?}",
    eprint = "hep-th/9703082",
    archivePrefix = "arXiv",
    reportNumber = "KUL-TF-97-10",
    doi = "10.1016/S0550-3213(97)00408-2",
    journal = "Nucl. Phys. B",
    volume = "503",
    pages = "565--613",
    year = "1997"
}

@article{Strominger:1995cz,
    author = "Strominger, Andrew",
    title = "{Massless black holes and conifolds in string theory}",
    eprint = "hep-th/9504090",
    archivePrefix = "arXiv",
    doi = "10.1016/0550-3213(95)00287-3",
    journal = "Nucl. Phys. B",
    volume = "451",
    pages = "96--108",
    year = "1995"
}

@book{Baumann:2014nda,
	Archiveprefix = {arXiv},
	Author = {Baumann, Daniel and McAllister, Liam},
	Doi = {10.1017/CBO9781316105733},
	Eprint = {1404.2601},
	Isbn = {978-1-107-08969-3, 978-1-316-23718-2},
	Month = {5},
	Primaryclass = {hep-th},
	Publisher = {Cambridge University Press},
	Series = {Cambridge Monographs on Mathematical Physics},
	Title = {{Inflation and String Theory}},
	Year = {2015}}

@article{Lanza:2022zyg,
    author = "Lanza, Stefano and Marchesano, Fernando and Martucci, Luca and Valenzuela, Irene",
    title = "{Large Field Distances from EFT strings}",
    eprint = "2205.04532",
    archivePrefix = "arXiv",
    primaryClass = "hep-th",
    doi = "10.22323/1.406.0169",
    journal = "PoS",
    volume = "CORFU2021",
    pages = "169",
    year = "2022"
}

@article{Lanza:2020qmt,
	Archiveprefix = {arXiv},
	Author = {Lanza, Stefano and Marchesano, Fernando and Martucci, Luca and Valenzuela, Irene},
	Doi = {10.1007/JHEP02(2021)006},
	Eprint = {2006.15154},
	Journal = {JHEP},
	Pages = {006},
	Primaryclass = {hep-th},
	Title = {{Swampland Conjectures for Strings and Membranes}},
	Volume = {02},
	Year = {2021}}

@article{Alim:2021vhs,
	Archiveprefix = {arXiv},
	Author = {Alim, Murad and Heidenreich, Ben and Rudelius, Tom},
	Eprint = {2108.08309},
	Month = {8},
	Primaryclass = {hep-th},
	Reportnumber = {ACFI-T21-09},
	Title = {{The Weak Gravity Conjecture and BPS Particles}},
	Year = {2021}}

@article{Lee:2019wij,
    author = "Lee, Seung-Joo and Lerche, Wolfgang and Weigand, Timo",
    title = "{Emergent strings from infinite distance limits}",
    eprint = "1910.01135",
    archivePrefix = "arXiv",
    primaryClass = "hep-th",
    reportNumber = "CERN-TH-2019-159",
    doi = "10.1007/JHEP02(2022)190",
    journal = "JHEP",
    volume = "02",
    pages = "190",
    year = "2022"
}

@article{Vafa:2005ui,
	Archiveprefix = {arXiv},
	Author = {Vafa, Cumrun},
	Eprint = {hep-th/0509212},
	Primaryclass = {hep-th},
	Reportnumber = {HUTP-05-A043},
	Slaccitation = {%%CITATION = HEP-TH/0509212;%%},
	Title = {{The String landscape and the swampland}},
	Year = {2005}}

@article{Ooguri:2006in,
	Archiveprefix = {arXiv},
	Author = {Ooguri, Hirosi and Vafa, Cumrun},
	Doi = {10.1016/j.nuclphysb.2006.10.033},
	Eprint = {hep-th/0605264},
	Journal = {Nucl.Phys.},
	Pages = {21-33},
	Primaryclass = {hep-th},
	Reportnumber = {CALT-68-2600, HUTP-06-A017},
	Slaccitation = {%%CITATION = HEP-TH/0605264;%%},
	Title = {{On the Geometry of the String Landscape and the Swampland}},
	Volume = {B766},
	Year = {2007}}

@book{Ibanez:2012zz,
	Author = {Ib{\'a}{\~n}ez, Luis E. and Uranga, Angel M.},
	Isbn = {9780521517522, 9781139227421},
	Publisher = {Cambridge University Press},
	Slaccitation = {%%CITATION = INSPIRE-1112474;%%},
	Title = {{String theory and particle physics: An introduction to string phenomenology}},
	Url = {http://www.cambridge.org/de/knowledge/isbn/item6563092/?site_locale=de_DE},
	Year = {2012}}

@article{Brennan:2017rbf,
	Archiveprefix = {arXiv},
	Author = {Brennan, T. Daniel and Carta, Federico and Vafa, Cumrun},
	Booktitle = {{Proceedings, Theoretical Advanced Study Institute in Elementary Particle Physics: Physics at the Fundamental Frontier (TASI 2017): Boulder, CO, USA, June 5-30, 2017}},
	Doi = {10.22323/1.305.0015},
	Eprint = {1711.00864},
	Journal = {PoS},
	Pages = {015},
	Primaryclass = {hep-th},
	Reportnumber = {IFT-UAM-CSIC-17-105},
	Slaccitation = {%%CITATION = ARXIV:1711.00864;%%},
	Title = {{The String Landscape, the Swampland, and the Missing Corner}},
	Volume = {TASI2017},
	Year = {2017}}

@article{Grimm:2018ohb,
	Archiveprefix = {arXiv},
	Author = {Grimm, Thomas W. and Palti, Eran and Valenzuela, Irene},
	Doi = {10.1007/JHEP08(2018)143},
	Eprint = {1802.08264},
	Journal = {JHEP},
	Pages = {143},
	Primaryclass = {hep-th},
	Slaccitation = {%%CITATION = ARXIV:1802.08264;%%},
	Title = {{Infinite Distances in Field Space and Massless Towers of States}},
	Volume = {08},
	Year = {2018}}

@article{Gendler:2020dfp,
	Archiveprefix = {arXiv},
	Author = {Gendler, Naomi and Valenzuela, Irene},
	Doi = {10.1007/JHEP01(2021)176},
	Eprint = {2004.10768},
	Journal = {JHEP},
	Pages = {176},
	Primaryclass = {hep-th},
	Title = {{Merging the weak gravity and distance conjectures using BPS extremal black holes}},
	Volume = {01},
	Year = {2021}}

@article{Grimm:2018cpv,
	Archiveprefix = {arXiv},
	Author = {Grimm, Thomas W. and Li, Chongchuo and Palti, Eran},
	Doi = {10.1007/JHEP03(2019)016},
	Eprint = {1811.02571},
	Journal = {JHEP},
	Pages = {016},
	Primaryclass = {hep-th},
	Reportnumber = {MPP-2018-260},
	Title = {{Infinite Distance Networks in Field Space and Charge Orbits}},
	Volume = {03},
	Year = {2019}}

@article{Corvilain:2018lgw,
	Archiveprefix = {arXiv},
	Author = {Corvilain, Pierre and Grimm, Thomas W. and Valenzuela, Irene},
	Doi = {10.1007/JHEP08(2019)075},
	Eprint = {1812.07548},
	Journal = {JHEP},
	Pages = {075},
	Primaryclass = {hep-th},
	Title = {{The Swampland Distance Conjecture for K\"ahler moduli}},
	Volume = {08},
	Year = {2019}}

@article{Baume:2019sry,
	Archiveprefix = {arXiv},
	Author = {Baume, Florent and Marchesano, Fernando and Wiesner, Max},
	Doi = {10.1007/JHEP04(2020)174},
	Eprint = {1912.02218},
	Journal = {JHEP},
	Pages = {174},
	Primaryclass = {hep-th},
	Reportnumber = {IFT-UAM/CSIC-19-161},
	Title = {{Instanton Corrections and Emergent Strings}},
	Volume = {04},
	Year = {2020}}

@article{Bergshoeff:2001pv,
	Archiveprefix = {arXiv},
	Author = {Bergshoeff, Eric and Kallosh, Renata and Ortin, Tomas and Roest, Diederik and Van Proeyen, Antoine},
	Doi = {10.1088/0264-9381/18/17/303},
	Eprint = {hep-th/0103233},
	Journal = {Class. Quant. Grav.},
	Pages = {3359--3382},
	Reportnumber = {UG-00-15, SU-ITP-01-09, IFT-UAM-CSIC-00-35, KUL-TF-01-06},
	Title = {{New formulations of D = 10 supersymmetry and D8 - O8 domain walls}},
	Volume = {18},
	Year = {2001}}

@article{Lanza:2021udy,
    author = "Lanza, Stefano and Marchesano, Fernando and Martucci, Luca and Valenzuela, Irene",
    title = "{The EFT stringy viewpoint on large distances}",
    eprint = "2104.05726",
    archivePrefix = "arXiv",
    primaryClass = "hep-th",
    doi = "10.1007/JHEP09(2021)197",
    journal = "JHEP",
    volume = "09",
    pages = "197",
    year = "2021"
}

@article{Candelas:1990rm,
    author = "Candelas, Philip and De La Ossa, Xenia C. and Green, Paul S. and Parkes, Linda",
    editor = "Yau, Shing-Tung",
    title = "{A Pair of Calabi-Yau manifolds as an exactly soluble superconformal theory}",
    reportNumber = "UTTG-25-90",
    doi = "10.1016/0550-3213(91)90292-6",
    journal = "Nucl. Phys. B",
    volume = "359",
    pages = "21--74",
    year = "1991"
}

@article{Ceresole:1995ca,
    author = "Ceresole, Anna and D'Auria, R. and Ferrara, S.",
    editor = "Gava, E. and Narain, K. S. and Vafa, C.",
    title = "{The Symplectic structure of N=2 supergravity and its central extension}",
    eprint = "hep-th/9509160",
    archivePrefix = "arXiv",
    reportNumber = "POLFIS-TH-10-95, CERN-TH-95-244",
    doi = "10.1016/0920-5632(96)00008-4",
    journal = "Nucl. Phys. B Proc. Suppl.",
    volume = "46",
    pages = "67--74",
    year = "1996"
}

@article{Marchesano:2022axe,
    author = "Marchesano, Fernando and Melotti, Luca",
    title = "{EFT strings and emergence}",
    eprint = "2211.01409",
    archivePrefix = "arXiv",
    primaryClass = "hep-th",
    doi = "10.1007/JHEP02(2023)112",
    journal = "JHEP",
    volume = "02",
    pages = "112",
    year = "2023"
}

@article{Candelas:1993dm,
    author = "Candelas, Philip and De La Ossa, Xenia and Font, Anamaria and Katz, Sheldon H. and Morrison, David R.",
    editor = "Greene, B. and Yau, Shing-Tung",
    title = "{Mirror symmetry for two parameter models. 1.}",
    eprint = "hep-th/9308083",
    archivePrefix = "arXiv",
    reportNumber = "CERN-TH-6884-93, UTTG-15-93, NEIP-93-005, OSU-M-93-1",
    doi = "10.1016/0550-3213(94)90322-0",
    journal = "Nucl. Phys. B",
    volume = "416",
    pages = "481--538",
    year = "1994"
}

@article{Palti:2019pca,
    author = "Palti, Eran",
    title = "{The Swampland: Introduction and Review}",
    eprint = "1903.06239",
    archivePrefix = "arXiv",
    primaryClass = "hep-th",
    reportNumber = "MPP-2019-53",
    doi = "10.1002/prop.201900037",
    journal = "Fortsch. Phys.",
    volume = "67",
    number = "6",
    pages = "1900037",
    year = "2019"
}

@article{Curio:2000sc,
    author = "Curio, Gottfried and Klemm, Albrecht and Lust, Dieter and Theisen, Stefan",
    title = "{On the vacuum structure of type II string compactifications on Calabi-Yau spaces with H fluxes}",
    eprint = "hep-th/0012213",
    archivePrefix = "arXiv",
    reportNumber = "HU-EP-00-58, AEI-2000-84",
    doi = "10.1016/S0550-3213(01)00285-1",
    journal = "Nucl. Phys. B",
    volume = "609",
    pages = "3--45",
    year = "2001"
}

@article{Kachru:1995wm,
    author = "Kachru, Shamit and Vafa, Cumrun",
    title = "{Exact results for N=2 compactifications of heterotic strings}",
    eprint = "hep-th/9505105",
    archivePrefix = "arXiv",
    reportNumber = "HUTP-95-A016",
    doi = "10.1016/0550-3213(95)00307-E",
    journal = "Nucl. Phys. B",
    volume = "450",
    pages = "69--89",
    year = "1995"
}

@article{Marchesano:2019ifh,
    author = "Marchesano, Fernando and Wiesner, Max",
    title = "{Instantons and infinite distances}",
    eprint = "1904.04848",
    archivePrefix = "arXiv",
    primaryClass = "hep-th",
    reportNumber = "IFT-UAM/CSIC-19-049",
    doi = "10.1007/JHEP08(2019)088",
    journal = "JHEP",
    volume = "08",
    pages = "088",
    year = "2019"
}

@article{Hosono:1994av,
    author = "Hosono, S. and Klemm, A. and Theisen, S.",
    editor = {Alekseev, Anton and Hietam{\"a}ki, Antero and Huitu, Katri and Morozov, Alexei and Niemi, Antti},
    title = "{Lectures on mirror symmetry}",
    eprint = "hep-th/9403096",
    archivePrefix = "arXiv",
    reportNumber = "HUTMP-94-01, LMU-TPW-94-02",
    doi = "10.1007/3-540-58453-6_13",
    journal = "Lect. Notes Phys.",
    volume = "436",
    pages = "235--280",
    year = "1994"
}

@article{Hosono:1993qy,
    author = "Hosono, S. and Klemm, A. and Theisen, S. and Yau, Shing-Tung",
    title = "{Mirror symmetry, mirror map and applications to Calabi-Yau hypersurfaces}",
    eprint = "hep-th/9308122",
    archivePrefix = "arXiv",
    reportNumber = "HUTMP-93-0801, LMU-TPW-93-22",
    doi = "10.1007/BF02100589",
    journal = "Commun. Math. Phys.",
    volume = "167",
    pages = "301--350",
    year = "1995"
}

@book{Cecotti:2015wqa,
    author = "Cecotti, Sergio",
    title = "{Supersymmetric Field Theories}: {Geometric Structures and Dualities}",
    isbn = "978-1-107-05381-6, 978-1-316-21359-9",
    publisher = "Cambridge University Press",
    month = "1",
    year = "2015"
}

@article{vandeHeisteeg:2022btw,
    author = "van de Heisteeg, Damian and Vafa, Cumrun and Wiesner, Max and Wu, David H.",
    title = "{Moduli-dependent species scale}",
    eprint = "2212.06841",
    archivePrefix = "arXiv",
    primaryClass = "hep-th",
    doi = "10.4310/bpam.2024.v1.n1.a1",
    journal = "Beijing J. Pure Appl. Math.",
    volume = "1",
    number = "1",
    pages = "1--41",
    year = "2024"
}

@article{vanBeest:2021lhn,
    author = "van Beest, Marieke and Calder\'on-Infante, Jos\'e and Mirfendereski, Delaram and Valenzuela, Irene",
    title = "{Lectures on the Swampland Program in String Compactifications}",
    eprint = "2102.01111",
    archivePrefix = "arXiv",
    primaryClass = "hep-th",
    doi = "10.1016/j.physrep.2022.09.002",
    journal = "Phys. Rept.",
    volume = "989",
    pages = "1--50",
    year = "2022"
}

@article{Marchesano:2023thx,
    author = "Marchesano, Fernando and Melotti, Luca and Paoloni, Lorenzo",
    title = "{On the moduli space curvature at infinity}",
    eprint = "2311.07979",
    archivePrefix = "arXiv",
    primaryClass = "hep-th",
    doi = "10.1007/JHEP02(2024)103",
    journal = "JHEP",
    volume = "02",
    pages = "103",
    year = "2024"
}

@article{Escobar:2018rna,
    author = "Escobar, Dagoberto and Marchesano, Fernando and Staessens, Wieland",
    title = "{Type IIA flux vacua and $\alpha'$-corrections}",
    eprint = "1812.08735",
    archivePrefix = "arXiv",
    primaryClass = "hep-th",
    reportNumber = "IFT-UAM/CSIC-18-131",
    doi = "10.1007/JHEP06(2019)129",
    journal = "JHEP",
    volume = "06",
    pages = "129",
    year = "2019"
}

@article{Eguchi:2007iw,
    author = "Eguchi, Tohru and Tachikawa, Yuji",
    title = "{Rigid limit in N=2 supergravity and weak-gravity conjecture}",
    eprint = "0706.2114",
    archivePrefix = "arXiv",
    primaryClass = "hep-th",
    reportNumber = "UT-07-21, YITP-07-35",
    doi = "10.1088/1126-6708/2007/08/068",
    journal = "JHEP",
    volume = "08",
    pages = "068",
    year = "2007"
}

@article{Kachru:1995fv,
    author = "Kachru, Shamit and Klemm, Albrecht and Lerche, Wolfgang and Mayr, Peter and Vafa, Cumrun",
    title = "{Nonperturbative results on the point particle limit of N=2 heterotic string compactifications}",
    eprint = "hep-th/9508155",
    archivePrefix = "arXiv",
    reportNumber = "CERN-TH-95-231, HUTP-95-A032",
    doi = "10.1016/0550-3213(95)00574-9",
    journal = "Nucl. Phys. B",
    volume = "459",
    pages = "537--558",
    year = "1996"
}

@article{Seiberg:1994rs,
    author = "Seiberg, N. and Witten, Edward",
    title = "{Electric - magnetic duality, monopole condensation, and confinement in N=2 supersymmetric Yang-Mills theory}",
    eprint = "hep-th/9407087",
    archivePrefix = "arXiv",
    reportNumber = "RU-94-52, IASSNS-HEP-94-43",
    doi = "10.1016/0550-3213(94)90124-4",
    journal = "Nucl. Phys. B",
    volume = "426",
    pages = "19--52",
    year = "1994",
    note = "[Erratum: Nucl.Phys.B 430, 485--486 (1994)]"
}

@article{Arkani-Hamed:2006emk,
    author = "Arkani-Hamed, Nima and Motl, Lubos and Nicolis, Alberto and Vafa, Cumrun",
    title = "{The String landscape, black holes and gravity as the weakest force}",
    eprint = "hep-th/0601001",
    archivePrefix = "arXiv",
    reportNumber = "HUTP-05-A0057",
    doi = "10.1088/1126-6708/2007/06/060",
    journal = "JHEP",
    volume = "06",
    pages = "060",
    year = "2007"
}

@article{tHooft:1974kcl,
    author = "'t Hooft, Gerard",
    editor = "Taylor, J. C.",
    title = "{Magnetic Monopoles in Unified Gauge Theories}",
    reportNumber = "CERN-TH-1876",
    doi = "10.1016/0550-3213(74)90486-6",
    journal = "Nucl. Phys. B",
    volume = "79",
    pages = "276--284",
    year = "1974"
}

@article{Polyakov:1974ek,
    author = "Polyakov, Alexander M.",
    editor = "Taylor, J. C.",
    title = "{Particle Spectrum in Quantum Field Theory}",
    reportNumber = "PRINT-74-1566 (LANDAU-INST)",
    journal = "JETP Lett.",
    volume = "20",
    pages = "194--195",
    year = "1974"
}

@article{FierroCota:2023bsp,
    author = "Fierro Cota, Cesar and Mininno, Alessandro and Weigand, Timo and Wiesner, Max",
    title = "{The minimal weak gravity conjecture}",
    eprint = "2312.04619",
    archivePrefix = "arXiv",
    primaryClass = "hep-th",
    reportNumber = "ZMP-HH/23-21",
    doi = "10.1007/JHEP05(2024)285",
    journal = "JHEP",
    volume = "05",
    pages = "285",
    year = "2024"
}

@article{Bastian:2020egp,
    author = "Bastian, Brice and Grimm, Thomas W. and van de Heisteeg, Damian",
    title = "{Weak gravity bounds in asymptotic string compactifications}",
    eprint = "2011.08854",
    archivePrefix = "arXiv",
    primaryClass = "hep-th",
    doi = "10.1007/JHEP06(2021)162",
    journal = "JHEP",
    volume = "06",
    pages = "162",
    year = "2021"
}

@article{Bastian:2023shf,
    author = "Bastian, Brice and van de Heisteeg, Damian and Schlechter, Lorenz",
    title = "{Beyond Large Complex Structure: Quantized Periods and Boundary Data for One-Modulus Singularities}",
    eprint = "2306.01059",
    archivePrefix = "arXiv",
    primaryClass = "hep-th",
    month = "6",
    year = "2023"
}

@article{Gopakumar:1997dv,
    author = "Gopakumar, Rajesh and Vafa, Cumrun",
    title = "{Branes and fundamental groups}",
    eprint = "hep-th/9712048",
    archivePrefix = "arXiv",
    reportNumber = "HUTP-97-A103, UCSB-97-23",
    doi = "10.4310/ATMP.1998.v2.n2.a7",
    journal = "Adv. Theor. Math. Phys.",
    volume = "2",
    pages = "399--411",
    year = "1998"
}

@article{Sherman-Morrison1,
author = {Jack Sherman and Winifred J. Morrison},
title = {{Adjustment of an Inverse Matrix Corresponding to Changes in the Elements of a Given Column or a Given Row of the Original Matrix}},
volume = {20},
journal = {The Annals of Mathematical Statistics},
number = {4},
publisher = {Institute of Mathematical Statistics},
pages = {620 -- 624},
year = {1949},
doi = {10.1214/aoms/1177729959},
URL = {https://doi.org/10.1214/aoms/1177729959}
}

@article{Sherman-Morrison2,
author = {Jack Sherman and Winifred J. Morrison},
title = {{Adjustment of an Inverse Matrix Corresponding to a Change in One Element of a Given Matrix}},
volume = {21},
journal = {The Annals of Mathematical Statistics},
number = {1},
publisher = {Institute of Mathematical Statistics},
pages = {124 -- 127},
year = {1950},
doi = {10.1214/aoms/1177729893},
URL = {https://doi.org/10.1214/aoms/1177729893}
}

@article{Castellano:2024gwi,
    author = "Castellano, Alberto and Marchesano, Fernando and Melotti, Luca and Paoloni, Lorenzo",
    title = "{The Moduli Space Curvature and the Weak Gravity Conjecture}",
    eprint = "2410.10966",
    archivePrefix = "arXiv",
    primaryClass = "hep-th",
    month = "10",
    year = "2024"
}

@article{Demirtas:2020ffz,
    author = "Demirtas, Mehmet and Kim, Manki and McAllister, Liam and Moritz, Jakob",
    title = "{Conifold Vacua with Small Flux Superpotential}",
    eprint = "2009.03312",
    archivePrefix = "arXiv",
    primaryClass = "hep-th",
    doi = "10.1002/prop.202000085",
    journal = "Fortsch. Phys.",
    volume = "68",
    pages = "2000085",
    year = "2020"
}

@article{Alvarez-Garcia:2020pxd,
    author = "\'Alvarez-Garc\'\i{}a, Rafael and Blumenhagen, Ralph and Brinkmann, Max and Schlechter, Lorenz",
    title = "{Small Flux Superpotentials for Type IIB Flux Vacua Close to a Conifold}",
    eprint = "2009.03325",
    archivePrefix = "arXiv",
    primaryClass = "hep-th",
    reportNumber = "MPP-2020-165",
    doi = "10.1002/prop.202000088",
    journal = "Fortsch. Phys.",
    volume = "68",
    pages = "2000088",
    year = "2020"
}

@article{Demirtas:2019sip,
    author = "Demirtas, Mehmet and Kim, Manki and Mcallister, Liam and Moritz, Jakob",
    title = "{Vacua with Small Flux Superpotential}",
    eprint = "1912.10047",
    archivePrefix = "arXiv",
    primaryClass = "hep-th",
    doi = "10.1103/PhysRevLett.124.211603",
    journal = "Phys. Rev. Lett.",
    volume = "124",
    number = "21",
    pages = "211603",
    year = "2020"
}

@article{Marchesano:2024tod,
    author = "Marchesano, Fernando and Melotti, Luca and Wiesner, Max",
    title = "{Asymptotic curvature divergences and non-gravitational theories}",
    eprint = "2409.02991",
    archivePrefix = "arXiv",
    primaryClass = "hep-th",
    doi = "10.1007/JHEP02(2025)151",
    journal = "JHEP",
    volume = "02",
    pages = "151",
    year = "2025"
}

@article{Herraez:2018vae,
    author = "Herraez, Alvaro and Ibanez, Luis E. and Marchesano, Fernando and Zoccarato, Gianluca",
    title = "{The Type IIA Flux Potential, 4-forms and Freed-Witten anomalies}",
    eprint = "1802.05771",
    archivePrefix = "arXiv",
    primaryClass = "hep-th",
    reportNumber = "IFT-UAM/CSIC-18-016, IFT-UAM-CSIC-18-016",
    doi = "10.1007/JHEP09(2018)018",
    journal = "JHEP",
    volume = "09",
    pages = "018",
    year = "2018"
}

@misc{kerr2019,
      title={Polarized relations on horizontal SL(2)s}, 
      author={Matt Kerr and Gregory Pearlstein and Colleen Robles},
      year={2019},
      eprint={1705.03117},
      archivePrefix={arXiv},
      primaryClass={math.AG},
      url={https://arxiv.org/abs/1705.03117}, 
}

@article{Mayr:2000as,
    author = "Mayr, P.",
    title = "{Phases of supersymmetric D-branes on Kahler manifolds and the McKay correspondence}",
    eprint = "hep-th/0010223",
    archivePrefix = "arXiv",
    reportNumber = "CERN-TH-2000-315",
    doi = "10.1088/1126-6708/2001/01/018",
    journal = "JHEP",
    volume = "01",
    pages = "018",
    year = "2001"
}

@article{Coudarchet:2023mmm,
    author = "Coudarchet, Thibaut and Marchesano, Fernando and Prieto, David and Urkiola, Mikel A.",
    title = "{Symmetric fluxes and small tadpoles}",
    eprint = "2304.04789",
    archivePrefix = "arXiv",
    primaryClass = "hep-th",
    doi = "10.1007/JHEP08(2023)016",
    journal = "JHEP",
    volume = "08",
    pages = "016",
    year = "2023"
}

@phdthesis{Castellano:2024bna,
    author = "Castellano, Alberto",
    title = "{The Quantum Gravity Scale and the Swampland}",
    eprint = "2409.10003",
    archivePrefix = "arXiv",
    primaryClass = "hep-th",
    school = "U. Autonoma, Madrid (main)",
    year = "2024"
}

@article{Grana:2021zvf,
    author = "Gra\~na, Mariana and Herr\'aez, Alvaro",
    title = "{The Swampland Conjectures: A Bridge from Quantum Gravity to Particle Physics}",
    eprint = "2107.00087",
    archivePrefix = "arXiv",
    primaryClass = "hep-th",
    doi = "10.3390/universe7080273",
    journal = "Universe",
    volume = "7",
    number = "8",
    pages = "273",
    year = "2021"
}

@article{Agmon:2022thq,
    author = "Agmon, Nathan Benjamin and Bedroya, Alek and Kang, Monica Jinwoo and Vafa, Cumrun",
    title = "{Lectures on the string landscape and the Swampland}",
    eprint = "2212.06187",
    archivePrefix = "arXiv",
    primaryClass = "hep-th",
    month = "12",
    year = "2022"
}

@article{VanRiet:2023pnx,
    author = "Van Riet, Thomas and Zoccarato, Gianluca",
    title = "{Beginners lectures on flux compactifications and related Swampland topics}",
    eprint = "2305.01722",
    archivePrefix = "arXiv",
    primaryClass = "hep-th",
    doi = "10.1016/j.physrep.2023.11.003",
    journal = "Phys. Rept.",
    volume = "1049",
    pages = "1--51",
    year = "2024"
}

@article{Harlow:2022ich,
    author = "Harlow, Daniel and Heidenreich, Ben and Reece, Matthew and Rudelius, Tom",
    title = "{Weak gravity conjecture}",
    eprint = "2201.08380",
    archivePrefix = "arXiv",
    primaryClass = "hep-th",
    reportNumber = "ACFI-T22-01",
    doi = "10.1103/RevModPhys.95.035003",
    journal = "Rev. Mod. Phys.",
    volume = "95",
    number = "3",
    pages = "035003",
    year = "2023"
}

\end{document}